\documentclass[%
 reprint,
superscriptaddress,
 amsmath,amssymb,
 aps,
pre,longbibliography
]{revtex4-1}

\usepackage{xcolor}
\usepackage{graphicx}
\usepackage{dcolumn}
\usepackage{bm}
\usepackage{hyperref}
\usepackage{physics}

\hypersetup{
  colorlinks=true
}

\newcommand{\p}{\partial}

\newcommand{\sgn}{\mathrm{sgn}}

\newcommand{\bs}{\boldsymbol}
\newcommand{\rmd}{\mathrm{d}}
\newcommand{\la}{\lambda}
\newcommand{\ud}{\mathrm{d}}

\def\m{\mu}\newcommand{\G}{\Gamma}
\def\C{{\mathbb C}}

\def\be{\begin{equation}}
\def\ee{\end{equation}}
\def\bea{\begin{eqnarray}}
\def\eea{\end{eqnarray}}

\def\frc#1#2{\frac{#1}{#2}}

\newcommand{\aver}[1]{\left\langle #1 \right\rangle}

\begin{document}


\title{Soliton Gas: Theory, Numerics and Experiments}


\author{Pierre Suret}
\email[Corresponding author : ]{Pierre.Suret@univ-lille.fr}
\affiliation{Univ. Lille, CNRS, UMR 8523 - PhLAM -  Physique des Lasers Atomes et Mol\'ecules, F-59000 Lille, France}

\author{Stephane Randoux}
\affiliation{Univ. Lille, CNRS, UMR 8523 - PhLAM -  Physique des Lasers Atomes et Mol\'ecules, F-59000 Lille, France}

\author{Andrey Gelash}
\affiliation{Laboratoire Interdisciplinaire Carnot de Bourgogne (ICB), UMR 6303 CNRS-Université Bourgogne Franche-Comté, 21078 Dijon, France}

\author{Dmitry Agafontsev}
\affiliation{Shirshov Institute of Oceanology of RAS,
Nakhimovskiy prosp. 36, Moscow, 117997, Russia}
\affiliation{Department of Mathematics, Physics and Electrical Engineering, Northumbria University, Newcastle upon Tyne, NE1 8ST, United Kingdom}

\author{Benjamin Doyon}
\affiliation{Department of Mathematics, King’s College London, Strand WC2R 2LS, London, United Kingdom}

\author{Gennady El}
\affiliation{Department of Mathematics, Physics and Electrical Engineering, Northumbria University, Newcastle upon Tyne, NE1 8ST, United Kingdom}

\date{\today}

\begin{abstract}

The concept of soliton gas was introduced in 1971 by V. Zakharov as an infinite collection of weakly interacting solitons in the framework of Korteweg-de Vries (KdV) equation. In this theoretical construction of a diluted soliton gas, solitons with random parameters are almost non-overlapping. More recently, the concept has been extended to dense gases in which solitons strongly and continuously interact.  The notion of soliton gas is inherently associated with integrable wave systems described by nonlinear partial differential equations like the KdV equation or the one-dimensional nonlinear Schr\"odinger equation that can be solved using the inverse scattering transform. Over the last few years, the field of soliton gases has received a rapidly growing interest from both the theoretical and  experimental points of view. In particular, it has been realized that the soliton gas dynamics underlies some fundamental nonlinear wave phenomena such as spontaneous modulation instability and the formation of rogue waves. The recently discovered deep connections of soliton gas theory with generalized hydrodynamics have  broadened the field and opened new fundamental questions related to the soliton gas statistics and thermodynamics. We review the main recent theoretical and experimental results in the field of soliton gas. The key conceptual tools of the field, such as the inverse scattering transform, the thermodynamic limit of finite-gap potentials and the Generalized Gibbs Ensembles are introduced and various open questions and future challenges are discussed.

\end{abstract}

\pacs{Valid PACS appear here}
\maketitle

\tableofcontents
\newpage

\section{Introduction}

Random  nonlinear  waves in dispersive media have been the subject of  
intense research in nonlinear physics for more than  half a century, most notably
in the contexts of water wave dynamics and nonlinear optics.
A significant portion of the work in this area has
been centred around wave turbulence---the theory of out of equilibrium random weakly nonlinear dispersive
waves in non-integrable systems \cite{Zakharov, nazarenko_wave_2011}. One of the  most important results of the wave turbulence theory is the analytical determination in  \cite{zakharov_weak_1965}  of  the power-law Fourier spectra analogous  to the Kolmogorov spectra  describing energy flux  through scales in dissipative hydrodynamic turbulence. 

More recently, a new theme in turbulence theory has emerged in
connection with the dynamics of strongly nonlinear random waves described by
integrable systems such as  the Korteweg-de Vries (KdV) and 1D nonlinear Schr\"odinger (NLS) equations.  This kind of random wave motion in nonlinear conservative systems,  dubbed   {\it integrable turbulence} \cite{Zakharov:09},  has attracted significant attention  from both  fundamental and applied perspectives.
 The interest in integrable turbulence is motivated by the inherent
randomness of many real-life systems (due to random initial and boundary conditions or to complex interaction mechanisms)  even though the underlying physical models may be amenable to the well-established mathematical techniques of integrable systems theory such as the inverse scattering transform  or finite-gap theory  \cite{novikov1984theory, Osborne}.   

The integrable turbulence framework is particularly pertinent to the description of modulationally unstable systems which can exhibit highly complex nonlinear behaviours  that can be adequately described  in terms of the turbulence theory concepts such as  probability distribution functions, ensemble averages, Fourier  spectra etc. \cite{Agafontsev:15, Agafontsev:16, Randoux2016Nonlinear, kraych2019statistical, copie2020physics, Suret2016Single}. We stress that the term `turbulence' in this context is understood as complex spatiotemporal  dynamics that require probabilistic description and are not related to the energy cascades through scales, the prime feature of  strong hydrodynamic and weak wave turbulence. 

Along with the fundamental, conceptual significance, the physical relevance of integrable turbulence is supported by recent laboratory experiments \cite{Walczak2015Optical, Suret2016Single, Narhi:16, Tikan2018Single, Redor:20, Suret:20, lebel2021single} and observations of natural wave phenomena, e.g. in ocean waves \cite{Costa:14, osborne_highly_2019}.

The main tool for the analysis of integrable nonlinear dispersive partial differential equations (PDEs) is the Inverse Scattering Transform (IST) \cite{gardner_method_1967} which is based on the reformulation of a nonlinear PDE as a compatibility condition of two {\it linear} problems (the so-called Lax pair): a stationary spectral (scattering) problem and an evolution problem---for the same auxiliary function.  Within the classical IST setting, formulated for the wave fields decaying sufficiently rapidly as $|x| \to \infty$ the scattering spectrum consists of two components: discrete and continuous, corresponding to two contrasting types of the wave motion: solitary waves (solitons) and dispersive radiation respectively. Importantly, integrable evolution preserves the IST spectrum in time.

Localized nonlinear solitary waves, termed solitons in the context of integrable systems,  are a ubiquitous  and fundamental feature of nonlinear dispersive wave propagation. They exhibit particle-like properties such as elastic, pairwise
interactions accompanied by certain phase/position shifts \cite{zabusky_interaction_1965} and have been extensively studied both 
theoretically \cite{ablowitz_inverse_1974, novikov1984theory, newell_solitons_1985} and experimentally \cite{remoissenet_waves_2013}.  
 The particle-like properties of solitons suggest some natural questions pertaining to
the realm of statistical mechanics, e.g. one can consider a {\it soliton gas} as an infinite
ensemble of interacting solitons characterised by random amplitude and phase
 distributions. Then, given the properties of the elementary,
`microscopic', soliton interactions the next natural step is the determination of the emergent, out of equilibrium macroscopic dynamics
(i.e. hydrodynamics or kinetics) of a soliton gas.  

 Due to the presence of an infinite number of conserved quantities, integrable systems do not reach the  thermodynamic equilibrium state characterized by the so-called Rayleigh-Jeans distribution of the modes (equipartition of energy). Consequently, the properties of soliton gases will be very different compared to the properties of classical
gases whose particle interactions are non-elastic. Additionally, the particle-wave duality of  solitons implies  that hydrodynamic description of a soliton gas should be complemented by the  characterisation of the associated nonlinear  turbulent wave field  in terms of the  probability density function, power spectrum, correlations etc.   
It has  transpired recently that soliton gas dynamics  are instrumental in the understanding of a number of important statistical nonlinear wave phenomena such as spontaneous modulational instability and the rogue wave formation in one-dimensional wave propagation~\cite{gelash2019bound}. 
 
\begin{figure}[h]
\centering
  \includegraphics[width=8.9cm]{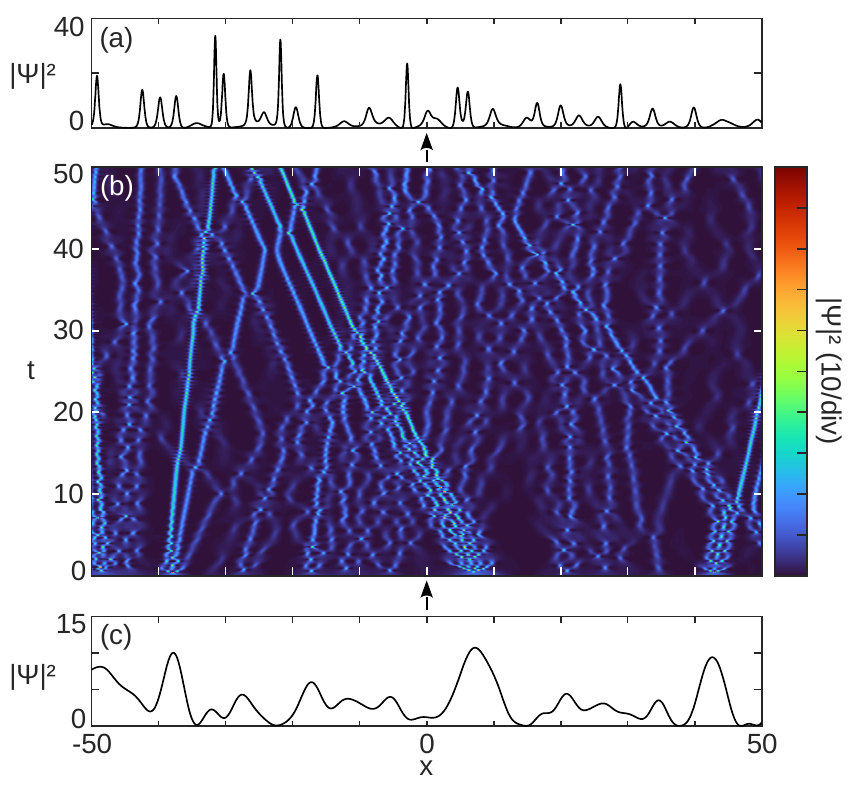}    
    \caption {Emergence of soliton gas in the long-time evolution of a partially coherent wave $\psi(x,t)$ in the focusing NLS equation (equation \eqref{eq:fNLS} with $\sigma=1$). (a) Intensity $|\psi(x,t=50)|^2$ (b) Spatiotemporal dynamics$ |\psi(x,t)|^2$ (c) Intensity $|\psi(x,t=0)|^2$ (initial condition)}
    \label{fig:PCWdynamics}  
  \end{figure}

 One can distinguish two basic mechanisms of the `spontaneous', uncontrollable  generation of a soliton gas.  One mechanism involves the process of soliton fission, where statistical soliton ensembles  emerge as the asymptotic outcome of long-time evolution of the so-called `partially coherent waves', which can be viewed as  collections of randomly distributed  broad pulses,  see Fig.~\ref{fig:PCWdynamics} and Refs.~\cite{Walczak2015Optical, Suret2016Single, Tikan2018Single}.  Alternatively, soliton ensembles can be initially generated from a non-random (e.g. periodic)  signal  and then undergo effective randomization due to elastic reflections from the boundaries and subsequent multiple collisions, see \cite{Redor:20} for the example of the soliton gas generation in a shallow-water wave tank.
  The second mechanism of the soliton gas generation is related to the already mentioned phenomenon of modulational instability, where the basic coherent nonlinear mode of an unstable system---the plane wave---is subjected to a random perturbation (a noise), resulting in the development of large-amplitude small-scale fluctuations of the wave field and the establishment at $t \to \infty$ of a stationary integrable turbulence \cite{Agafontsev:15}.   It was shown in \cite{gelash2019bound} that for the wave systems described by the focusing nonlinear Schr\"odinger (fNLS) equation such integrable turbulence exhibits the properties of a dense bound-state (non-propagating) soliton gas.
 
 Soliton gas can also be synthesized  in a controllable manner directly, e.g. by programming a water tank wavemaker according to the IST-prescribed random multi-soliton solution of the relevant integrable equation, see \cite{Suret:20}.

The theoretical concept of soliton gas  was introduced  by V.E.  Zakharov in  1971   \cite{zakharov1971kinetic}, where an approximate kinetic equation for KdV solitons was constructed by evaluating  the effective adjustment to the soliton's  velocity in a {\it rarefied gas} due to the infrequent interactions (collisions) between individual  solitons, accompanied by the well-defined phase shifts. The central object in the soliton gas theory is the density of states (DOS)---
the function describing the distribution of solitons with respect to the spectral parameter and the positions of the soliton's centres. When soliton gas is uniform (i.e. in an equilibrium state) the DOS is stationary and space-independent. In a non-uniform (non-equilibrium) gas the spatiotemporal evolution of the DOS on a large (Eulerian) scale is described by a continuity equation following from the isospectrality of integrable dynamics. 
 
In a rarefied gas  solitons are treated as isolated  point-like  quasi-particles.   
 In contrast, in  a {\it dense} soliton  gas the solitons  exhibit significant overlap and, as a result, are continuously involved in a strong nonlinear interaction with each other.  It is clear that, in a dense gas the particle interpretation of individual solitons  becomes less  transparent and the wave aspect of the collective soliton  dynamics comes to the fore. Indeed, a consistent generalization of Zakharov's kinetic equation for  KdV solitons  to the case of  a dense soliton gas  has been achieved in \cite{el_thermodynamic_2003} in the framework of the nonlinear wave modulation (Whitham) theory \cite{whitham_linear_1999}.  It was proposed in \cite{el_thermodynamic_2003}  that the KdV soliton gas can be modelled by the  thermodynamic type solitonic limit of the  multiphase, finite-gap KdV solutions and their modulations \cite{flaschka_multiphase_1980} (these solutions represent nontrivial generalization of solitons in problems with periodic boundary conditions). The resulting spectral kinetic equation  has the form of a nonlinear integro-differential equation consisting of the continuity equation for the DOS (equation \eqref{kin_eq0}) and the linear integral equation of state \eqref{eq_state_kdv} relating the effective, average velocity of the `tracer' soliton in the gas with its DOS. The structure of the kinetic equation derived in \cite{el_thermodynamic_2003} has motivated a fundamental conjecture that generally,  in a dense gas the net effect of soliton interactions can be formally evaluated using the same phase-shift argument  that was used in  the original rarefied gas theory \cite{zakharov1971kinetic}.  This conjecture, termed  {\it collision rate ansatz}, has enabled an effective phenomenological theory  of  a dense soliton gas  for the focusing fNLS   equation \cite{el_kinetic_2005} and more recently, for the defocusing NLS and integrable shallow water waves equations supporting bidirectional soliton propagation  \cite{congy_soliton_2021}.  The phenomenological soliton gas theory for the fNLS equation proposed in \cite{el_kinetic_2005} has been rigorously  confirmed and substantially extended in \cite{el_spectral_2020} within the framework of the thermodynamic limit of spectral finite-gap  solutions of the fNLS equation and their modulations. This latter work has revealed a number of  new soliton gas phenomena due to a  very different structure of the spectral phase space of the fNLS equation compared to the KdV equation. In particular, the generalization of soliton gas, termed {\it breather gas}, was introduced  by considering a special family of fNLS solitonic solutions on a non-zero unstable background.  
 Another peculiar type of soliton gas termed in \cite{el_spectral_2020} {\it soliton condensate} can be viewed as the densest possible ensemble of solitons constrained by a given spectral domain. Properties of soliton condensates for the KdV equation and their relation to the fundamental coherent structures in dispersive hydrodynamics such as rarefaction and dispersive shock waves were investigated in \cite{congy_dispersive_2022}.

Apart from the above line of  research  on  soliton gases  inspired by the Zakharov 1971 work  and summarized in the recent review \cite{el_soliton_2021} there have been many other developments---theoretical, numerical and experimental ---exploring  various aspects of  soliton gas/soliton turbulence  dynamics in both integrable and nonintegrable classical wave systems (see e.g.  \cite{meiss_drift-wave_1982, Schwache:97,  schmidt_non-thermal_2012, turitsyna_laminarturbulent_2013, dutykh_numerical_2014, Akhmediev2016breather, giovanangeli_soliton_2018, marcucci_topological_2019-1}).  In particular, recent numerical results \cite{slunyaev_role_2016, gelash2018strongly}  suggest that  the soliton gas theory could be instrumental for the development of the statistical description of the of rogue wave formation.  
Additionally, soliton gases have been recently attracting a growing interest from the mathematical community.  Various nontrivial mathematical properties of  the  kinetic equation for soliton gas were studied in \cite{el_kinetic_2011, bulchandani_classical_2017,  kuijlaars_minimal_2021, ferapontov_kinetic_2022}. Beyond the kinetic, Euler scale, description, recent rigorous studies \cite{girotti_rigorous_2021, girotti_soliton_2022}  were devoted to the  construction of asymptotic solutions of the KdV and modified KdV equations respectively, describing  {\it realizations} of a special class soliton gases within the framework of {\it primitive potentials} \cite{dyachenko2016primitive}, via the consideration of $N$-soliton solutions  in the limit $N \to \infty$,

Finally we mention recent major developments in a closely related area of  {\it generalized hydrodynamics}  (GHD)(see \cite{castro-alvaredo_emergent_2016, bertini_transport_2016, doyon_lecture_2020} and references therein), where the equations analogous to those arising in the spectral kinetic theory of soliton gas  became pivotal for the understanding of large-scale, emergent hydrodynamic properties of integrable quantum and classical many-body systems. The relation between spectral theory of soliton gas and the GHD has been recently established in \cite{bonnemain_2022} which enabled the formulation of  
 the {\it thermodynamics} (free energy, entropy, temperature) of the KdV soliton gas.
 
 \medskip
The goal of this Perspective article is to present the state of the art in the modern theoretical and experimental soliton gas research,  highlighting the  connections with other areas of nonlinear physics and mathematics and outlining the  avenues for future investigations.

\medskip
The structure of the article is as follows. In Section~\ref{sec:concept_sg} we introduce the concept of soliton gas, from rarefied to dense, and present a straightforward phenomenological approach to the construction of the  spectral kinetic equation for integrable systems with known two-soliton interactions. In Section~\ref{sec:spec_theory} we proceed with outlining the results of rigorous spectral theory of soliton gas based on the thermodynamic limit of finite-gap potentials and their modulations for the KdV and fNLS equations. In Section~\ref{Sec:IST}, we summarize the basic concept of IST and the recent progress allowing the numerical computation of $N$-soliton solutions with $N$ large. In Section~\ref{Sec:Exp}, we review the experimental results on soliton gases. In Section~\ref{sec:Applications}, we show how soliton gas theory can be used to understand and predict integrable turbulence phenomena. In Section~\ref{sec:GHD} we review the key results of GHD and their links with SG. Finally, in Section~\ref{sec:open}, we review fundamental open questions and perspectives of this field of research.
      
\section{The concept of soliton gas}
\label{sec:concept_sg}

\subsection{Solitons in integrable systems}
\label{sec:sol_int}

We first outline the basic properties of solitons using the KdV equation as a prototypical example. We consider the KdV equation in the   form
\begin{equation}\label{KdV}
 u_t\ +\ 6u  u u_x\ +\ u_{xxx}\ =\ 0 \, .
\end{equation}
Equation \eqref{KdV} belongs to the class of  completely  integrable equations and, for a  broad class of initial conditions, its integrability  is realised via the inverse scattering transform (IST) method \cite{gardner_method_1967} sometimes called Nonlinear Fourier Transform.
The inverse scattering theory  associates a  single soliton solution  of the KdV equation  with a point of discrete spectrum $\la=\la_1<0$ of the Schr\"odinger operator
\be\label{schr}
\mathcal{L}= -\partial_{xx}^2  - u(x,t) \, . 
\ee
 Assuming $u \to 0$ as $x \to \pm \infty$, the KdV soliton solution corresponding to an eigenvalue $\la_1 = -\eta^2_1$, $\eta_1>0$, is given by
\be\label{kdv1sol}
u_{\rm s}(x,t; \eta_1, x_1^0)=2\eta_1^2 \hbox{sech}^2 [\eta_1(x-4\eta_1^2 t - x_1^{0})],
\ee
where $2\eta_1^2$ is the soliton amplitude,  $4\eta_1^2$ its speed, and  $x_1^0$ its initial position or  ``phase''. Note that soliton has finite width $\sim 1/\eta_{1}$, which affects the notion of the interaction range, particularly for small-amplitude solitons. In what follows we will be referring to  $\eta$ as a spectral parameter  with the understanding that $\eta = \sqrt{-\la}$. Along with the simplest single-soliton solution \eqref{kdv1sol}, the KdV equation supports $N$-soliton solutions $u_N(x,t)$ characterised by $N$ discrete spectral parameters 
$0< \eta_N < \eta_{N-1} < \dots <\eta_1$ and the set of the so-called norming constants that could be interpreted in terms of the initial positions  of solitons --- the analogs of $\{x_i^0 | i=1, \dots, N\}$  in \eqref{kdv1sol}  (note that the actual position of a soliton within the $N$-soliton solution depends nontrivially on all norming constants).  Thus, $N$-soliton solution can be viewed as a nonlinear superposition of $N$ single-soliton solutions, the notion supported by the asymptotic behavior at $t \to \pm \infty$, when $u_N(x,t)$  assumes the form of rank-ordered soliton trains, $u_N (x,t) \xrightarrow[\footnotesize{t \to \pm \infty}]{}  \sum_i^N u_{\rm s}(x,t; \eta_i, x_i^\pm)$, with appropriately chosen phases $x_{i}^\pm$ depending on the configuration at $t=0$, see \cite{drazin_solitons_1989, novikov1984theory, ablowitz_inverse_1974}. 

It should be stressed that general solutions to the KdV equation exhibit, along with solitons, a dispersive radiation component corresponding to the continuous spectrum of the Schr\"odinger operator \eqref{schr}. However, the soliton gas construction considered here involves only discrete spectrum. 

\medskip
The integrable structure of the KdV equation has profound implications for the dynamics of soliton interactions.

\begin{enumerate}
\item  The KdV evolution preserves the IST spectrum,  $\partial_t \eta_j=0$, implying that  soliton collisions are `elastic' i.e. solitons remain unchainged (retaining the amplitude, speed and the waveform \eqref{kdv1sol}) upon  interactions. In other words, the solution exhibiting $N$  solitons at $t \to - \infty$ will exhibit exactly the same $N$ solitons (modulo their positions) at $t \to +\infty$; 

\item The  collision of two solitons with spectral parameters $\eta_i$ and $\eta_j$, $i \ne j$ results in the asymptotic  shifts of their positions at $t \to + \infty$ relative to the respective free propagation  trajectories from $t \to - \infty$. These position shifts correspond to the phase shifts of the discrete spectrum norming constants and are given by
\be \label{shift_kdv}
\Delta_{ij} \equiv \Delta (\eta_i, \eta_j)= \frac{\sgn(\eta_i-\eta_j)}{\eta_i}   \log\left|\frac{\eta_i + \eta_j}{\eta_i-\eta_j} \right|,
\ee
so that the taller soliton acquires shift forward and the smaller one -- shift backwards.

\item Solitons interact pairwise so that the resulting phase shift $\Delta_i$ of a given  soliton with spectral parameter $\eta_i$  after its interaction with  $M$ solitons with parameters $\eta_j$, $j \ne i$, is equal to the sum of the individual phase shifts,
\be\label{delta_total}
\Delta_i= \sum \limits_{j =1, j\ne i}^M \Delta_{ij}.
\ee
Thus the interaction of $N$ solitons  can be factorized, with respect to the phase shifts, into superposition of $2$-soliton interactions, i.e. multi-particle effects are absent.
\end{enumerate}
It is important to stress that  the collision phase shifts are the far-field effects. Mathematically they are the artefacts of the asymptotic representation of the exact two-soliton solution of the KdV equation  as a sum of two individual solitons: $u_2(x,t; \eta_1, \eta_2) \simeq u_{\rm s}(x+\Delta_{12},t; \eta_1)+u_{\rm s}(x+\Delta_{21},t; \eta_2)$, which
is only valid if solitons are sufficiently separated (the long-time asymptotics). The interaction of solitons is a complex nonlinear process  \cite{lax_integrals_1968} and the resulting wave field $u(x,t)$ in the interaction region  cannot be represented as a superposition of the phase-shifted one-soliton solutions. 
We note that the above properties of soliton collisions (the preservation of  soliton parameters and pairwise phase/position shifts) are not exclusive to KdV but are generic features of other integrable systems supporting soliton propagation.

\smallskip
For the NLS equation 
\be\label{eq:fNLS}
i \psi_t + \psi_{xx} +2 \sigma |\psi|^2 \psi=0, \quad \psi \in \mathbb{C}, 
\ee
in the focusing regime, $\sigma=+1$, the single-soliton solution is characterised by a discrete complex eigenvalue  $\lambda_1 =a+ib$  and c.c., of the linear scattering operator called the  Zakharov-Shabat operator \cite{Zakharov:1972Exact}, the fNLS analogue of the Schr\"odinger operator \eqref{schr}.  The fNLS soliton is given by
\begin{equation}\label{fnls_soliton}
  \psi_{\rm s} (x,t)= 2b \, \frac{e^{-2i[ax + 2(a^2-b^2)t]+i\phi_0}}{\hbox{cosh}[2b(x+4at-x_0)]},
   \end{equation}
where $x_0$ is the initial position of the soliton and $\phi_0$ is the initial phase.  One can see that  the fNLS soliton represents a localised wavepacket with the envelope propagating with the group velocity  $c_g=-4 a= -4 \hbox{Re} \lambda_1$ and the carrier wave having the phase velocity $c_p=2(b^2-a^2)/a = -2\hbox{Re} (\lambda_1^2)/\hbox{Re} \lambda_1$.  In contrast with KdV equation, the amplitude and velocity of the fNLS soliton are two independent parameters. 

 Similar to other integrable  models, the solitons of the fNLS equation interact pairwise and experience both position and (genuine) phase shifts upon the interaction.  Unlike the KdV equation the fNLS solitons are bi-directional but the position shifts in the overtaking
and head-on soliton collisions  are given by the same expression, 
\be \label{fnls_phase_shift}
\Delta(\lambda, \mu) = \frac{\hbox{sgn} [\hbox{Re} (\mu - \lambda)]}{ \hbox{Im} \lambda}\ln \left|\frac{\mu-\lambda^*}{\mu-\lambda}\right|. 
\ee
In some other bidirectional integrable systems such as the Kaup-Boussinesq equations describing shallow water waves and the resonant NLS equation having applications
in magnetohydrodynamics of cold collisionless plasma the soliton collisions are anisotropic, i.e. the head-on and overtaking position shifts are described by different  expressions, see \cite{congy_soliton_2021}.

\subsection{Rarefied soliton gas}
\label{sec:rar_sg}

Following the historical paper of Zakharov \cite{zakharov1971kinetic} we first introduce soliton gas phenomenologically, as an infinite random ensemble of KdV solitons  distributed on $\mathbb{R}$ with some non-zero spatial density $\alpha$.   For $\alpha \ll 1$ (rarefied gas) one can approximate such a soliton gas by an  infinite superposition of the single-soliton solutions \eqref{kdv1sol}, 
\be \label{rar_u}
u(x,t) \approx \sum \limits_{i = 1} ^{\infty} 2\eta_i^2 \hbox{sech}^2 [\eta_i(x - 4\eta_i^2 t  - x_i^0)]   
\ee
with certain distribution of the solitonic spectral parameters $\eta_j \in \Gamma \subset \mathbb{R}^+$, and random initial  phases $x_j^{0} \in \mathbb{R}$ distributed according to  Poisson  with density parameter $\alpha$.  We will be initially assuming that  $\Gamma$ is a fixed, simply-connected interval (in which case without loss generality one can set $\Gamma=[0,1]$ but we shall keep general notation in the anticipation of future generalizations).

Due to the small spatial density, most of the individual solitons in a rarefied  gas \eqref{rar_u} overlap only in the regions of their exponential tails, except for the rare events of soliton collisions and neglecting the effects related to the possible presence of small-amplitude, wide solitons. Thus, each realization of the random process \eqref{rar_u}  satisfies the KdV equation \eqref{KdV} almost everywhere  on $ \mathbb{R}$. 

 Within the above phenomenological construction we introduce the  
 {\it density of states} (DOS) $ f(\eta; x,t) \ge 0$  such that $f(\eta_0; x_0, t_0)\rmd\eta \rmd x$ is the number of solitons found at $t=t_0$   
 in the element  $ [\eta_0, \eta_0+ \rmd\eta] \times [x_0, x_0 + \rmd x]$ of the  spectral phase space $\mathfrak{S}=\Gamma \times \mathbb{R}$.  It is assumed that the interval $[x_0, x_0 + \rmd x]$ contains a large number of solitons.  
For an equilibrium (spatially homogeneous, or uniform) soliton gas the DOS does not depend on space and time, $f(\eta; x,t) \equiv f(\eta)$.
 The  total spatial density of the soliton gas is computed as
\begin{equation}\label{kappa}
\alpha = \int_\Gamma f(\eta) \rmd \eta \, .
\end{equation}

\begin{figure}
\centering
\includegraphics[width=8.9cm]{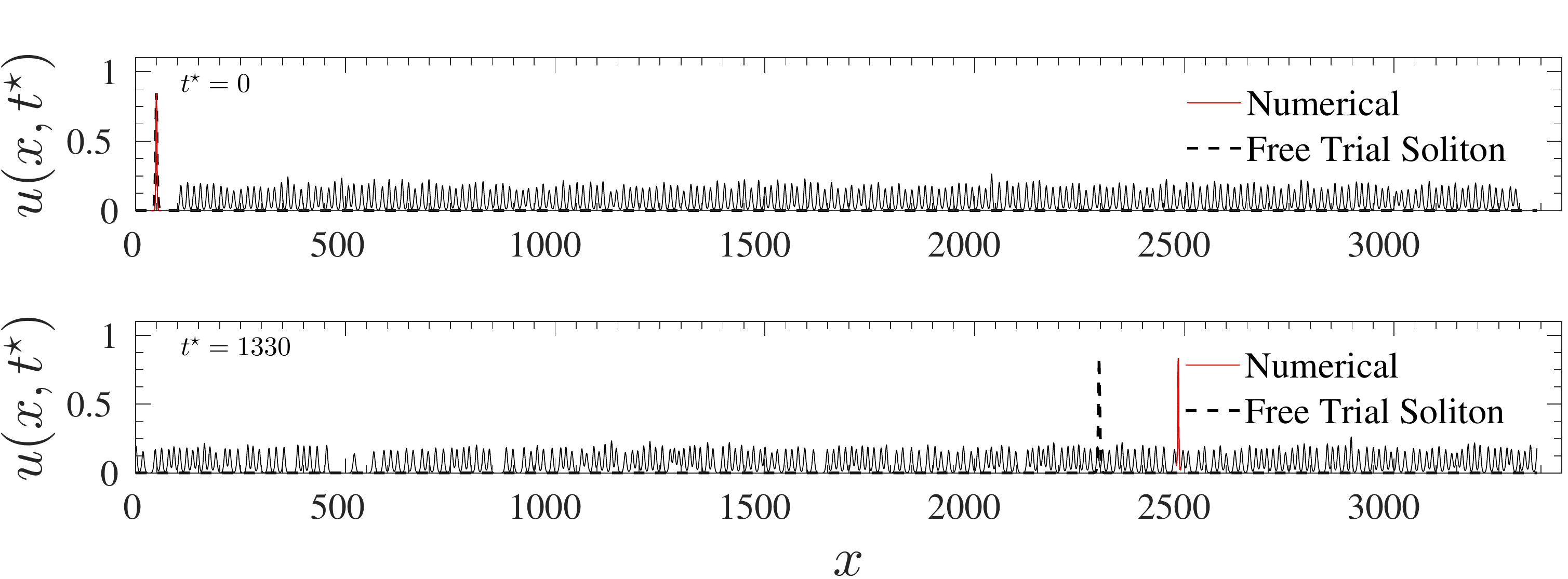}
    \caption{(adapted from \cite{carbone_macroscopic_2016}) Comparison for the propagation of a free soliton with the spectral parameter $\eta=\eta_1$ in a void (black dashed line) with the propagation of the trial soliton with the same spectral parameter $\eta_1$ (red solid line) through a rarefied soliton gas with the DOS supported on a narrow spectral interval around $\eta_0 < \eta_1$. One can see that the trial soliton  propagates faster in the gas due to the interactions with smaller solitons. Reproduced with permission.}
    \label{fig:trial_sol1}  
  \end{figure}

\medskip
 In a rarefied gas $\alpha \ll 1$ and the  total spatial shift of a soliton with spectral parameter $\eta \equiv \eta_1$ (we shall call it $\eta_1$-soliton) acquired over the time interval  $\rmd t$, due to the interactions with `$\mu$-solitons' having spectral parameters $\mu \in \Gamma$, $\mu \ne \eta$, is approximately evaluated  as
\be \label{collision_rar}
\Delta_1 \approx \int_\Gamma [\Delta(\eta_1, \mu) |s_0(\eta_1) - s_0(\mu)| f(\mu)  \rmd \mu ] \rmd t,
\ee 
where $s_0(\eta)$ is the speed of an isolated, non-interacting, soliton (it is assumed that in a rarefied gas the collision rate  is at leading order defined by the free soliton velocities).
For the KdV equation $s_0(\eta)=4\eta^2$ and $\Delta(\eta, \mu)$ is given by equation \eqref{shift_kdv}. This simple argument was used  in  \cite{zakharov1971kinetic} to derive the expression for the effective (mean) velocity $s(\eta)$ of a `trial' soliton in a spatially uniform (equilibrium) KdV soliton gas:
\begin{equation}\label{eq_state_kdv_dilute} 
s(\eta) \approx 4\eta^2+\frac{1}{\eta}\int _\Gamma \log
\left|\frac{\eta + \mu}{\eta-\mu}\right|f(\mu)[4\eta^2- 4 \mu^2]\rmd \mu\, .
\end{equation}
See Fig.~\ref{fig:trial_sol1} for the numerical simulations illustrating the effect of soliton interactions  on the effective velocity of a trial soliton propagating through a soliton gas.

For a weakly non-homogeneous  (out of equilibrium) gas  we have $f(\eta) \rightarrow f(\eta; x, t)$, $s(\eta) \rightarrow s(\eta; x, t)$, where the $(x,t)$-variations of $f$ and $s$ occur on macroscopic, Euler, scales, much larger than the typical scales associated with variations of the wave field $u(x,t)$ in individual solitons. 

Now, isospectrality of the KdV evolution within the IST framework  implies  the  continuity equation for the phase space density (the DOS),
\be \label{kin_eq0}
\partial_tf+\partial_x(sf) = 0,
\ee
which, together with  \eqref{eq_state_kdv_dilute}, provides the spectral hydrodynamic/kinetic description of  a rarefied  KdV soliton gas.

\subsection{Dense soliton gas}
\label{sec:dense_SG}

\begin{figure}
\centering
  \includegraphics[width= 4cm]{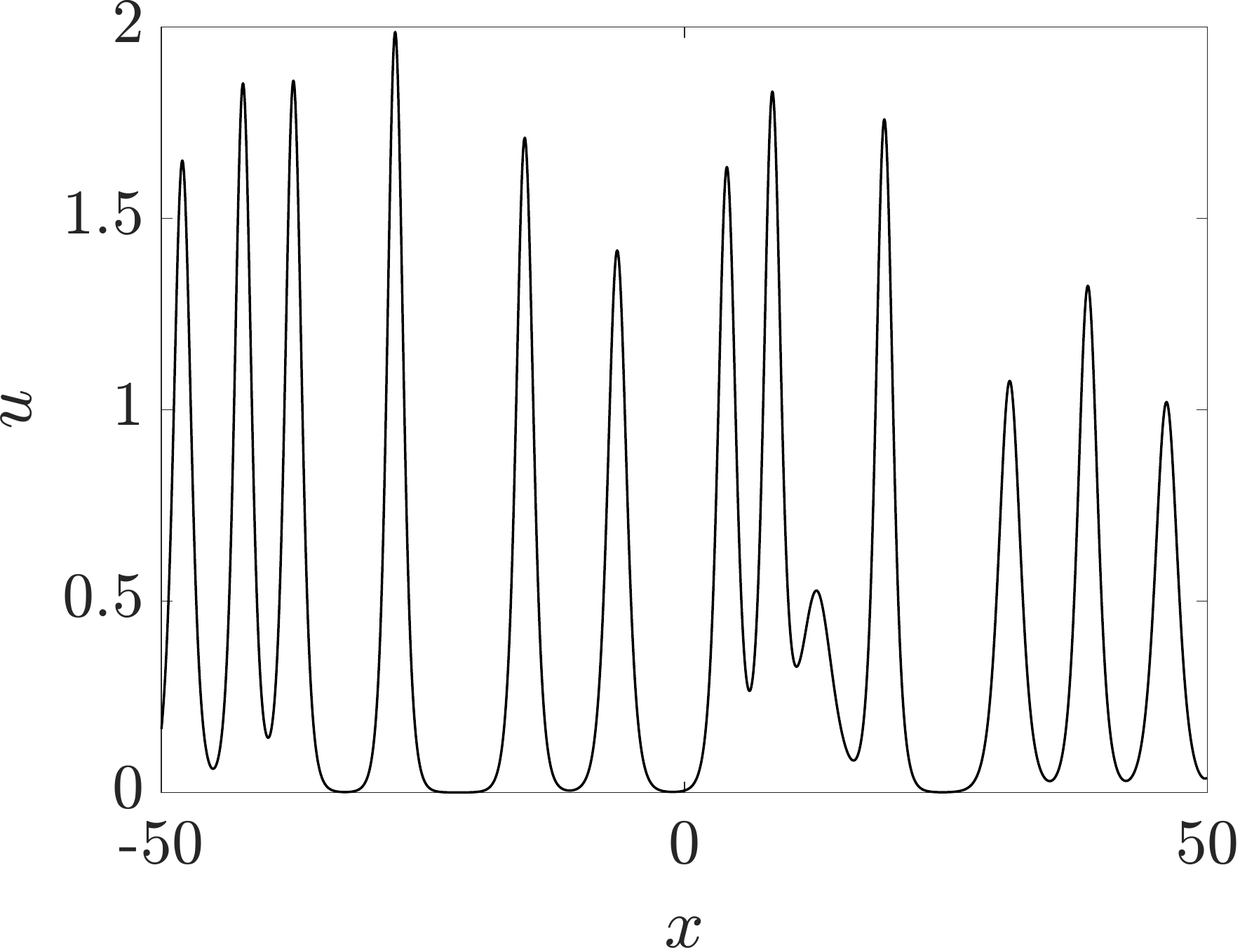} \quad    \includegraphics[width= 4cm]{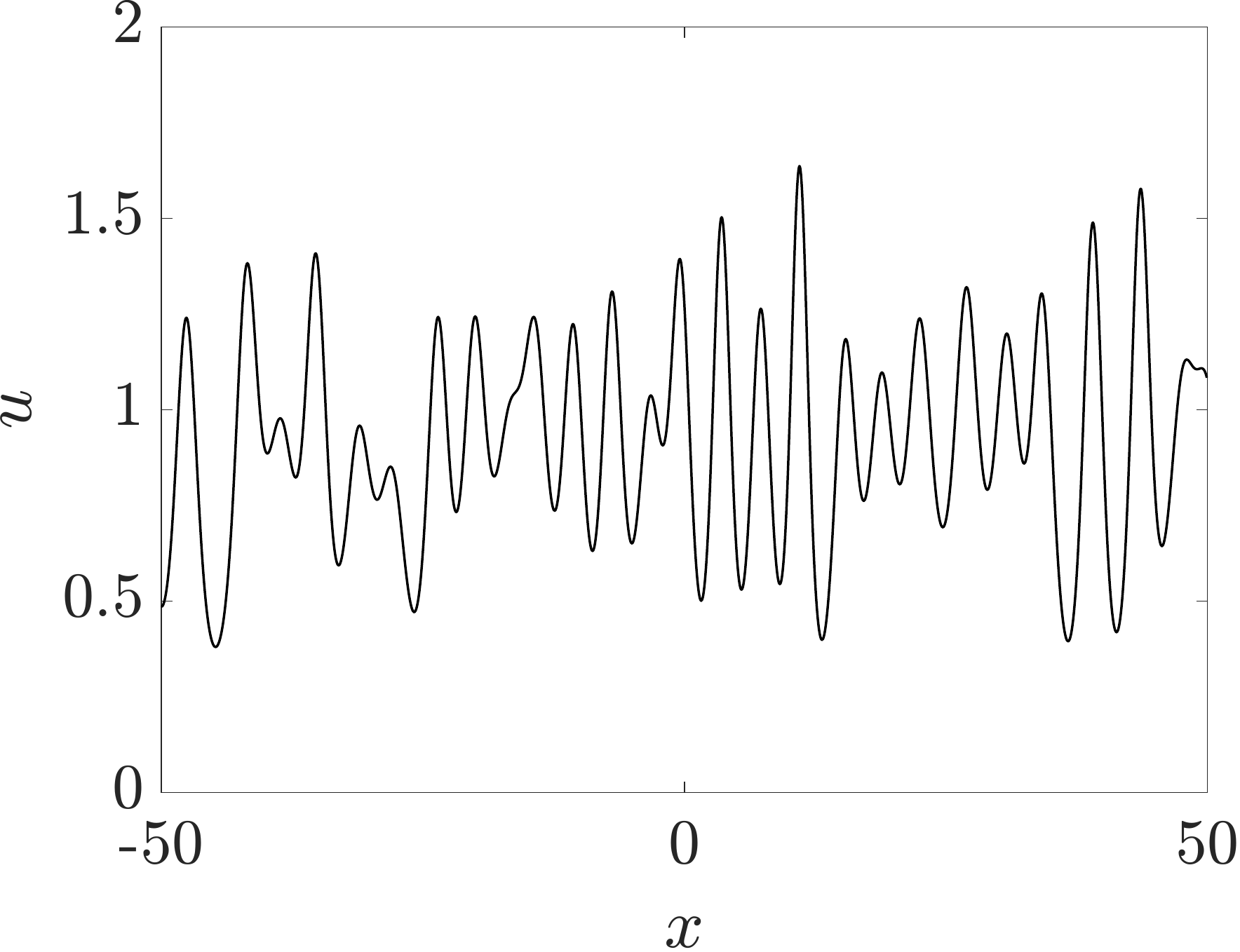} 
    \caption{Rarefied (left) vs dense (right) KdV soliton gases with the same spectral density $\phi(\eta)=\eta/{\sqrt{1-\eta^2}}$ but different spatial densities: $\alpha \simeq  0.1$ (left) and $\alpha \simeq 0.3$ (right).}
    \label{fig:rar_dense}  
  \end{figure}

If the KdV soliton gas is sufficiently dense, the simple heuristic construction of the previous section based on the assumption of short-range interactions between solitons becomes, strictly speaking, invalid as solitons in such a gas strongly overlap and hence, are involved in a continual nonlinear interaction so that the corresponding KdV solution can nowhere  be represented as a linear superposition of individual solitons as in \eqref{rar_u}, cf. Fig.~\ref{fig:rar_dense} (b).  In particular, the approximation \eqref{collision_rar} for the total phase shift based on the free soliton velocities ceases to be valid.

A general (dense or rarefied) KdV soliton gas at equilibrium can be defined as a non-decaying random  solution of the KdV equation whose realizations can be approximated, on any sufficiently  large interval $[x_0-L/2, x_0+L/2]$, by an appropriate  $N$-soliton solution with  $N \sim L \gg 1$  so that:  (i) the gas has finite spatial density $\alpha$ following the thermodynamic limit ($L, N \to \infty$,  $N/L  \to  \alpha $); (ii) the soliton spectral parameters  $\eta_i$,  $j=1, 2, \dots, N$, are distributed on some finite interval $\Gamma \in \mathbb{R}^+$ with  density $\phi(\eta) > 0$ defined for $N \to \infty$ via 
\be \label{phi_sol}
\eta_{j+1}-\eta_j \sim \frac{1}{\phi(\eta_j)N}, \quad
\int_\Gamma \phi(\eta) \rmd \eta =1,
\ee
 so that $\phi(\eta)$ does not depend on the chosen realization of soliton gas and on  the reference point $x_0$. The set of discrete spectral parameters $\{ \eta_i\}$ in $N$-soliton solutions is complemented by the associated  set of norming constants, whose phases can be interpreted for diluted gases in terms of the spatial locations of individual solitons within the $N$-soliton solution, see \cite{drazin_solitons_1989, novikov1984theory, ablowitz_inverse_1974}.  Randomness enters this  soliton gas construction in two ways: (i) spectral---via interpreting $\phi(\eta) \rmd \eta $ as a probability measure on $\Gamma$; (ii) spatial---by assuming that the phases of the norming constants are random values uniformly distributed on some fixed interval.  The above definition of soliton gas as the thermodynamic limit of $N$-soliton solutions, while lacking full mathematical rigour,  is physically intuitive and sufficient for the majority of practical (numerical or experimental) considerations, where one inevitably deals with finite numbers of solitons. It can also be readily generalized to other integrable equations (see Section~\ref{Sec:IST} for the  implementation of this construction in the context of the fNLS equation).

\subsubsection{Density of states}
\label{sec:DOS}
\begin{figure}
\centering
  \includegraphics[width=4cm]{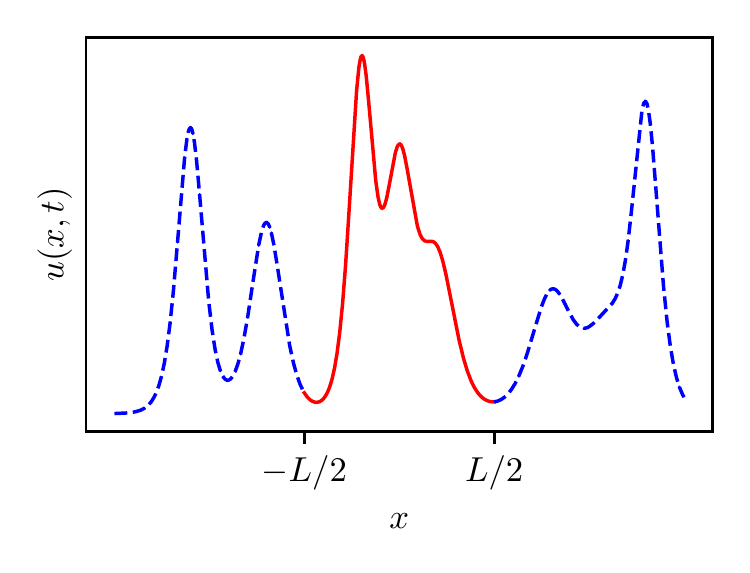} \includegraphics[width=4cm]{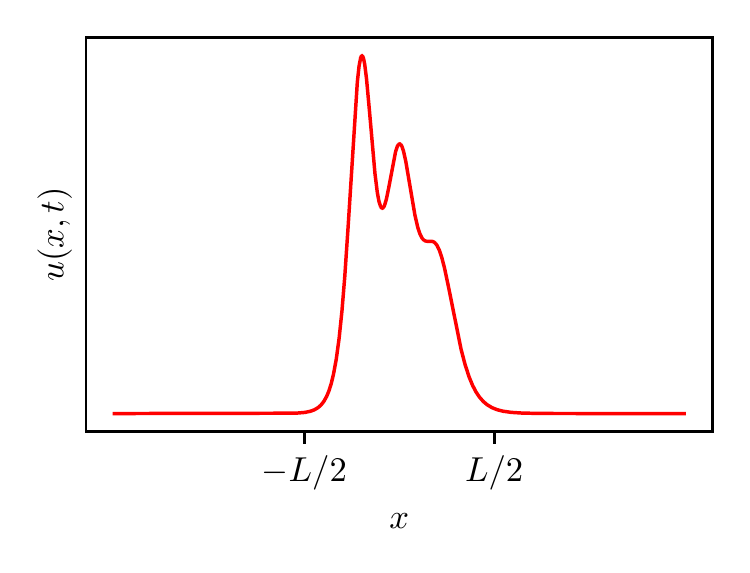}  
   \includegraphics[width=4cm]{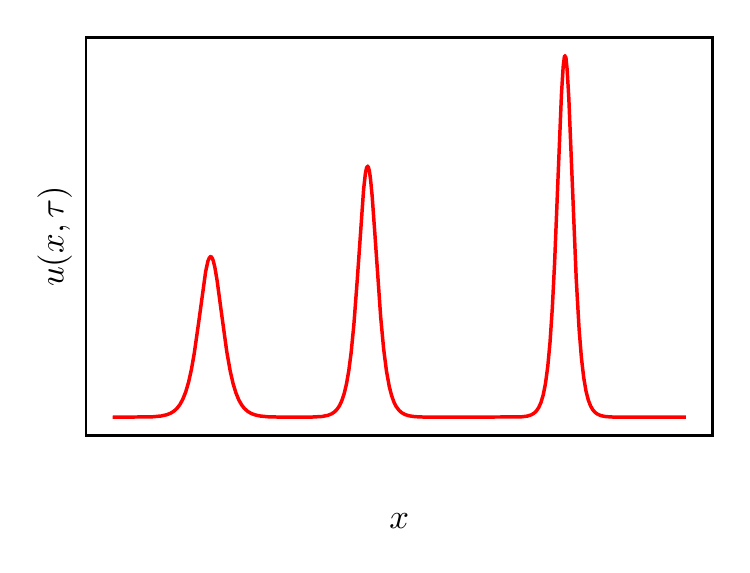}  
        \caption{Schematic illustrating the $N$-soliton approximation of the  `windowed' portion $U_L(x)$ of a KdV soliton gas at some $t=t^*$ (upper row) and its evolution into the ordered soliton train at $t=\tau \gg t^*$ (lower plot).}
    \label{fig:sol_windowing}  
  \end{figure}

The phenomenological definition of the DOS $f(\eta)$ introduced in Section~\ref{sec:rar_sg}  for rarefied soliton gas can be meaningfully  interpreted  in the context of a dense  gas where solitons strongly interact and cannot be identified as individual localized wave structures. Consider a typical realization  of a uniform soliton  gas at some  $t=t^*$ for $x \in [x_0-L/2, x_0+L/2]$, $L \gg1$. We now impose  zero boundary conditions for $x \notin [x_0-L/2, x_0+L/2]$, and consider  $U_L(x)=\chi_{{[x_0-L/2, x_0+L/2]}}u(x,t^*)$, where $\chi_{[a,b]}$ is the indicator function.  Replacing $U_L$ by the approximating $N$-soliton solution $u_N $ with $N \gg 1$ \footnote{To avoid boundary effects one can assume that the transition to zero at the edges of the `window' $\chi_{{[x_0, x_0+L]}}$  is smooth but sufficiently rapid (e.g. exponential) so that such a
`windowed' portion $U_L(x)$ of a soliton gas can be more faithfully approximated by the 
$N$-soliton solution  for some $N \gg 1$, i.e. the continuous spectrum can be neglected.}  one determines the density $\phi(\eta)$ of the discrete spectrum points  $\eta_j \in \Gamma$, $j=1, \dots, N$ \eqref{phi_sol} and the spatial density of the gas $\alpha \sim N/L$. {One simple way to realize this construction practically is to use  $U_L(x)$ as the initial condition for KdV and evolve it in  time, see Fig.~\ref{fig:sol_windowing}. One then infers $\phi(\eta)$, $\Gamma$ and $\alpha$ from the analysis of the  distribution of soliton amplitudes $a_i=2 \eta_i^2$ in the resulting soliton train at sufficiently large $t$.}  

\smallskip
We introduce the {\it partial DOS}  $f_L(\eta) =\varkappa \phi(\eta)$, where the  coefficient $\varkappa$ is determined from the normalization 
$\int_{x_0-L/2}^{x_0+L/2}\int_{\Gamma}f_L(\eta) \rmd \eta \rmd x=N$, consistent with the DOS definition  in Section~\ref{sec:rar_sg}. Then the limit  $\lim \limits_{L \to \infty} f_{L}(\eta) \equiv f_{\rm eff}(\eta)$ can be viewed as the {\it effective DOS} of a dense soliton gas. Using \eqref{kappa} and \eqref{phi_sol} we find the normalization constant $\varkappa = \alpha$ so that 
\be \label{fnorm}
f_{\rm eff} (\eta) = \alpha \phi(\eta)\, .
\ee
Shown in Fig~\ref{fig:rar_dense} are  realizations of  two soliton gases with the same $\phi(\eta)$ and different $\alpha$'s. We note that the criterion $\alpha \ll 1$ for rarefied gas is understood in the formal, asymptotic sense since the actual, numerical,  value of $\alpha$ depends on the definition of the (mesoscopic) unit interval of $x$. 

As we shall see in Section~\ref{sec:spec_theory}, in a more general soliton gas setting based on finite-gap theory one has
$\varkappa \equiv \varkappa(\eta)$, which is interpreted as the `scaled spectral wavenumber' responsible for the spatial density of solitons  with a given spectral parameter. However, the above phenomenological setting with $\varkappa$ constant is a useful approximation that is particularly relevant to the numerical realization of soliton gases (see \cite{gelash2019bound, congy_dispersive_2022} and Section~\ref{Sec:IST}) and identifying their connection with GHD (see \cite{bonnemain_2022} and Section~\ref{sec:GHD}).

\subsubsection{Kinetic equation}
\label{ssect:kinetic}

For a weakly non-uniform gas we assume scale separation, where the gas is considered to be at local equilibrium over intermediate,  mesoscopic, scale involving sufficiently large numbers of solitons, while appreciable $(x,t)$-variations of the DOS occur on a larger, macroscopic, Euler, scale. We note that this scale separation is at heart of GHD, where the mesoscopic scale is associated with the notion of `fluid cells', where the entropy is locally maximized with respect to the infinite number of conserved quantities \cite{castro-alvaredo_emergent_2016, doyon_lecture_2020}, see Section~\ref{sec:GHD}.

The generalization of Zakharov's kinetic equation to the case of a dense gas was derived in \cite{el_thermodynamic_2003} (see Section \ref{subsec:kdv} below). It involves the same continuity equation \eqref{kin_eq0} for the  DOS but the approximate expression \eqref{eq_state_kdv_dilute} for the tracer velocity is replaced by the exact integral {\it equation of state}: 
\begin{equation}\label{eq_state_kdv} 
s(\eta)=4\eta^2+\frac{1}{\eta}\int _\Gamma \ln
\left|\frac{\eta + \mu}{\eta-\mu}\right|f(\mu)[s(\eta)-s(\mu)]\rmd\mu\, ,
\end{equation}
where we have dropped for brevity the $(x,t)$-dependence for $f(\eta)$ and $s(\eta)$.

In simple terms \eqref{eq_state_kdv} represents an extrapolation of the rarefied gas properties to a dense gas, realised by replacing $s_0(\eta) \to s(\eta)$ in the collision rate expression \eqref{collision_rar}.  This observation has led in \cite{el_kinetic_2005} to the 
phenomenological prescription for the construction of the soliton gas equation of state involving  the free soliton velocity $s_0(\eta)$ and the phase shift expression $\Delta(\eta, \mu)=\sgn[(s_0(\eta) - s_0(\mu)] G(\eta, \mu)$,  specific to each integrable system:
\begin{equation}\label{eq_state_gen} 
s(\eta)= s_0(\eta)+\int _\Gamma G(\eta, \mu)f(\mu)[s(\eta)-s(\mu)]\rmd\mu\, .
\end{equation}

\medskip
In the fNLS case, the solitonic spectrum $\{\lambda_j\}$ in the associated linear (Zakharov-Shabat) scattering problem is complex (see Section~\ref{subsec:ISTtheory} below) so that
the DOS $f(\lambda)$  is generally supported on some compact Schwarz symmetric 2D set $\Lambda \subset \mathbb{C}$ so it is sufficient to consider only the upper half plane part $\Lambda^+$ (here Schwarz symmetry means  that if $\lambda \in \mathbb{C}$
 is  a  point of the spectrum then so is the c.c. point $ \lambda^*$).   
 Then, using $s_0(\la)=-4 \hbox{Re} \lambda$ for the free-soliton velocity  and the expression \eqref{fnls_phase_shift}  for the two-soliton scattering shift, the kinetic equation for the fNLS soliton gas assumes the form \cite{el_kinetic_2005}
\begin{equation}\label{FNLS_kin}
\begin{split}
&f_t + (fs)_x= 0, \\
&s(\lambda; x, t) = -4 \hbox{Re} \lambda + \\
&\frac{1}{ \hbox{Im} \lambda}   \iint \limits_{\Lambda^+}  \ln \left|\frac{\mu- \lambda^*}{\mu-\lambda}\right| [s(\lambda; x, t) - s(\mu; x, t)] f(\mu; x, t) \rmd \xi \rmd\zeta,
\end{split}
\end{equation}
where $\mu = \xi +i \zeta$ and $\Lambda^+ \subset \mathbb{C}^+ \setminus i \mathbb{R}^+$.

The special case when  all discrete spectrum points are located on the imaginary axis, 
$\Lambda^+ \subset i \mathbb{R}^+$, corresponds to non-propagating  multisoliton solutions called bound states \cite{Zakharov:1972Exact}. This case  requires a separate consideration. since for the corresponding bound state soliton gas  $\hbox{Re} \lambda =0$ the equation of state in \eqref{FNLS_kin} immediately yields $s(\lambda) = 0$ resulting in the equilibrium DOS, $f_t=0$. 

\subsubsection{Conserved quantities}
One of the fundamental properties of integrable dynamics  is the availability of 
an infinite set of  conservation laws 
\be\label{cons_law}
\partial_tP_n+\partial_xQ_n=0, \quad  n= 1, 2, \dots\; , 
\ee
where the  $P_n$ and $Q_n$ are functions of the field variable $u$ and its derivatives. 
For the KdV equation, of particular interest are the first three conserved densities: 
\be \label{kruskal_norm}
P_1=u, \quad P_2=u^2, \quad P_3 = \frac{u_x^2}{2} -  u^3 \, , 
\ee
typically associated with  the  ``mass'', ``momentum'' and ``energy'' conservation.  Their counterparts for the fNLS equation (equation \eqref{kappa}  with $\sigma=+1$) have the form \cite{Zakharov:1972Exact}
\be    \label{eq:pq}  P_1 = {|\psi|^2},\ \  P_2=  \text{Im} (\psi_x \psi^*), \ \ P_3 =  |\psi|^4 -
    |\psi_x|^2.                            
\ee
 
For non-equilibrium soliton gas dynamics conservation equations \eqref{cons_law} are replaced by their  averaged analogs:
\begin{equation}\label{stoch_Whitham}
\partial_t \langle P_n[u ] \rangle + \partial_x \langle Q_n[u]  \rangle = 0, \quad
n= 1, 2, \dots,
\end{equation}
where $\langle \cdot \rangle$ denotes ensemble averaging, and the $x,t$-variations in \eqref{cons_law} occur on much larger scales than in \eqref{cons_law}.

In contrast with the discrete set of conservation laws \eqref{cons_law} for the original equation, kinetic equation \eqref{kin_eq0} possesses a continuum of conserved quantities.  Indeed,  \eqref{kin_eq0} implies that for any $h(\eta) \ne 0$, $\int_\Gamma h(\eta)f (\eta; x, t) \rmd \eta$ is a density of the conserved quantity   with  $\int_\Gamma h(\eta)f (\eta; x, t) s(\eta; x, t)\rmd \eta$ being the corresponding flux density.  For the KdV equation, the densities of the special ``Kruskal'' series  \eqref{stoch_Whitham}  are given by \cite{el_thermodynamic_2003, el_critical_2016}
\be \label{pn_aver}
\langle P_n [u] \rangle = C_n \int_\Gamma \eta^{2n-1} f(\eta) \rmd \eta, ,  \ \ n=1,2, \dots, 
\ee
where the coefficients $C_n$ depend on the normalization of the conserved densities. For the physical densities \eqref{kruskal_norm} we have
\be \label{Cn}
C_1=4, \quad C_2={16}/{3}, \quad C_3={32}/{5}.
\ee
Expressions \eqref{pn_aver}, \eqref{Cn} are readily obtained by considering a  large portion of a homogeneous soliton gas: $u_L= \chi_{[0, L]} u(x,t)$ with $L \gg \alpha^{-1}$ at some arbitrary $t=t^*$. Assuming ergodicity one can replace the ensemble average  $\langle P_n \rangle$  by the spatial average $L^{-1}\int_{x_0}^{x_0+L} P_n [u_L]\rmd x$ which is a conserved quantity and
 can be evaluated  over  the long-time asymptotic solution: $u_L \sim  \sum_i u_s(x,t; \eta_i)$  as $t \to \infty$. 

A fundamental restriction imposed on the  DOS  $f(\eta)$ follows  from non-negativity of the variance  $
  \mathcal{A}\ =\ \sqrt{\langle {u^2} \rangle \ -\ \langle {u} \rangle ^2}\ \geqslant\ 0,
$
or equivalently, recalling \eqref{pn_aver}, \eqref{Cn}
\begin{equation}\label{variance+}
    \int_\Gamma \eta ^{3}f(\eta)\,\ud\eta - 3\left(\int_\Gamma \eta f(\eta )\,\ud\eta\right)^2 \geqslant\ 0\; .
\end{equation}

For the fNLS equation the averaged conserved densities can also be expressed in terms of moments of the DOS as
\cite{tovbis_periodic_2022}

\begin{equation}    \label{eq:tovbis}
\langle P_n[\psi]\rangle = C_n \iint \limits_{\Lambda^+} {\rm Im}(\lambda^n) f(\lambda) \rmd \xi \rmd \zeta, \quad n=1,2, \dots
\end{equation}
where $\la = \xi + i \zeta$ and the coefficients $C_n$ for the physical conserved quantities \eqref{eq:pq} are  
\be
C_1 = 4 , \quad C_2= -4, \quad C_3 = {16}/{3}.
\ee

\section{Spectral theory of soliton gas}
\label{sec:spec_theory}

\subsection{General framework}
\label{sec:gen_fram}

The phenomenological kinetic theory of soliton gas described  in the previous section is essentially based on the interpretation of  solitons as quasi-particles experiencing short-range pairwise interactions accompanied by the well-defined phase/position shifts. As  was already stressed, although this theoretical framework is justifiable in the case  of rarefied gas, it is less satisfactory for a dense gas where solitons experience significant overlap and continual nonlinear interactions so that they could become indistinguishable as separate entities. This suggests that a more consistent theoretical approach involving the wave aspect of the soliton's ``dual identity''  is necessary. In this section we outline a general mathematical framework for the spectral theory of soliton gas based on the thermodynamic limit of nonlinear multiphase solutions of integrable equations.  This approach has been first developed in \cite{el_thermodynamic_2003} for KdV equation and more recently applied to the description of fNLS soliton and breather gases \cite{el_spectral_2020}.

%
%
 With the KdV equation  as the simplest prototypical example in mind we consider the family of multiphase  solutions of the form  
 \begin{equation} \label{nonlin_multiphase}
u(x,t) = F_N(\theta_1, \dots, \theta_N), \quad \theta_j=k_j x -\omega_j t + \theta_j^0, 
\end{equation}
where $k_j$ and $\omega_j$, $j=1, \dots, N$ are the wavenumbers and frequencies (generally incommensurable), and  the function $F_N$ is $2 \pi$-periodic with respect to each phase component $\theta_j \in [-\pi,  \pi)$, $\theta_j^0$ being initial phases. (In the context of the NLS equation \eqref{eq:fNLS} the representation \eqref{nonlin_multiphase} is valid for $|\psi|$).
We stress that the existence of multiphase quasiperiodic solutions  \eqref{nonlin_multiphase} to a nonlinear dispersive equation is a unique property of integrable systems.
Such solutions are typically expressed in terms of  Riemann theta-functions, see e.g. \cite{Osborne} but we won't be using their specific form here.
 
It has been discovered in 1970-s that multiphase solutions to integrable equations have remarkable spectral properties defined within the (quasi-)periodic analogue of IST called the finite-gap theory, see  \cite{novikov1984theory, Osborne}. 
The fundamental result of the finite gap theory is that the IST spectrum $\mathcal{S}_N$ of the $N$-phase solution \eqref{nonlin_multiphase}  lies in the union of $N+1$ disjoint  bands $\gamma_j =[\lambda_{2j-1}, \lambda_{2j}]$, $j=1, \dots, N+1$, 
\be \label{lax_spectrum}
\la \in  \mathcal{S}_N \equiv \cup _{i=1}^{N+1} \gamma_i , \quad \gamma_i \cap \gamma_j= \emptyset, \ \ i \ne j
\ee 
separated by $N$ finite gaps $c_j=(\lambda_{2j}, \lambda_{2j+1})$. The number of spectral gaps $N$ is called the {\it genus}.  

For the preliminary discussion of this section it is convenient to assume that the  spectrum $\mathcal{S}_N$ is real-valued.  This is the case for the  (unidirectional) KdV equation and (bidirectional) defocusing NLS equation (equation  \eqref{eq:fNLS} with $\sigma=-1$). The case of complex band spectrum, $\mathcal{S}_N \subset \mathbb{C}$, arises for the fNLS equation, and this case will be considered separately in Section \ref{sec:fNLS}. 
We also note that one of the spectral bands could be semi-infinite (as is the case for the KdV equation), then $\gamma_{N+1}=[\lambda_{2N+1}, + \infty)$. 
Thus the  spectrum of a finite-gap solution (also called finite-gap potential) is fully parametrised by the state vector ${\bs \lambda}={(\lambda_1, \lambda_2, \dots, \lambda_\mathcal{D})}$, where   $\mathcal{D}=2N+1$ or $\mathcal{D}=2N+2$ depending on the presence or absence of the semi-infinite band.

One of the important outcomes of the finite-gap theory are the {\it nonlinear dispersion relations} (NDRs) linking the physical parameters of the multiphase solution \eqref{nonlin_multiphase} such as the wavenumbers, the  frequencies and the mean  with the components of the $\mathcal{D}$-dimensional spectral state vector ${\bs \lambda}$.
In particular, for the $N$-component wavenumber and  frequency vectors ${\bs k}=(k_1, \dots, k_N)$ and ${\bs \omega}= (\omega_1, \dots, \omega_N)$ in \eqref{nonlin_multiphase}  the NDRs can be  represented as
\be\label{n_gap_disp}
k_j = {K}_j({\bs \lambda}), \qquad \omega_j=  \Omega_j ({\bs \lambda}),   \quad j=1, \dots, N,
\ee  
where ${K}_j({\bs \lambda}), \ \Omega_j({\bs \lambda})$ are typically expressed in terms of complete hyperelliptic integrals, see e.g. \cite{Osborne, flaschka_multiphase_1980, el_spectral_2020}.

By manipulating the endpoints of spectral bands $\lambda_j$ one can modify the waveform of the solution \eqref{nonlin_multiphase}. In particular, by collapsing all  spectral bands into double points, 
$\la_{2j-1}, \la_{2j} \to \la^j_*$, $j=1, \dots, N$,  the $N$-gap solution transforms into $N$-soliton solution with the discrete spectrum eigenvalues $\la_*^j$ \cite{novikov1984theory} (for the KdV equation $\la_*^j = - \eta_j^2$, see Section~\ref{sec:sol_int}). This solitonic transition corresponds to the limit
$k_j, \omega_j \to 0, (\omega_j/k_j) \to s_0(\la_*^j)=\mathcal{O}(1)$, where $s_{0}(\la)$ is the velocity of a free soliton corresponding to the discrete spectral eigenvalue $\la_*^j$.  (We note that any linear combination of wavenumbers with integer coefficients is also a wavenumber, so by $\{ k_j\}_{j=1}^{N}$  we always assume a particular, ``fundamental'', set of the wavenumbers that vanish in the solitonic limit). Thus, finite-gap potentials represent periodic or quasiperiodic generalizations of multisoliton solutions.  Importantly, finite-gap potentials, unlike $N$-soliton solutions, are non-decaying functions with  nonzero mean, which makes them natural  building blocks for the construction of equilibrium, spatially uniform, soliton gases. Another advantage of using finite-gap solutions for the soliton gas construction is the presence of a natural probability measure---the uniform measure on the $N$-dimensional phase torus $\mathbb{T}^N$ of \eqref{nonlin_multiphase}, i.e. each phase $\theta_j^0 \in [-\pi, \pi)$ is assumed to be a random value uniformly distributed on the period. Assuming incommensurability of the wavenumbers $k_j$ and frequencies $\omega_j$ this measure gives rise to the ergodic random process with realizations defined by \eqref{nonlin_multiphase}.

The dynamics of weakly non-uniform finite-gap potentials are described by the Whitham modulation theory \cite{whitham_linear_1999, a_m_kamchatnov_nonlinear_2000, flaschka_multiphase_1980}, which prescribes slow  evolution of the spectrum, $\la_j(x,t)$, on the spatiotemporal scales much larger  than those associated with `rapid' variations  of the  wave \eqref{nonlin_multiphase}  itself. The modulation system
 inherently includes  wave conservation equations
\be \label{mod_eq}
\partial_t k_j + \partial_x \omega_j =0, \quad j=1, \dots , N, 
\ee
where $k_j({\bs \la})$ and $\omega_j({\bs \la})$ are given by the NDRs \eqref{n_gap_disp}.

We now define equilibrium soliton gas via the thermodynamic limit of finite-gap potentials \cite{el_spectral_2020}. Namely, we consider a sequence of finite-gap potentials \eqref{nonlin_multiphase} such that
 \be \label{term_gen}
\quad N \to \infty: \ \    k_1, \dots, k_N \to 0,   \ \ \sum \limits_{j=1}^N k_j \to 2 \pi\alpha = \mathcal{O}(1),
 \ee
 with a similar behaviour for the frequency components $\omega_j$ so that $\omega_j/k_j = \mathcal{O}(1)$. 
The limit \eqref{term_gen} suggests the following asymptotic scaling for the fundamental wavenumbers and frequencies as $N \to \infty$:
 \be \label{kom_scale}
N \to \infty: \quad k_j \sim \omega_j \sim N^{-1}.  
\ee

 It can be shown quite generally that under the limit \eqref{term_gen} the uniform distribution for the initial phases $\theta_j^0 \in [-\pi, \pi)$, $j=1, \dots, N$ transforms  to the Poisson distribution with the density parameter $\alpha$ on $\mathbb{R}$ for the `position phases' $l_j^0 \equiv \theta_j^0/k_j$  \cite{el_soliton_2021}. We associate the limiting  random process $\lim_{N \to \infty}F_N({\bs \theta})$ satisfying \eqref{term_gen} with soliton gas. By construction this process is ergodic. As we shall see, the above  definition is consistent with the phenomenological construction of soliton gas in Section~\ref{sec:dense_SG}. 
 
   As we show in the next section, the thermodynamic limit \eqref{term_gen} is achieved by imposing a special  band-gap distribution (scaling) for the spectrum $\mathcal{S}_N$ for $N \gg 1$.  Generally the spectral bands are required to be exponentially narrow compared to the gaps although the super-exponential and sub-exponential scalings are also possible and these corresponds to the noninteracting (ideal) soliton gas and soliton condensate respectively. In all cases we shall call the corresponding limit as $N \to \infty$ of a function  $ F({\bs \lambda})$ defined on $\mathcal{S}_N$ the thermodynamic  limit of $G$.

 

The density of states $f(\la)$ is then defined via the  thermodynamic limit of the partial sum
\be \label{tlim_k}
\frac{1}{2 \pi}\sum \limits_{j=1}^{M \le N} k_j \to  \int \limits_{\la_{\min}}^{\la} f(\la')\rmd \la',
\ee
 where $\la $ is a continuous spectral variable interpolating the discrete positions  $\la_j^*$ of  the band centres; as a matter of fact $\int_{\la_{\min}}^{\la_{\max}} f(\la')\rmd \la'= \alpha$, cf. \eqref{kappa}.

 Similarly, we have
 \be \label{tlim_om}
\frac{1}{2 \pi}\sum \limits_{j=1}^{M \le N} \omega_j  \to  \int \limits_{\la_{\min}}^{\la} v (\la')\rmd \la',
\ee
where $v(\la)$ is the spectral flux density; then $s(\la)=v(\la)/f(\la)$ has the meaning of the soliton gas transport velocity that can also be interpreted as an average tracer soliton velocity in the gas. 

Applying the thermodynamic limit \eqref{tlim_k}, \eqref{tlim_om} to the  NDRs  \eqref{n_gap_disp} for finite-gap solutions one  obtains the NDRs for an equilibrium soliton gas and hence, the equation of state  $s(\la)=\mathcal{F}[f(\la)]$, which typically has the form of a linear  integral equation \eqref{eq_state_gen}. Further, assuming $f\equiv f(\la;x,t)$, $s \equiv s(\la;x,t)$ and applying the thermodynamic limit to the modulation equations \eqref{mod_eq}, we obtain the continuity equation \eqref{kin_eq0} for DOS in a non-equilibrium gas.

\subsection{Korteweg-de Vries equation}
           \label{subsec:kdv}
           
We now present the results of  the application of the above general spectral construction  to the KdV equation \eqref{KdV} following Refs. \cite{el_thermodynamic_2003, el_soliton_2021}
The key input ingredient of the theory are the discrete NDRs \eqref{n_gap_disp} for finite-gap potentials. The specific expressions for the KdV NDRs can be found elsewhere (see e.g. \cite{flaschka_multiphase_1980, el_thermodynamic_2003, el_soliton_2021}),  here we only discuss their thermodynamic limit  as $N \to \infty$.

First we recall that the $N$-soliton limit of an $N$-gap solution is achieved by collapsing all the finite bands $\gamma_j$ in the spectral set $\mathcal{S}_N$ \eqref{lax_spectrum} into double points corresponding to the soliton discrete spectral values.
It was proposed in \cite{el_thermodynamic_2003} that the special infinite-soliton limit of the spectral $N$-gap KdV solutions, termed the thermodynamic limit, provides spectral description the KdV soliton gas. The thermodynamic limit  is achieved by assuming a special band-gap distribution (scaling) of the spectral set $\mathcal{S}_N$  for $N \gg 1$ on the fixed interval $[\la_{1}, \la_{2N+1}]$ (e.g. $[-1, 0]$). Specifically, we require the spectral bands $\gamma_j$ to be exponentially narrow compared to the gaps $c_j$
so that for $N \to \infty$ the spectral set  $\mathcal{S}_N$ is asymptotically characterized by two continuous positive functions: the density $\phi(\eta)$ of the lattice points $\eta_j \in \G \subset \mathbb{R}^+$ defining the band centers via $-\eta_j^2=(\la_{2j}+\la_{2j-1})/2$, and the logarithmic bandwidth distribution $\tau(\eta)$ defined for $N \to \infty$ by
\begin{equation} \label{therm_scale}
  \eta_j-\eta_{j+1}  \sim  \frac{1}{N \phi(\eta_j)}, \quad  \tau(\eta_j) \sim -\frac{1}{N} \ln (\la_{2j}-\la_{2j-1}).
\end{equation}
Additionally, invoking  the asymptotic behaviors \eqref{kom_scale} we introduce the interpolating functions $\kappa(\eta)$, $\nu(\eta)$ for the scaled wavenumbers and frequencies  
\begin{equation}\label{kom_scale1}
k_j \sim \frac{\kappa(\eta_j)}{N}, \quad \omega_j \sim \frac{\nu(\eta_j)}{N}.
\end{equation}
Then the definitions \eqref{tlim_k} and \eqref{tlim_om} of the DOS and the spectral flux density imply
\begin{equation}
f(\eta)= \frac{1}{2\pi}\kappa(\eta) \phi(\eta), \quad v(\eta)= \frac{1}{2\pi}\nu(\eta) \phi(\eta),
\end{equation}
where we have used a more convenient in the  KdV context spectral variable $\eta$ instead of $\la=-\eta^2$.
Now, considering  the KdV finite-gap NDRs~\eqref{n_gap_disp} subject to the thermodynamic scaling \eqref{therm_scale} , \eqref{kom_scale1} and letting $N \to \infty$, yields the integral equations \cite{el_thermodynamic_2003, el_soliton_2021}:
\begin{equation}\label{ndr_kdv1}
\begin{split}
&\int_{\Gamma} \ln \left|\frac{\mu +\eta}{\mu-\eta}\right|
  f(\mu)\rmd\m+f(\eta)\sigma(\eta)= \eta,   \\
  &\int_{\Gamma} \ln \left|\frac{\mu+\eta}{\mu-\eta}\right|
  v(\mu)\rmd\mu+v(\eta)\sigma(\eta)= 4  \eta^3  
  \end{split}
\end{equation}
for all $\eta\in \Gamma $ (if $\Gamma$ is a fixed, simply-connected compact interval one can set $\Gamma=[0,1]$ without loss of generality).
Here the {\it spectral scaling function}
$\sigma: \Gamma \to [0,\infty)$ is a continuous non-negative function that encodes the Lax spectrum of the soliton gas  via $\sigma(\eta)=\tau(\eta)/\phi(\eta)$.  Equations~\eqref{ndr_kdv1}  are the KdV soliton gas NDRs.

Eliminating $\sigma(\eta)>0$ from the NDRs~\eqref{ndr_kdv1} yields the  equation of state \eqref{eq_state_kdv} for the KdV soliton gas.
 Next, for a non-homogeneous soliton gas $f(\eta)\equiv f(\eta;x,t)$, $v(\eta) \equiv v(\eta;x,t)$, and the application of the thermodynamic limit to the modulation equations \eqref{mod_eq}  yields the continuity equation \eqref{kin_eq0} for the DOS. Indeed, \eqref{mod_eq} imply
 \be \label{wc_sum}
\left(\frac{1}{2 \pi}\sum \limits_{j=1}^{M \leq N} k_j \right)_t + \left(\frac{1}{2 \pi}\sum \limits_{j=1}^{M \leq N} \omega_j \right)_x =0.
 \ee
 for $M=1, \dots, N$.
 Applying the thermodynamic limit \eqref{tlim_k}, \eqref{tlim_om} to \eqref{wc_sum} we obtain 
 the kinetic equation $f_t+(fs)_x=0$ as required.
 
 Thus, the spectral kinetic equation \eqref{kin_eq0}, \eqref{eq_state_kdv} for soliton gas represents the thermodynamic limit of the KdV-Whitham   modulation system \cite{el_thermodynamic_2003}. We note that  condition $\sigma>0$ used in the derivation of the equation of state \eqref{eq_state_kdv} implies the restriction
 $\eta^{-1}\int_{\Gamma} \ln \left|\frac{\mu +\eta}{\mu-\eta}\right|
  f(\mu) \rmd\m <1$ on the admissible DOS $f(\eta)$, complementing the earlier formulated restriction \eqref{variance+}.
 The limiting case $\sigma=0$  corresponds to  the special soliton gas termed {\it soliton condensate}, see Section~\ref{sec_reductions} below.

\subsection{Focusing nonlinear Schr\"odinger equation}
\label{sec:fNLS}

The spectral theory of soliton gas for the fNLS equation was developed in \cite{el_spectral_2020}. It follows the same general framework of the thermodynamic limit of finite-gap potentials outlined in Section~\ref{sec:gen_fram} resulting in the kinetic equation \eqref{FNLS_kin} for the dense gas of fundamental fNLS solitons. However,  due to the fact that the finite-gap spectral set  for the fNLS equation lies in the complex plane, $ \la \in \mathbb{C}$, the spectral theory of  fNLS soliton gas admits a much broader range of scenarios than the KdV theory.  In particular, it covers the case of {\it breather gases}, including infinite random ensembles of interacting Akhmediev, Kuznetsov-Ma and Peregrine breathers \cite{roberti_numerical_2021}. Another highly nontrivial object is the gas of bound state fNLS solitons (bound states are $N$-soliton solutions with all discrete spectral parameters $\la_j$, $j=1, \dots, N$ having the same, possibly zero, real part \cite{Zakharov:1972Exact}).  The latter was shown in \cite{gelash2019bound} to represent an accurate model for the nonlinear stage of the development of spontaneous (noise induced) modulation instability, see Section~\ref{subsec:MI}. 

The soliton gas theory for the fNLS equation is more technically involved than in the KdV case. Here we only present the NDRs for the fNLS soliton gas, a counterpart of the KdV  NDRs
\eqref{ndr_kdv1}:
\begin{equation}\label{dr_soliton_gas}
\begin{split}
& \int _{\G^+}\ln \left|\frac{\m-\bar\la}{\m-\la}\right|
f(\m)|\rmd\m|+\sigma(\la)f(\la) =  \text{Im} \la,   \\
& \int _{\G^+}\ln \left|\frac{\m-\bar\la}{\m-\la}\right| v(\m)|\rmd\m|+  \sigma(\la) v(\la) = -4 \text{Im} \la \, \text{Re}\la, 
 \end{split}
\end{equation}
where $\Gamma^+$ is the upper part of the 1D Schwarz-symmetric curve $\Gamma \subset \mathbb{C}$ --- the spectral support of the DOS $f(\lambda)$
(in the general 2D case the integration with respect to arc length of $\Gamma^+$ in \eqref{dr_soliton_gas} is replaced by the integration over a 2D compact domain $\Lambda^+ \subset \mathbb{C}^+$:
$\int \limits_{\G^+} \dots |\rmd \mu| \to \iint \limits_{\Lambda^+}\dots \rmd \xi \rmd\zeta\, $,
where $\mu = \xi +i \zeta$).  

Eliminating the spectral scaling function $\sigma(\lambda)$ from the NDRs \eqref{dr_soliton_gas}  we obtain the equation of state in the kinetic equation \eqref{FNLS_kin}. The continuity equation  in \eqref{FNLS_kin} is derived via the thermodynamic limit of the modulation equations \eqref{mod_eq}, similar to the KdV case. See Ref.~\cite{el_spectral_2020} for details.

 Concluding this Section, we note that the requirement for the spectral support $\Gamma^+$ (or $\Lambda^+$) to be a compact set is purely technical and can be dropped so that the integrals in the NDRs and the equation of state can be taken over semi-infinite spectral domains: the only important requirement is the sufficiently fast decay of the DOS ensuring the existence of the integrals. The same is true for the KdV SG equations.

\subsection{Polychromatic  soliton gases and soliton condensates}
\label{sec_reductions}

Integration of the spectral kinetic equation \eqref{kin_eq0}, \eqref{eq_state_gen} for  soliton gas  in any generality represents a challenging mathematical problem. One can, however, consider some physically interesting particular cases that admit effective analytical treatment. The most obvious one is the case of spectrally polychromatic gases studied in \cite{el_kinetic_2011}. 
The DOS of a polychromatic soliton gas represents  a linear combination of the `monochromatic'  components in the form of Dirac delta-functions centered at distinct spectral points $\zeta_j \in \Gamma$ (note that $\Gamma$ can be real or complex domain, depending on the original dispersive equation)
 \be\label{u_delta1}
 f(\la; x, t) = \sum \limits_{j=1}^{M} w_j (x,t)\delta(\la - \zeta_j),
 \ee
 where $w_j(x,t)>0$ are the components' weights, and  
 $\{\zeta_{j}\}_{j=1}^M \subset \Gamma$, $(\zeta_j \ne \zeta_k \iff  j \ne k)$.  Substitution of  \eqref{u_delta1} into the kinetic equation \eqref{kin_eq0}, \eqref{eq_state_gen} reduces it to a system of hyperbolic hydrodynamic conservation laws
 \begin{equation}\label{cont}
(w_i)_{t}+(w_{i}s_{i})_{x} =0,\qquad i=1,\dots ,M\,,  
\end{equation}%
 where the component densities $w_i(x,t)$ and the transport velocities $s_{j}(x,t)\equiv s(\zeta_j, x, t)$ are related 
algebraically:
 \begin{equation}\label{s_alg}
 s_{j}= s_{0j} + \sum \limits_{m=1, m \ne j}^M G_{jm} w_m(s_{j}- s_{m}), \quad j=1, 2, \dots M.
 \end{equation}
Here we used the notation
$s_{0j} \equiv s_0(\zeta_{j}) , \quad G_{jm} \equiv G(\zeta_{j}, \zeta_{m}), \quad j \ne m$. One should also mention an important restriction $\sum \limits_{m=1, m \ne j}^M G_{jm} w_m <1$ equivalent to the condition of positivity of the spectral scaling function $\sigma$ in the thermodynamic limit construction.

 We note that the delta-function representation \eqref{u_delta1}  is a mathematical idealisation, which has a formal sense in the context of the integral equation of state \eqref{eq_state_gen},  but cannot be applied to the original dispersion relations  where it appears  in both the integral and the secular terms (cf. \eqref{ndr_kdv1} for the KdV equation).  In a physically realistic description  the delta-functions in \eqref{u_delta1} should be replaced by some narrow distributions around the spectral points $\zeta_{j}$, {i.e. we first take
the thermodynamic limit $N\to\infty$ and then allow the distributions to become sharply peaked.}  

For $M=2$ system \eqref{s_alg} can be solved  to give explicit expressions for $s_{1,2}(w_1, w_2)$:
  \be \label{s_12}
  \begin{split}
 s_1= s_{01} + \frac{ G_{12} w_2 (s_{01}-s_{02})}{1-(G_{12} w_2+ G_{21} w_1)}, \\
 s_{2}= s_{02}-\frac{ G_{21} w_1 (s_{01} - s_{02})}{1-(G_{12} w_2+ G_{21} w_1)}. 
 \end{split}
\ee
As was shown in \cite{el_kinetic_2005} (see also \cite{kamchatnov_dynamics_2022}) the two-component system \eqref{cont}, \eqref{s_12} is equivalent to the so-called Chaplygin gas equations that occur  in certain theories of cosmology (see e.g. \cite{bento_generalized_2002}), and to the Born-Infeld equations arising in nonlinear electromagnetic field theory \cite{born_foundations_1934}, \cite{whitham_linear_1999}.

It was shown in \cite{el_kinetic_2011} that the  system \eqref{cont}, \eqref{s_alg} for any $M \in \mathbb{N}$ possesses $M$ Riemann invariants and belongs to the  special class of  linearly degenerate, semi-Hamiltonian  systems of hydrodynamic type \cite{ferapontov_integration_1991}. Linear degeneracy of \eqref{cont}, \eqref{s_alg} implies the absence of the wavebreaking and shock formation for  generic initial-value problems with smooth Cauchy data \cite{rozhdestvenskii_systems_1983}. On the other hand, it  implies that  the solution to a Riemann (the evolution of an initial discontinuity) problem for  polychromatic soliton gas  will be given by  a combination of differing constant states $w_i=const_j$, separated by {\it contact discontinuities} propagating with classical shock speeds found from  the Rankine-Hugoniot conditions for the conservation laws \eqref{cont},\eqref{s_alg}. Such weak solutions were constructed  in  Refs.~\cite{el_kinetic_2005, carbone_macroscopic_2016, congy_soliton_2021, el_soliton_2021} for various soliton gases. More general solutions are available via the  hodograph transform, see \cite{el_kinetic_2011, kamchatnov_dynamics_2022}.

Another special class of soliton gases is presented by  soliton condensates whose properties are dominated by the collective effect of soliton interactions while the individual soliton dynamics are completely suppressed. Soliton condensates were first introduced in \cite{el_spectral_2020} for the fNLS equation and then thoroughly studied in \cite{congy_dispersive_2022} for the KdV equation. Spectrally, a soliton condensate is realized by vanishing the spectral scaling function, $\sigma(\eta) \to 0$, in the soliton gas NDRs (equations \eqref{ndr_kdv1}  for KdV or \eqref{dr_soliton_gas} for fNLS). For the KdV case the condensate NDRs are then given by \cite{congy_dispersive_2022}
\begin{equation} \label{NDR_cond0}
  \int_\Gamma \ln \left|\frac{\m+\eta}{\m-\eta}\right|
  f(\m)d\m = \eta,  \quad \int_\Gamma \ln \left|\frac{\m+\eta}{\m-\eta}\right|
  v(\m)d\m = 4  \eta^3.
\end{equation}
For the simplest case $\Gamma=[0,q]$ these are solved by
\begin{equation}\label{uv0}
  f(\eta)= \frac{\eta}{\pi\sqrt{q^2-\eta^2}},\quad
  v(\eta)=  \frac{6\eta(2\eta^2-1)} {\pi \sqrt{q^2-\eta^2}}.
\end{equation}
The counterpart fNLS solution of the NDRs \eqref{dr_soliton_gas} with $\la \in \Gamma^+=[0, iq]$ and $\sigma=0$ is given by  \cite{el_spectral_2020}
\be \label{cond_nls_dos}
f(\la) = \frac{-i\la}{\pi\sqrt{q^2+\la^2}}, \quad v(\la)=0\, ,
\ee
---and describes the DOS $f(\la)$ in the non-propagating, bound state ($s=v/f=0$) soliton condensate. By choosing a different 1D support $\Gamma^+ \subset \mathbb{C}^+$ one can construct other types of fNLS soliton condensates. E.g. if $\Gamma^+ = \{\xi+i\eta \;|\; \xi^2+\eta^2=1,\; \eta>0 \}$ (a cemi-circle) then the correposnding condensate DOS $f(\lambda) = \frac{{\rm Im}\lambda}{\pi}$ \cite{el_spectral_2020}. Such a `circular' soliton condensate propagates with the speed $s(\lambda)=-8 \text{Re} \lambda$---twice the speed of a free fNLS soliton.

Concluding this section, we mention an important generalization of the spectral theory of KdV soliton condensates developed in \cite{congy_dispersive_2022} by assuming the spectral support $\Gamma$ in \eqref{NDR_cond0} to be a union of $N+1$ finite disjoint intervals, termed `s-bands', $\Gamma =[0,\beta_1]\cup [\beta_2,\beta_3]\cup [\beta_{2j}, \beta_{2j+1}], \ j=0, \dots N$, with  $\beta_j=\beta_j(x,t)$. It was shown in \cite{congy_dispersive_2022} that the kinetic equation \eqref{kin_eq0}, \eqref{eq_state_kdv} then implies that the endpoints $\beta_j$ of the s-bands vary according to the genus $N$ KdV-Whitham equations \cite{flaschka_multiphase_1980}, providing the connection of non-equilibrium soliton gas  with the fundamental objects  of dispersive hydrodynamics such as rarefaction and dispersive shock waves \cite{El2016Dispersive}. The fNLS counterpart of the KdV theory of generalized soliton condensates is in progress.


\section{IST approaches to synthesis and analysis of soliton gas}\label{Sec:IST}

As shown in the previous Sections, the IST and finite-gap theory lay the foundations for the theory of soliton gas, demonstrating that soliton collisions are elastic and providing exact relations for the shifts in soliton positions, ultimately leading to the kinetic equation. 
Here, on the example of the fNLS equation, we discuss how the IST method allows one to observe the wave field of soliton gas in practice by generating such fields in numerical simulations or experiments from known soliton parameters. 
Also, we discuss the numerical techniques to solving the (opposite) direct scattering problem and determining the complete set of soliton parameters -- eigenvalues and norming constants -- from numerically or experimentally observed wave fields. 
Combined, the solutions of these two problems form a complete recipe for the IST synthesis and analysis of soliton gas.

For a rarefied soliton gas, the wave field can be constructed as an arithmetic sum of wave fields of single solitons with eigenvalues and positions chosen in accordance with the desired DOS. 
The dynamical and statistical properties of such gases have been studied using the weak interaction model, two-soliton interaction models and direct numerical simulations~\cite{gerdjikov1996asymptotic,uzunov1996description,pelinovsky2013two,dutykh2014numerical,shurgalina2016nonlinear}. 
Dense soliton gas requires full consideration of the interaction of solitons. 
In this Section, we describe an approach to the construction of soliton gas wave field used in the recent numerical and experimental studies~\cite{gelash2018strongly,gelash2019bound,Suret:20,agafontsev2021rogue}, which is based on exact dense $N$-soliton solutions. 
Although multi-soliton solutions are localized in space, for a large number $N$ of solitons, edge effects can be neglected and the central part of the wave field can be considered as a continuous section of soliton gas. 
By changing the soliton norming constants, it is possible to influence the distribution of solitons in the physical space, even though the exact mathematical link between the norming constants and the soliton spatial density (or, more generally, the DOS) is still missing.

Note that the explicit formulas for exact multi-soliton solutions have been known for decades, see e.g.~\cite{novikov1984theory}, however, their practical application was impossible due to numerical errors in the form of extreme gradients that appeared already starting from $N\sim 10$ solitons. 
The main source of these errors is the roundoff during a large number of arithmetic operations with exponentially small and large numbers. 
A solution to this problem has been found only recently in~\cite{gelash2018strongly} with a specific implementation of the dressing method combined with high-precision arithmetic computations, making it possible to successfully generate wave fields containing hundreds of solitons. 

As for the direct scattering procedure, there are several well established methods for the computation of soliton eigenvalues, see e.g. the Fourier collocation and Boffetta–Osborne methods. 
In the present paper, we focus on a highly challenging problem of the accurate identification of soliton norming constants, which is hampered by several types of numerical instabilities and has been solved only very recently in~\cite{gelash2020anomalous,mullyadzhanov2019direct}. 

In this Section, we consider the fNLS equation in the form,
\begin{equation}
\label{f_nlse_IST}
    i \psi_t + \frac12 \psi_{xx} + |\psi|^2 \psi=0,
\end{equation}
following the studies~\cite{gelash2018strongly,gelash2019bound,gelash2020anomalous,gelash2021solitonic,agafontsev2023bound} on the application of numerical IST and also other literature where the coefficients used in Eq.~(\ref{f_nlse_IST}) are conventional.


\subsection{IST method formalism}
\label{subsec:ISTtheory}

The IST method is based on the correspondence between an integrable nonlinear PDE and a specific auxiliary system of two linear equations (Lax pair), which consists of a stationary eigenvalue problem and an evolutionary problem for the same auxiliary function. 
The considered PDE is then obtained from the Lax pair as a compatibility condition. 
Using this compatibility condition, one can prove the fundamental property of the auxiliary system that its eigenvalue spectrum does not change with the evolution of wave field~\cite{novikov1984theory}.

For the fNLS equation~(\ref{f_nlse_IST}), the Lax pair is known as the Zakharov--Shabat system~\cite{Zakharov:1972Exact} for a two-component vector wave function $\mathbf{\Phi}(x, \lambda) = (\phi_1,\phi_2)^\mathrm{T}$, 
\begin{subequations}\label{ZH system}
\begin{eqnarray}
\label{ZH system_a}
\mathbf{\Phi}_x = \begin{pmatrix}\ -i\lambda & \psi \\ -\psi^* & i\lambda \end{pmatrix}\mathbf{\Phi},
\\
\label{ZH system_b}
\mathbf{\Phi}_t =\begin{pmatrix}\ -i\lambda^2 + \frac{i}{2} |\psi|^2 & \lambda \psi + \frac{i}{2} \psi_x \\ -\lambda \psi^* + \frac{i}{2} \psi^*_x & i\lambda^2 - \frac{i}{2} |\psi |^2 \end{pmatrix} \mathbf{\Phi},
\end{eqnarray}
\end{subequations}
where the superscript $\mathrm{T}$ stands for the matrix transpose and $\lambda = \xi + i \eta$ is a complex-valued spectral parameter. 
The first equation~(\ref{ZH system_a}) is equivalent to the eigenvalue problem for $\lambda$ written via the Lax operator $\widehat{\mathcal{L}}$ as 
\begin{equation}\label{L operator}
	\widehat{\mathcal{L}}\mathbf{\Phi} = \lambda \mathbf{\Phi},
	\quad
	\widehat{\mathcal{L}} = i \begin{pmatrix}\ 1 & 0 \\ 0 & -1 \end{pmatrix}\frac{\partial}{\partial x} - i\begin{pmatrix}\ 0 & \psi \\ \psi^* & 0 \end{pmatrix}.
\end{equation}
One can check that the fNLS equation, i.e., Eq. (\ref{eq:fNLS}) with $\sigma=1$, can be obtained in the anti-diagonal elements of the compatibility condition,
\begin{equation}
\mathbf{\Phi}_{xt} = \mathbf{\Phi}_{tx}.
\end{equation}

Note that, for the KdV equation~(\ref{KdV}), the equivalent Lax operator represents the self-adjoint Schr\"odinger operator \eqref{schr}, for which the spectral theory is well developed in quantum mechanics, see e.g.~\cite{landau1958quantum}. 
For the fNLS equation, the Lax operator is not self-adjoint, meaning that its eigenvalues can be located in the entire complex plane, though it is sufficient to consider only the upper half of it, $\eta = \mathrm{Im}\,\lambda\ge 0$. 
The latter follows from the fact that for every solution $\mathbf{\Phi} = (\phi_1,\phi_2)^\mathrm{T}$ of the Zakharov--Shabat system, which corresponds to an eigenvalue $\lambda$, there exists a counterpart $\tilde{\mathbf{\Phi}} = (-\phi_2^{*},\phi_1^{*})^\mathrm{T}$ corresponding to the complex-conjugate eigenvalue $\lambda^{*}$. 
Despite these differences, there are many similarities in the spectral theory of the operator~(\ref{L operator}) and the Schr\"odinger operator, and we encourage the reader to keep in mind this analogy, according to which the wave field $\psi$ of the fNLS equation is considered as a potential, and the vector function $\mathbf{\Phi}$ as a wave function.
In what follows, we will consider only a rapidly decaying potentials $\psi(x)$. 

Similarly to quantum mechanics, the scattering problem~(\ref{ZH system_a}) for the wave function $\mathbf{\Phi}$ can be introduced with the following asymptotics at infinity (the so-called ``right'' scattering problem, in contrast to the ``left'' scattering problem, see e.g.~\cite{LambBook1980}), 
\begin{eqnarray}
	\lim_{x\to -\infty}\biggl\{ \mathbf{\Phi} &-& \begin{pmatrix} e^{-i \lambda x} \\ 0 \end{pmatrix}\biggr\} = 0, \label{ScatteringProblem1}\\
	\lim_{x \to +\infty}\biggl\{\mathbf{\Phi} &-& \begin{pmatrix} a(\lambda)\, e^{-i \lambda x} \\ b(\lambda)\, e^{i \lambda x} \end{pmatrix}\biggr\} = 0. \label{ScatteringProblem2}
\end{eqnarray}
These asymptotics represent a two-component generalisation of the ``right''  scattering problem for the Schr\"odinger operator. 
The scattering coefficients $a(\lambda)$ and $b(\lambda)$ have the meaning that a wave $(a\, e^{-i \lambda x} ,0)^\mathrm{T}$ comes from the right side of the potential $\psi(x)$ and then splits into the transmitted wave $(e^{-i \lambda x} ,0)^\mathrm{T}$ at $x\to -\infty$ and the reflected wave $(0,b\, e^{i \lambda x})^\mathrm{T}$ at $x\to +\infty$. 
Hence, the quantity $r=b/a$ represents the so-called reflection coefficient. 
Note that the alternative choice of the asymptotics corresponding to the ``left'' scattering problem is also common in the IST constructions, see e.g.~\cite{LambBook1980}. 

The eigenvalue spectrum of the scattering problem consists of the eigenvalues $\lambda$ corresponding to bounded solutions $\mathbf{\Phi}$ of the Zakharov--Shabat system with asymptotics~(\ref{ScatteringProblem1})-(\ref{ScatteringProblem2}). 
Such solutions exist for real-valued spectral parameter, $\lambda = \xi\in\mathbb{R}$, and also for complex-valued $\lambda$, $\eta=\mathrm{Im}\,\lambda>0$, if and only if $a(\lambda) = 0$. 
For rapidly decaying potentials $\psi(x)$, the latter part of the eigenvalue spectrum usually consists of a finite number of discrete points $\lambda_n$, $a(\lambda_n)=0$, $n=1,...,N$ (discrete spectrum), and the overall eigenvalue spectrum contains also the real line $\lambda=\xi\in\mathbb{R}$ (continuous spectrum), see~\cite{novikov1984theory}. 
The full set of the scattering data represents a combination of the discrete $\{\lambda_n, \rho_n \}$ and continuous $\{ r \}$ spectra,
\begin{eqnarray}
	&& \bigg\{
	\lambda_n \,\,|\,\, a(\lambda_n) = 0, \,\, \mathrm{Im}\,\lambda_n > 0
	\bigg\}, \nonumber\\
	&& \rho_n = \frac{b(\lambda_n)}{a'(\lambda_n)}, \quad
	r(\xi) = \frac{b(\xi)}{a(\xi)}, \label{ScatteringData}
\end{eqnarray}
where $a'(\lambda)$ is complex derivative of $a(\lambda)$ with respect to $\lambda$, $\rho_n$ are the so-called norming constants associated with the eigenvalues $\lambda_n$, and $r(\xi)$ is the reflection coefficient defined at the real line $\xi\in\mathbb{R}$. 
Most importantly, the time evolution of the scattering data~(\ref{ScatteringData}) is trivial, 
\begin{eqnarray}
	&& \forall n: \lambda_n=\mathrm{const}, \quad\quad \rho_n (t) = \rho_n (0) e^{2i \lambda_n^2 t}, \nonumber\\
	&& r(\xi,t) = r(\xi,0)e^{2i \xi^2 t}, \label{ScattData(t)}
\end{eqnarray}
and the wave field $\psi(x,t)$ can be recovered from it at any moment of time with the IST by solving the integral Gelfand-Levitan-Marchenko (GLM) equations~\cite{novikov1984theory}. 
However, in the general case, the latter procedure can only be done numerically, asymptotically at large time, or in the semi-classical approximation~\cite{lewis1985semiclassical,jenkins2014semiclassical}.

Note that the function $a(\lambda)$ is analytic in the upper half of the $\lambda$-plane and has simple zeros at the eigenvalue points, $a(\lambda_n)=0$ (we do not consider the degenerate case when an eigenvalue point represents a multiple zero), see e.g.~\cite{novikov1984theory,faddeev2007hamiltonian}. 
Meanwhile, the analyticity is not always the case for the function $b(\lambda)$. 
However, in numerical simulations or experiments, the wave field $\psi(x)$ is always confined to a finite region of space, i.e., it has compact support, and in this case  the function $b(\lambda)$ is also analytic in the upper half of the $\lambda$-plane~\cite{novikov1984theory,faddeev2007hamiltonian}. 
This property of $a(\lambda)$ and $b(\lambda)$ is essential for algorithmic implementations of the direct scattering transform discussed below.

In the physical space, the continuous spectrum with non-zero reflection coefficient $r(\xi)$ corresponds to nonlinear dispersive waves, while the discrete eigenvalues $\lambda_{n}$ together with the norming constants $\rho_n$ -- to solitons. 
In particular, the eigenvalues $\lambda_{n} = \xi_n + i \eta_n$ contain information about the soliton amplitudes, $A_n=2\eta_n$, and group velocities, $V_n=-2\xi_n$, while the soliton norming constants -- about their positions in space $x_{n}^{\mathrm{IST}}\in\mathbb{R}$ and complex phases $\theta_n^{\mathrm{IST}}\in[0,2\pi)$. 
In a weakly nonlinear case, the discrete spectrum disappears and the function $r(\xi)$ tends to a conventional Fourier spectrum of the wave field $\psi(x)$, so that the IST is often considered as a nonlinear analogue of the Fourier transform.

In the (opposite) reflectionless case $r(\xi)=0$, the dispersive waves are absent and the IST procedure can be performed analytically by solving the GLM equations, leading to an exact $N$-soliton solution ($N$-SS) $\psi_{(N)}(x,t)$. 
There is also an alternative procedure for the construction of $N$-SS called the dressing method~\cite{novikov1984theory,zakharov1978relativistically}, also known as the Darboux transformation~\cite{akhmediev1991extremely,matveev1991darboux}. 
The dressing method allows one to add solitons to the resulting solution recursively by one at a time using a special algebraic construction~\cite{novikov1984theory,zakharov1978relativistically,matveev1991darboux}. 
The numerical implementation of this construction turns out to be much more stable and resource-efficient than solving the GLM equations, making it possible to build multi-soliton wave fields containing large number of solitons~\cite{gelash2018strongly}.

The dressing procedure starts from the trivial potential of the fNLS equation, $\psi_{(0)}(x) = 0$ for $x\in\mathbb{R}$, and the corresponding matrix solution of the Zakharov--Shabat system~(\ref{ZH system}),
\begin{eqnarray}\label{Psi0}
	\mathbf{\Phi}^{(0)}(x,\lambda) = \begin{pmatrix}\ e^{-i\lambda x} & 0 \\ 0 & e^{i\lambda x} \end{pmatrix};
\end{eqnarray}
here we fix time, $t=0$, for definiteness. 
At the $n$-th step of the recursive method, the $n$-soliton potential $\psi_{(n)}(x)$ is constructed via the $(n-1)$-soliton potential $\psi_{(n-1)}(x)$ and the corresponding matrix solution $\mathbf{\Phi}^{(n-1)}(x,\lambda)$ as
\begin{eqnarray}\label{psi_n}
	\psi_{(n)}(x) = \psi_{(n-1)}(x) + 2i(\lambda_n-\lambda^*_n)\frac{q^*_{n1}q_{n2}}{|\mathbf{q_n}|^2},
\end{eqnarray}
where the vector $\mathbf{q}_{n}=(q_{n1},q_{n2})^{\mathrm{T}}$ is determined by $\mathbf{\Phi}^{(n-1)}(x,\lambda)$ and the scattering data of the $n$-th soliton $\{\lambda_{n},C_{n}\}$, 
\begin{eqnarray}\label{qn}
	\mathbf{q}_{n}(x) = [\mathbf{\Phi}^{(n-1)}(x,\lambda^*_n)]^{*}\cdot 
	\left(\begin{array}{c} 1 \\C_n \end{array}\right).
\end{eqnarray}
Here $C_n$, $n=1,...,N$, are the soliton norming constants in the dressing method formalism, see the discussion below.
The corresponding matrix solution $\mathbf{\Phi}^{(n)}(x,\lambda)$ of the Zakharov--Shabat system is calculated via the so-called dressing matrix $\boldsymbol{\sigma}^{(n)}(x,\lambda)$, 
\begin{eqnarray}
	\mathbf{\Phi}^{(n)}(x,\lambda) &=& \boldsymbol{\sigma}^{(n)}(x,\lambda)\cdot \mathbf{\Phi}^{(n-1)}(x,\lambda), \label{dressing Psi}\\
	\sigma^{(n)}_{ml}(x,\lambda) &=& \delta_{ml} + \frac{\lambda_n-\lambda_n^*}{\lambda - \lambda_n}\frac{q_{nm}^{*}q_{nl}}{|\mathbf{q_n}|^2}, \label{dressing matrix}
\end{eqnarray}
where $m,l=1,2$ and $\delta_{ml}$ is the Kronecker delta. 
The time dependency is recovered using the time-evolution of the norming constants, 
\begin{eqnarray}
	C_n(t) = C_n(0) e^{-2i\lambda^2_n t}, \label{DM-norming-constants-evolution}
\end{eqnarray}
and repeating the dressing procedure at each time $t$.

The norming constants $C_n$ are related to the IST norming constants $\rho_n$ as follows~\cite{aref2016control,gelash2020anomalous} (this equation is valid for pure multi-soliton solutions only),
\begin{eqnarray}
	C_n(t) = \frac{1}{\rho_n(t)} \prod_{k=1}^{N} (\lambda_n - \lambda_k^*) \times \prod_{j \ne n}^{N} \frac{1}{\lambda_n - \lambda_j}, \label{rhok_Ck_conn}
\end{eqnarray} 
and can be parameterized in terms of soliton positions $x^{\mathrm{DM}}_{n}$ and phases $\theta^{\mathrm{DM}}_{n}$,
\begin{eqnarray}
	C_{n} = -\exp\bigg[2i\lambda_{n}x^{\mathrm{DM}}_{n} + i\theta^{\mathrm{DM}}_{n}\bigg]. \label{Ck_param}
\end{eqnarray}
Note that the IST norming constants $\rho_n$ have an alternative parameterization via IST positions $x^{\mathrm{IST}}_{n}$ and phases $\theta^{\mathrm{IST}}_{n}$, which coincide with the dressing method positions $x^{\mathrm{DM}}_{n}$ and phases $\theta^{\mathrm{DM}}_{n}$ and also with the observed in the physical space positions and phases only for the one-soliton solution~(\ref{fnls_soliton}). 
In presence of other solitons or dispersive waves, all three types of positions and phases may differ significantly from each other; see e.g. the discussion in~\cite{gelash2021solitonic} and the references wherein.


\subsection{IST synthesis of soliton gas wave field}
\label{subsec:denseSGgeneration}

The discussed method for the numerical construction of soliton gas wave field is based on the computation of $N$-SS for a large number of solitons $N$ using the straightforward algorithmic implementation of the dressing method. 
The high-precision arithmetics is applied to accurately resolve operations with exponentially small and large numbers coming from the elements of vectors $\mathbf{q}_{n}$ in Eqs.~(\ref{Psi0})-(\ref{dressing matrix}). 
The required number of digits grows with $N$ non-trivially and depends on specific choice of the soliton eigenvalues and norming constants, but usually stays in the hundreds for $N\simeq 100$ and thousands for $N\simeq 1000$; see~\cite{gelash2018strongly,gelash2019bound,gelash2021solitonic} for detail. 
Note that, while this inherent difficulty of the dressing method and other schemes based on the IST theory cannot be entirely avoided, the recently developed optimizations~\cite{prins2021accurate} can substantially reduce the necessary number of digits.

For the fNLS equation, soliton gas is characterized by the distribution of soliton eigenvalues (amplitudes and velocities) and soliton norming constants (positions and phases). 
Soliton eigenvalues are generally problem-specific and cannot be easily changed without modifying the context in which the soliton gas is studied. 
Soliton phases can usually be chosen as random values uniformly distributed over the interval $[0,2\pi)$; in this case, evolution over time, see Eq.~(\ref{DM-norming-constants-evolution}), preserves this distribution. 
In what follows, we focus on the study of dense soliton gases that are in equilibrium and have wave fields that are statistically homogeneous in space. 
This poses two problems: (i) how to achieve a high spatial density of solitons and (ii) how to construct multi-soliton wave fields, which are statistically homogeneous over a wide region in the physical space for random soliton phases. 

As has been observed empirically in~\cite{gelash2018strongly, gelash2019bound}, if soliton positions are distributed within the interval $x^{\mathrm{DM}}_{n} \in [-L_0/2, L_0/2]$ and $L_0$ approaches zero, then the characteristic size of the corresponding $N$-SS in the physical space shrinks to some finite non-zero limit and the soliton spatial density reaches its maximum value. 
However, for $L_0=0$ the $N$-SS becomes symmetric, $\psi(x)=\psi(-x)$. 
To avoid this artificial symmetry, one can use sufficiently small intervals $L_{0}\simeq 1$, so that the symmetry is not observed and the characteristic size of the $N$-SS remains close to the size in the limiting case $L_0=0$. 

In~\cite{gelash2018strongly}, a method has been developed for the construction of statistically homogeneous soliton gas wave fields, which starts from the computation of $N$-SS wave fields using rather arbitrary soliton positions from a small interval $x^{\mathrm{DM}}_{n} \in [-L_0/2, L_0/2]$, $L_{0}\simeq 1$. 
Then, these wave fields are put into a sufficiently large box $x \in [-L_p/2, L_p/2]$ where they are small near the edges, 
$$
    |\psi(\pm L_p)|\lesssim 10^{-16}\max|\psi(x)|,
$$
so that one can treat this box as a periodic one and simulate the time evolution of the constructed solutions inside it using the direct numerical simulation of the fNLS equation. 
If soliton velocities are random, then after some time the wave fields spread over the box $L_p$ and the system arrives to a statistically steady state, in which its basic statistical functions no longer depend on time. 
Then this state is used as a model of statistically homogeneous soliton gas of spatial density $N/L_{p}$ in an infinite space; in~\cite{gelash2018strongly}, it has been confirmed that for large enough number of solitons $N$ and box size $L_p$ the results depend on them only in the combination $N/L_{p}$. 
Note that such ``periodization'' of solitons requires the periodic box $L_{p}$ to be significantly larger than the characteristic size of the initial $N$-SS, decreasing the maximum soliton density that can be achieved with the described method.

In terms of the finite band theory, the periodic evolution in time replaces $N$-SS by $N$-band periodic solutions having exponentially narrow bands compared to the gaps, as the characteristic soliton width is much smaller than the box size $L_p$. 
This allows one to neglect the difference between the two types of solutions, similarly as it is done in Section~\ref{sec:gen_fram}, where, vice versa, the soliton gas is considered as a limit of finite-band solutions. 

\begin{figure}[t]\centering
	\includegraphics[width=8.8cm]{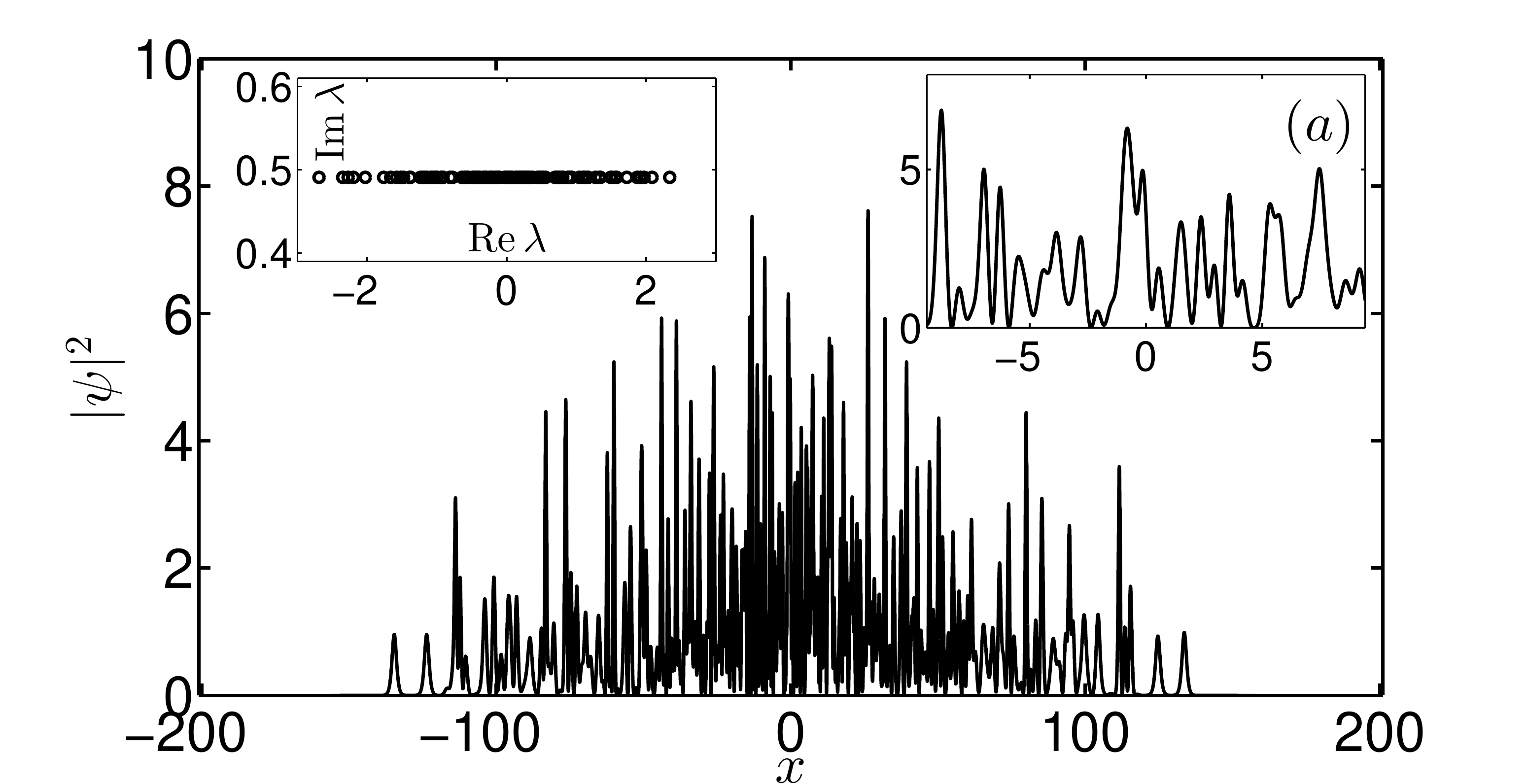}\\
        \includegraphics[width=8.8cm]{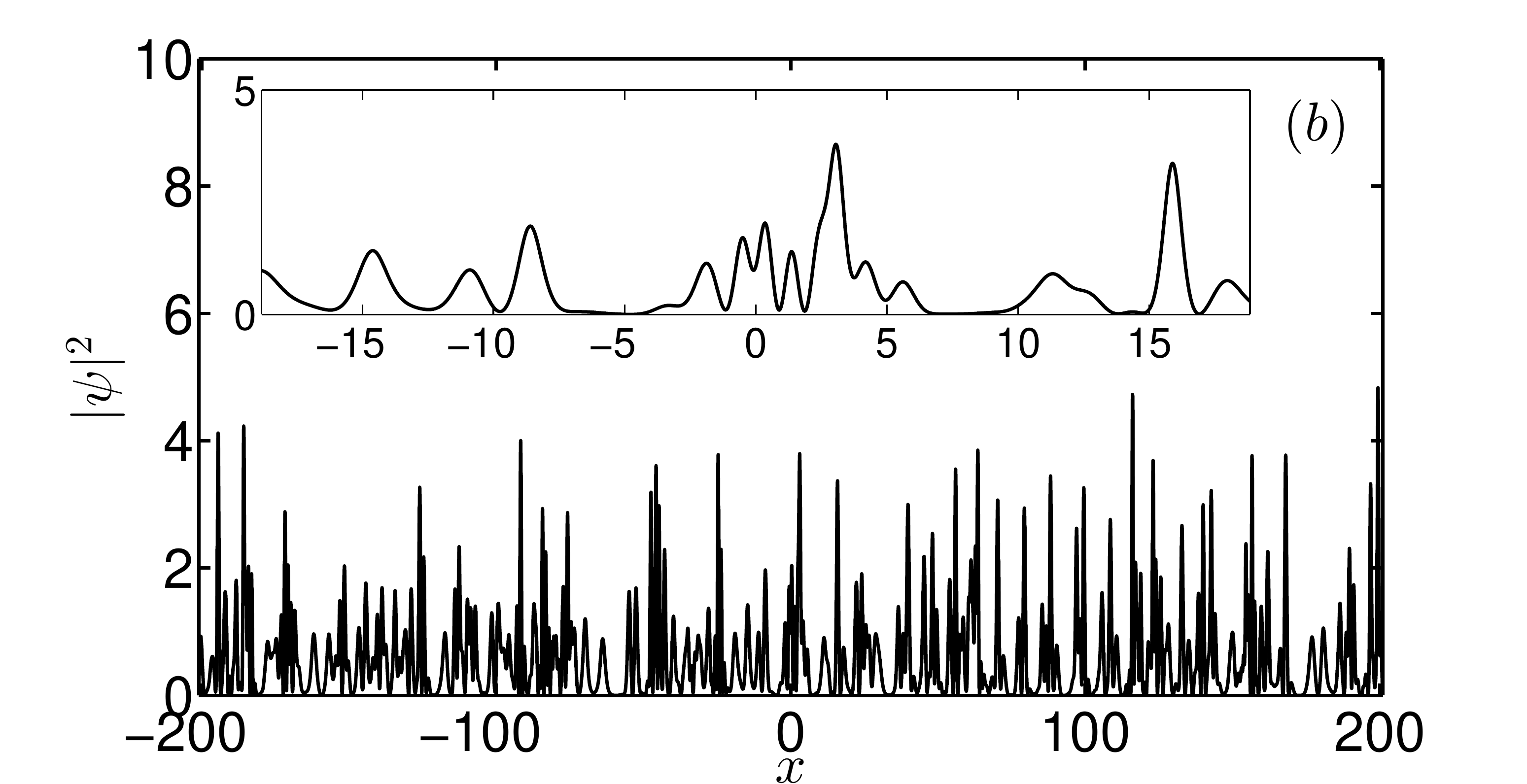}\\
	\includegraphics[width=8.8cm]{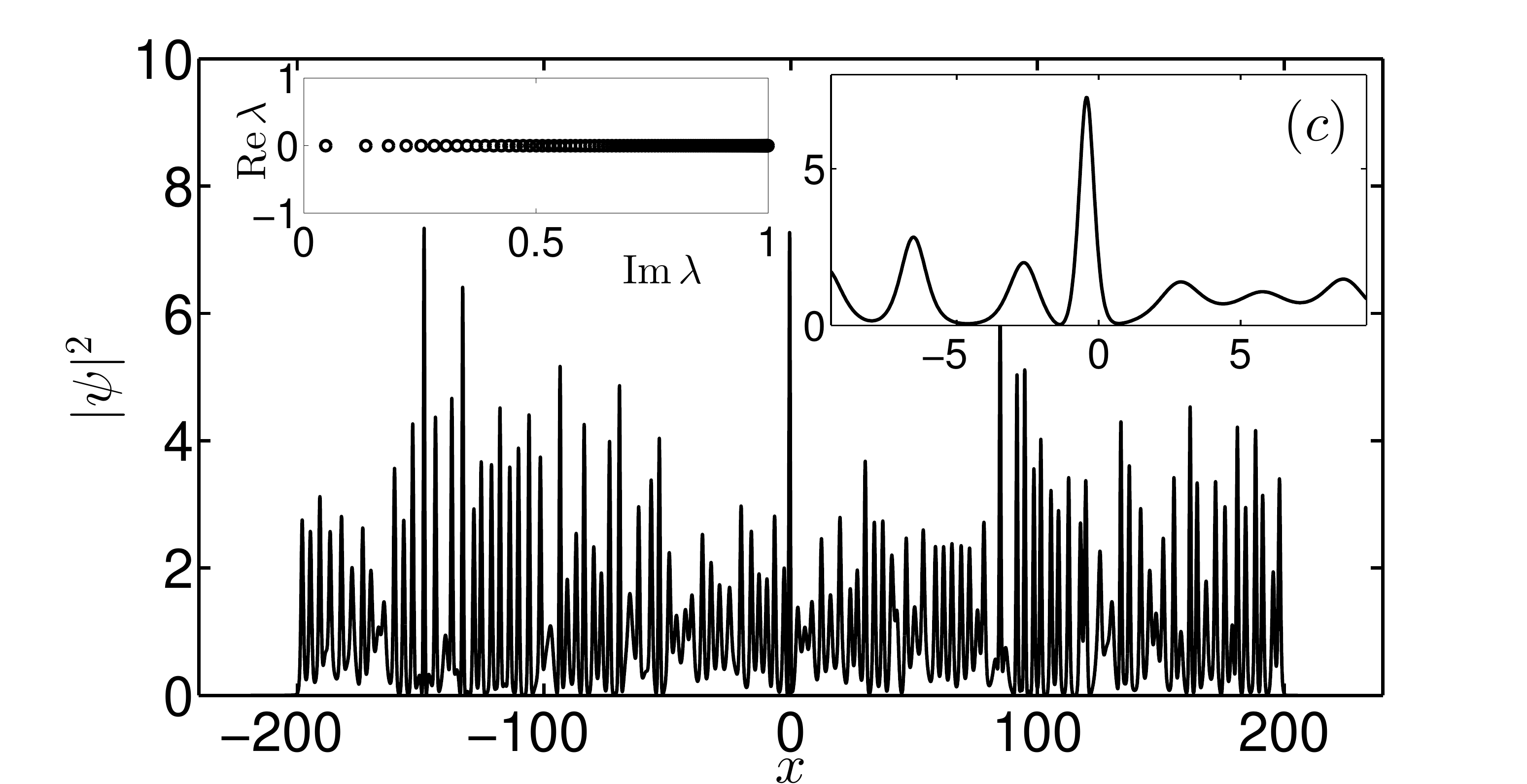}
	
        \caption{\small 
            (Adapted from~\cite{gelash2018strongly} and~\cite{gelash2019bound}) 
            (a,c) Wave fields of $128$-SS constructed from solitons having uniformly distributed positions $x^{\mathrm{DM}}_{n} \in [-L_0/2, L_0/2]$, $L_{0}=2$, and phases $\theta^{\mathrm{DM}}_{n} \in [0, 2\pi)$: (a) solitons have equal amplitudes $A_{n}=\pi/3.2\approx 1$ and Gaussian-distributed velocities with zero mean and standard deviation $V_{0} = 2$, $V_{n}\sim\mathcal{N}(0, V_{0}^{2})$, and (c) solitons have amplitudes distributed according to the Bohr-Sommerfeld quantization rule for a rectangular box, see Eq.~(\ref{eq:EIGfixed}) below, and zero velocities $V_n=0$. 
            (b) The wave field from panel (a) after placing it into the periodic box $x \in [-L_p/2, L_p/2]$, $L_p=128\pi$, and simulating the time evolution within the fNLS equation up to the final time $t=200$. 
            Right insets in panels (a,c) and the inset in panel (b) show zoom of the wave fields. 
            Left insets in panels (a,c) demonstrate soliton eigenvalues (note the swapped notations between the axes).
        }
\label{Fig:NLSE_NSS}
\end{figure}

Figure~\ref{Fig:NLSE_NSS} illustrates the ``periodization'' method on the example of a single $128$-SS generated from solitons having uniformly distributed positions $x^{\mathrm{DM}}_{n} \in [-L_0/2, L_0/2]$, $L_{0}=2$, and phases $\theta^{\mathrm{DM}}_{n} \in [0, 2\pi)$, equal amplitudes $A_{n}=\pi/3.2\approx 1$ and Gaussian-distributed velocities with zero mean and standard deviation $V_{0} = 2$, $V_{n}\sim\mathcal{N}(0, V_{0}^{2})$. 
The initial $128$-SS has characteristic size in the physical space $\delta X\simeq 280$, and in average the wave field is greater at the center than closer to the edges of the solution, see Fig.~\ref{Fig:NLSE_NSS}(a). 
Then, this solution is placed into the periodic box $x \in [-L_p/2, L_p/2]$, $L_p=128\pi$, and the evolution is simulated until the final time $t=200$, when the wave field in average becomes fairly uniform, see Fig.~\ref{Fig:NLSE_NSS}(b). 
As has been verified in~\cite{gelash2018strongly}, the soliton eigenvalues $\Lambda_n$ calculated at the final simulation time with the Fourier collocation method~\cite{YangBook2010} almost coincide with the eigenvalues $\lambda_n$ of the initial $128$-SS with the relative differences between the two $|\lambda_n - \Lambda_n|/|\lambda_n|$ of $10^{-9}$ order.

The described above ``periodization'' method can be applied only when the soliton velocities are distributed over some finite interval of values. 
In~\cite{gelash2019bound}, it has been observed that for the special case of bound-state soliton gas, i.e., when all solitons have the same velocity (which one can set to zero for simplicity), a certain distribution of soliton eigenvalues (i.e., amplitudes) leads to a statistically homogeneous multi-soliton wave fields in a wide region of the physical space for sufficiently small soliton positions $|x^{\mathrm{DM}}_{n}|\lesssim 1$ and random soliton phases; see Fig.~\ref{Fig:NLSE_NSS}(c). 
The figure shows $128$-SS constructed from solitons having zero velocity $V_n=0$ and uniformly distributed positions and phases over the intervals $x^{\mathrm{DM}}_{n} \in [-L_0/2, L_0/2]$, $L_{0}=2$, and $\theta^{\mathrm{DM}}_{n} \in [0, 2\pi)$. 
The amplitudes are distributed according to the Bohr-Sommerfeld quantization rule, which is deduced from the solution of the direct scattering problem for the rectangular box potential; see Sec.~\ref{subsec:MI} for detail. 
The resulting wave field turns out to be statistically homogeneous $\langle|\psi(x)|^2\rangle\approx 1$ over more than $70$\% of its characteristic size in the physical space for random soliton phases~\cite{gelash2019bound}. 
Cutting out the remaining $30$\% at the edges where the wave field is not statistically homogeneous, one can use this $70$\% part as a model of statistically homogeneous bound-state soliton gas. 
As discussed in Sec.~\ref{subsec:MI}, this soliton gas accurately models the long-time statistically stationary state of the noise-induced modulational instability of the plane wave solution. 

We believe that there are other distributions of soliton amplitudes leading to the statistically homogeneous multi-soliton wave fields in a wide region of the physical space for sufficiently small soliton positions $|x^{\mathrm{DM}}_{n}|\lesssim 1$ and random soliton phases. 
The general question of constructing multi-soliton bound-state wave fields with a given profile $\langle|\psi(x)|^2\rangle = P(x)$ in the physical space and a given set of amplitudes $A_{n}$ by using random soliton phases and a specific distribution of soliton positions represents a challenging problem for future studies.


\subsection{Direct scattering transform analysis}

In this subsection, we discuss the direct scattering transform (DST) analysis, which allows one to study the nonlinear composition of numerically or experimentally observed wave fields. 
Focusing only on the discrete spectrum (soliton eigenvalues and norming constants), we assume that the wave field in question is given in a simulation box $x\in [-L/2,L/2]$ and outside this box it equals zero. 
If the actual boundary conditions are different, then one can assume that the box $L$ is large enough compared to the characteristic sizes of nonlinear structures, so that the difference in the boundary conditions and the resulting edge effects can be neglected. 
Note that in this formulation the scattering coefficients $a(\lambda)$ and $b(\lambda)$ are analytic functions in the upper half of the $\lambda$-plane, that is essential for the algorithms discussed below. 

In what follows, we describe the DST procedure presented in the recent study~\cite{agafontsev2023bound}. 
This procedure, based on the standard DST methods~\cite{BofOsb1992, Burtsev1998,YangBook2010} supplemented by the latest studies~\cite{mullyadzhanov2019direct,gelash2020anomalous,mullyadzhanov2021magnus} for the accurate calculation of the norming constants, contains several steps which are discussed below. 


First, if there is a discontinuity of the wave field at $x=\pm L/2$, then it is smoothed using a smoothing window of the same size as the characteristic soliton width. 
It is assumed that the number of solitons inside the box $L$ is large and that these discontinuities, together with their smoothing, do not introduce significant inaccuracies in the results.

Second, an approximate location of the soliton eigenvalues is found using the standard Fourier collocation method~\cite{YangBook2010}. 
Being fast and fairly accurate, this method is based on the Fourier decomposition of the wave field, which artificially shifts the continuous spectrum eigenvalues to the upper half of the $\lambda$-plane due to the implied periodization. 
Also, it does not distinguish between the eigenvalues of discrete and continuous spectra, leading to the problem of identifying low-amplitude solitons. 

Third, to cope this this problem, the wave field is considered in two larger boxes $x\in[-3L/4, 3L/4]$ and $x\in[-L, L]$ by filling with zeros $\psi=0$ the intervals where the wave field is not defined. 
Then, the Fourier collocation method is executed in both boxes and only the eigenvalues coinciding in both calculations are selected as belonging to the discrete spectrum. 
While the latter provides a good approximation of the soliton eigenvalues, i.e., zeros of the coefficient $a(\lambda)$, this approximation is still insufficient for the accurate calculation of the norming constants, which requires knowledge of roots $a(\lambda_n)=0$ to hundreds of digits~\cite{gelash2020anomalous}. 
That is why the calculated eigenvalues are then used as seeding values for a high-accuracy root-finding procedure.

The fourth and the final step of the described DST procedure consists in application of the standard second-order Boffetta--Osborne method~\cite{BofOsb1992} on a fine interpolated grid using high-precision arithmetic operations, as suggested in~\cite{mullyadzhanov2019direct, gelash2020anomalous}. 
The Boffetta--Osborne method is based on the calculation of the so-called extended $4 \times 4$ scattering matrix $\mathbf{S}$, which translates the solution of the Zakharov--Shabat system $\mathbf{\Phi}$ together with its derivative $\mathbf{\Phi}' = \partial \mathbf{\Phi} / \partial \lambda$ from $x=-L$ to $x=L$,
\begin{equation}
    \begin{pmatrix} \mathbf{\Phi}(L) \\ \mathbf{\Phi}'(L) \end{pmatrix} = 
    \underbrace{ \begin{pmatrix} \Sigma & 0 \\ \Sigma' & \Sigma \end{pmatrix} }_{\mathbf{S}}
    \begin{pmatrix} \mathbf{\Phi}(-L) \\ \mathbf{\Phi}'(-L) \end{pmatrix}. \label{Smatrix}
\end{equation}
Here $\Sigma(\lambda)$ is $2\times 2$ matrix, such that $\mathbf{\Phi}(L) = \Sigma(\lambda) \mathbf{\Phi}(-L)$, and the scattering coefficients are connected with the elements of matrix $\mathbf{S}$ as
\begin{eqnarray}
    a(\lambda) &=& S_{11} e^{2 i \lambda L}, \quad b(\lambda) = S_{21}, \nonumber\\
    a'(\lambda) &=& [ S_{31} + i L (S_{11}+S_{33}) ]\, e^{2 i \lambda L}. \label{abViaS}
\end{eqnarray}
Note that instead of the standard second-order Boffetta--Osborne method one can use the higher-order methods obtained with the Magnus expansion; see~\cite{mullyadzhanov2019direct,mullyadzhanov2021magnus} for detail. 
A fine spatial grid and the high-precision arithmetic operations are necessary to (i) neglect the round-off errors when calculating the wave function $\mathbf{\Phi}$ of the Zakharov--Shabat system, (ii) avoid the anomalous errors in computation of the norming constants, and (iii) suppress the numerical instability of the wave scattering through a large potential, see~\cite{mullyadzhanov2019direct,gelash2020anomalous} for detail. Also note that when avoiding the anomalous errors, one can supplement the DST procedure with the bidirectional algorithm and its improvements, see \cite{prins2019soliton}, to decrease the necessary number of digits in the high-precision operations.

The Boffetta--Osborne method allows one to find the scattering coefficients $a(\lambda)$ and $b(\lambda)$ for any value of $\lambda$ by the direct numerical integration of the Zakharov--Shabat system on the interval $[-L/2,L/2]$ with boundary conditions~(\ref{ScatteringProblem1}). 
Note that $a(\lambda)$ and $b(\lambda)$ are analytic functions in the upper-half of the $\lambda$-plane, as the potential $\psi(x)$ has a compact support. 
Then, with the help of the Newton method, one can find roots $a(\lambda_n)=0$ with the necessary precision by using the eigenvalues obtained by the Fourier collocation method as seeding values. 
Finally, the norming constants are calculated according to their definition~(\ref{ScatteringData}) using the extended scattering matrix $\mathbf{S}$, see Eq.~(\ref{abViaS}), to find the derivative $a'(\lambda)$. 

\begin{figure}[t]\centering
	\includegraphics[width=8.9cm]{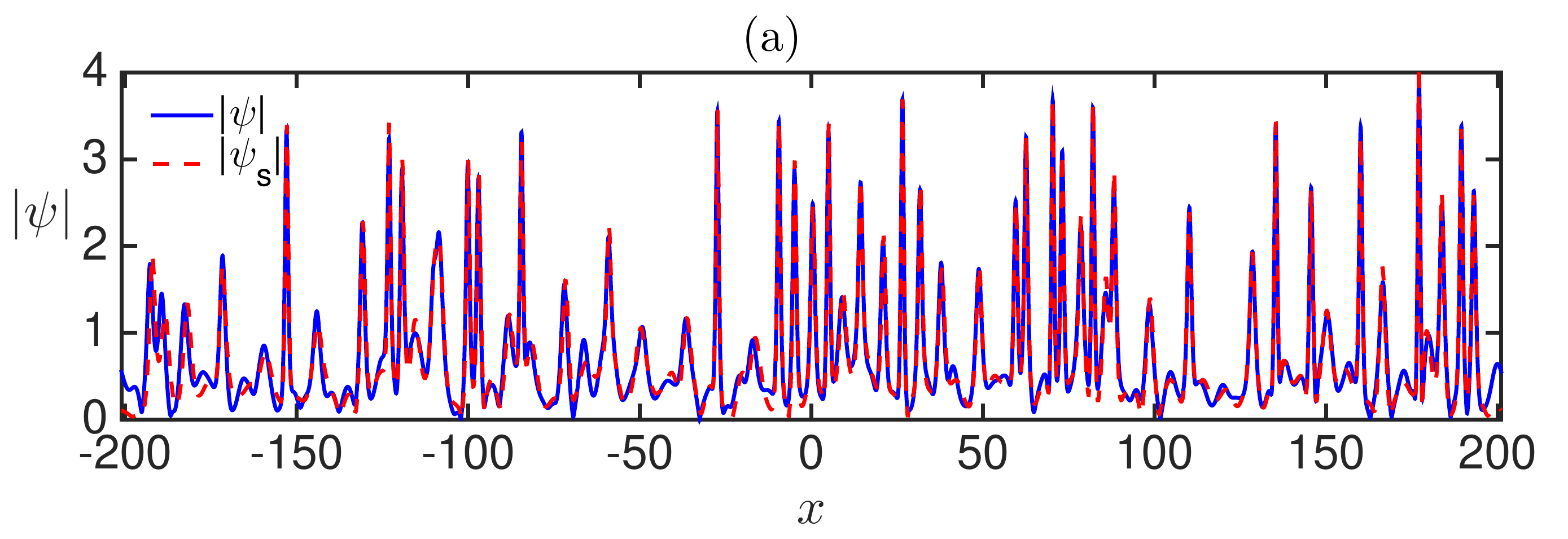}\\
        \includegraphics[width=8.9cm]{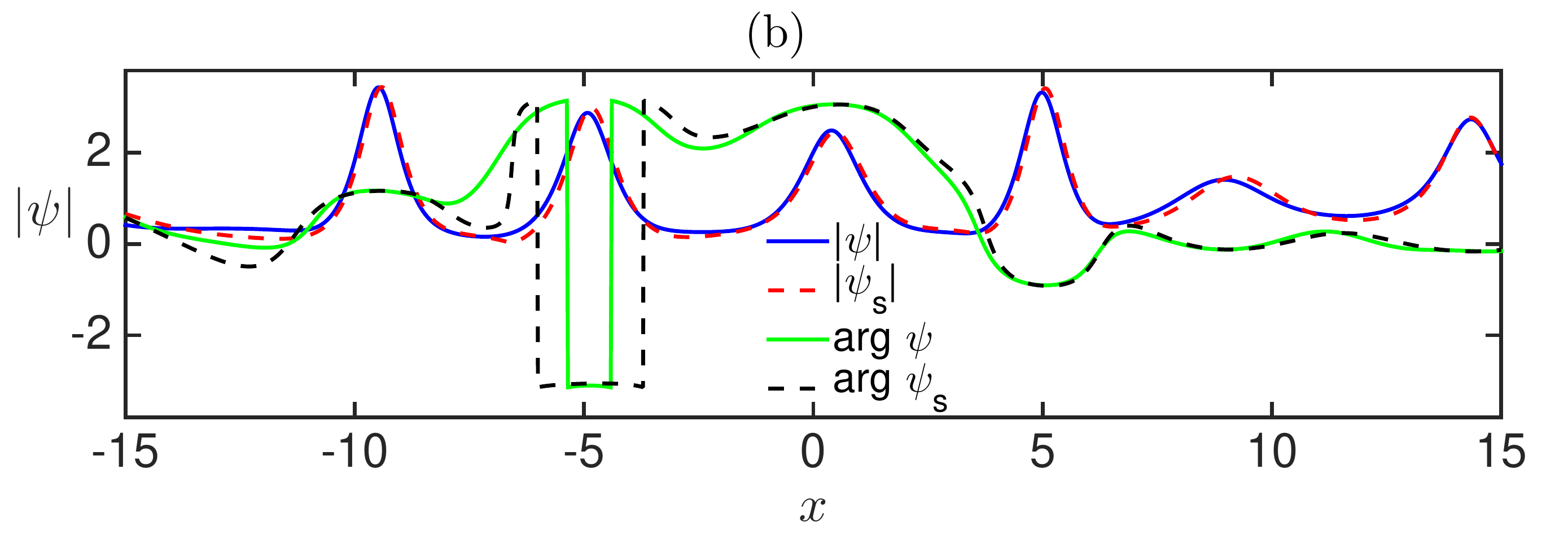}\\
	\includegraphics[width=4.2cm]{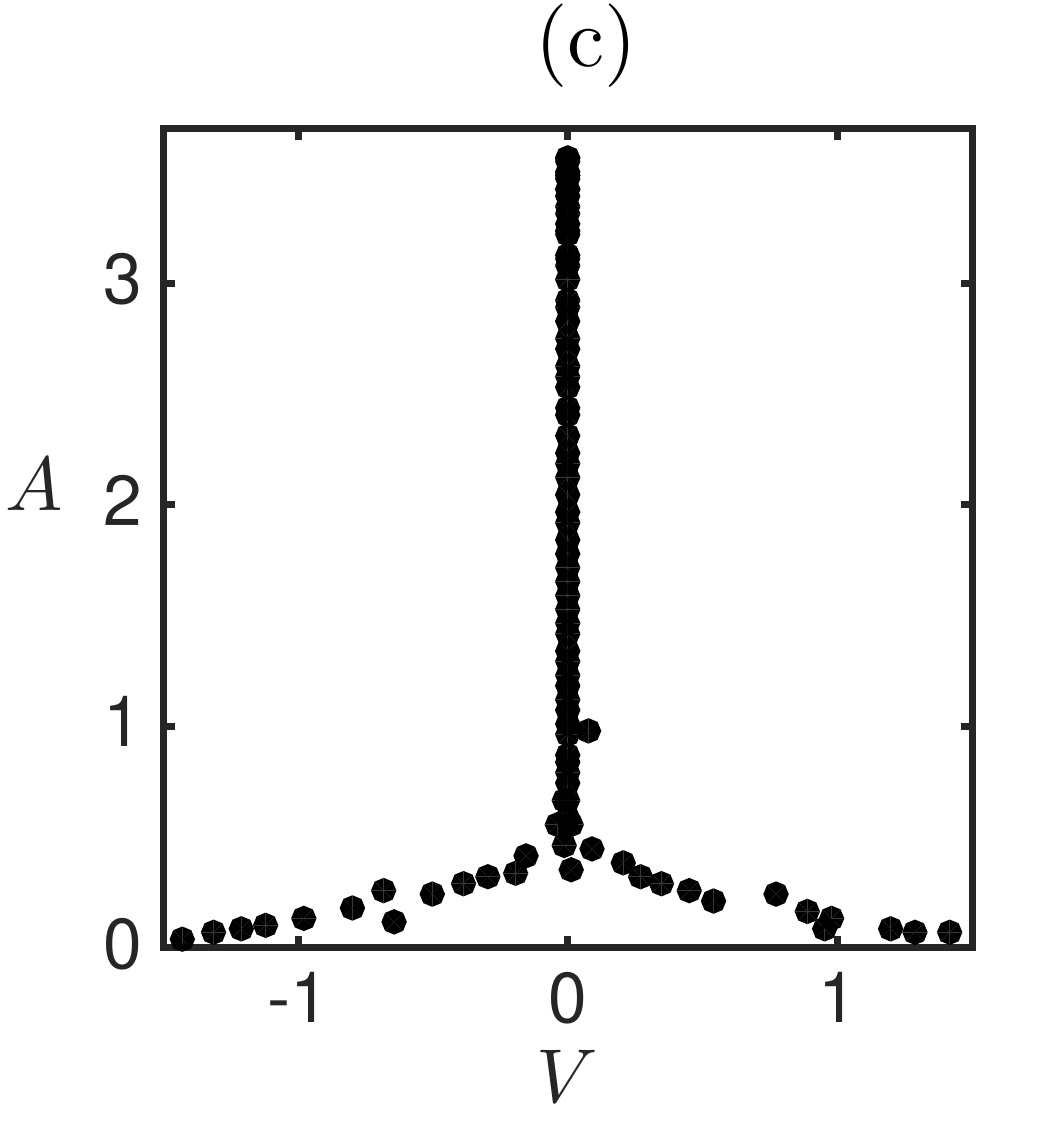}
	\includegraphics[width=4.2cm]{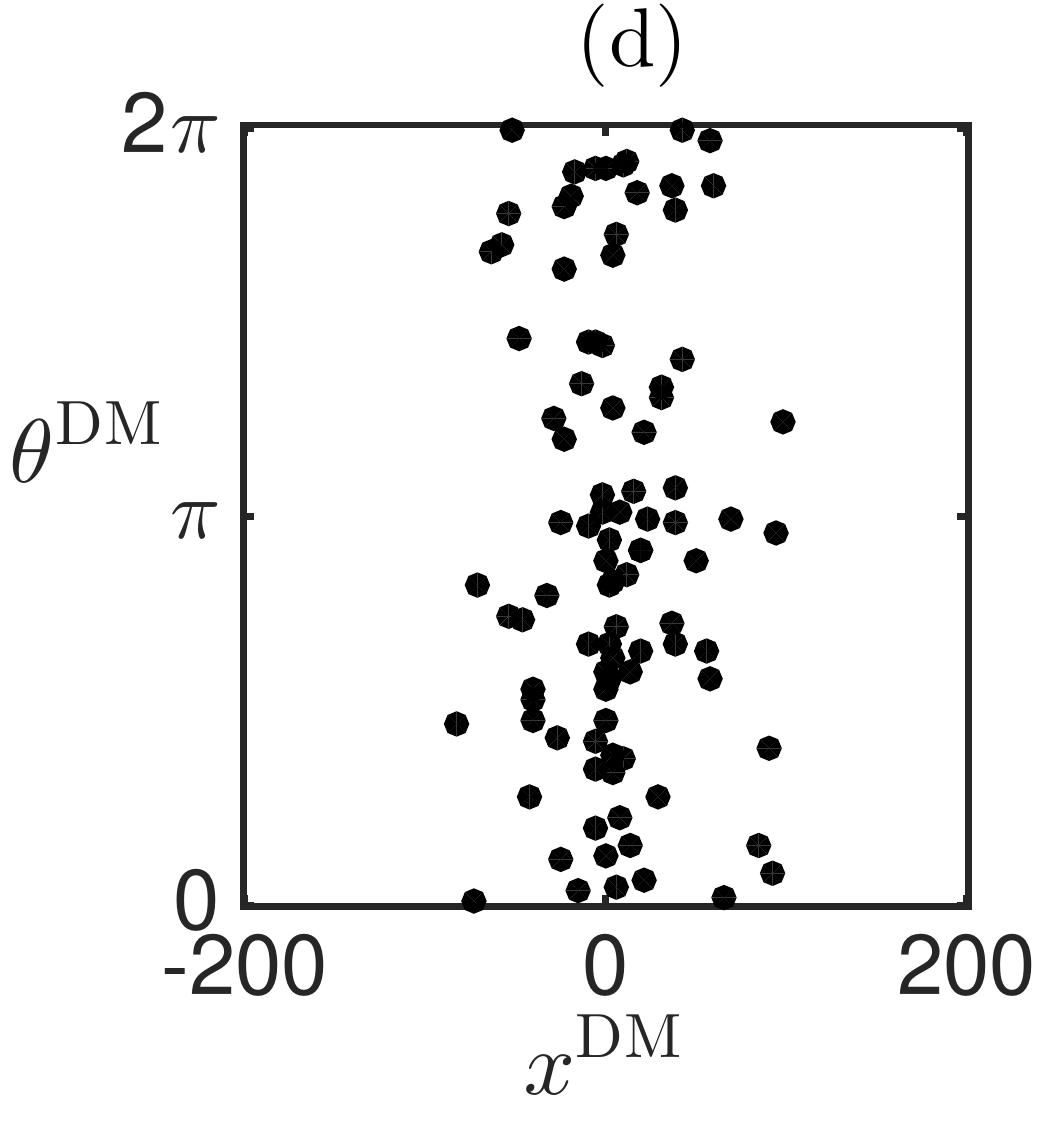}
        
	\caption{\small 
            (Adapted from~\cite{agafontsev2023bound}) Numerical DST analysis of a periodic wave field that was ``grown'' from small statistically homogeneous in space noise within the fNLS equation, supplemented by a small linear pumping term, until the intensity, averaged over the simulation box, reached unity, $\overline{|\psi|^2} = 1$; see~\cite{agafontsev2023bound} for details. 
            Panel (a) shows the absolute values of the ``grown up'' wave field $|\psi|$ (solid blue) and the multi-soliton solution $|\psi_{s}|$ (dashed red); $\psi_{s}$ is constructed using the soliton parameters obtained in the DST procedure. 
            Panel (b) represents zoom of panel (a), also demonstrating the complex phases of the ``grown up'' wave field (solid green) and the multi-soliton solution (dashed black). 
            The dots in panels (c) and (d) illustrate soliton amplitudes $A_n$, velocities $V_n$, positions $x^{\mathrm{DM}}_n$ and phases $\theta^{\mathrm{DM}}_n$.
	}
\label{Fig:NLSE_DST_example}
\end{figure}

Figure~\ref{Fig:NLSE_DST_example} illustrates the performance of this DST procedure on the example of a periodic wave field that was ``grown'' from small statistically homogeneous in space noise within the fNLS equation, supplemented by a small linear pumping term, until the intensity, averaged over the simulation box, reached unity, $\overline{|\psi|^2} = 1$; see~\cite{agafontsev2023bound} for details. 
The solid blue and green lines in Fig.~\ref{Fig:NLSE_DST_example}(a,b) show the amplitude $|\psi|$ and complex phase $\mathrm{arg}\,\psi$ of the ``grown up'' wave field, while the dots in Fig.~\ref{Fig:NLSE_DST_example}(c,d) demonstrate the calculated soliton amplitudes $A_n$, velocities $V_n$, positions $x^{\mathrm{DM}}_n$ and phases $\theta^{\mathrm{DM}}_n$. 
Using these soliton parameters, one can construct the corresponding exact multi-soliton solution as discussed in the previous subsection; it turns out that this solution approximates the original wave field very well, as illustrated by the dashed red and black lines in Fig.~\ref{Fig:NLSE_DST_example}(a,b). 

Note that the average intensity of the multi-soliton solution $\psi_{s}$ in Fig.~\ref{Fig:NLSE_DST_example} equals $99$\% of that of the original ``grown up'' wave field $\psi$. 
Also, most of the solitons of this solution have zero velocities, forming a bound state. 
In~\cite{agafontsev2023bound}, such a situation is observed if the initial noise amplitude and the pumping coefficient are small enough. 
If this is not the case, then the ``grown up'' wave fields with intensity of unity order $\overline{|\psi|^2} \sim 1$ also represent solitons-dominated states, which are not bound as these solitons have different velocities. 
Moreover, as shown in the paper, during the growth stage the solitonic part of the wave field becomes the dominant one very early when the average intensity is still small, $\overline{|\psi|^2} \simeq 0.1$, and the dispersion effects are leading in the dynamics. 
These observations indicate that the soliton gas model can be applicable even to weakly nonlinear cases, so that a soliton gas can be a very common object in nature.


\section{Experiments}
\label{Sec:Exp}

From the experimental point of view, a few attempts to generate and to observe soliton gases have been made in some optical fiber experiments performed at the end of the 1990's \cite{Schwache:97,Steinmeyer:95,Mitschke:96}. The soliton gas was generated by the synchronous injection of laser pulses inside a passive ring cavity. No direct observation but only averaged measurements of the Fourier power spectrum and of the second-order autocorrelation function characterizing the optical soliton gas have been reported in these pioneering experiments. Moreover, the dynamics of the ring resonator was so complex that many features ranging from purely temporal chaos to spatio-temporal chaos or turbulence were observed in this fiber system \cite{Mitschke:96}. 

Analysing ocean waves recordings, Costa et al. have reported the observation of random wavepackets in shallow water waves in 2014 \cite{Costa:14}. The wavepackets have been analyzed using numerical tools of nonlinear spectral analysis \cite{Osborne:19} and interpreted as being composed of random solitons that might be associated with KdV soliton gas. One year later, large ensembles of interacting and colliding solitons have been observed in a laboratory environment \cite{Perrard:15}. The experimental system was a water cylinder deposited on a heated channel and levitating on its own generated vapor film owing to the Leidenfrost effect. Multiple soliton propagation was observed at the surface of the water cylinder and the Fourier analysis that was made in an attempt to characterize the multiple coherent structures revealed a “soliton turbulence-like spectrum”. Note also that a striking transition between weak turbulence and solitonic regimes has been evidenced in the hydrodynamic experiments reported in ref. \cite{Hassaini:17}. In these experiments, water waves have been generated by exciting horizontally a water container by using an oscillating table. The weak turbulence regime observed at low forcing and/or large depth was shown to abruptly evolve into a solitonic regime at larger forcing and/or small depth. Remarkably, these results establish a possible link between the field of integrable turbulence and the field of wave turbulence.

In the recent laboratory experiments reported in ref. \cite{Redor:19}, Redor et al. have taken advantage of the process of fission of a sinusoidal wave train to generate a bidirectional shallow water soliton gas in a 34-m long flume. The space-time observations have revealed complex dynamics where large numbers of colliding solitons retained their profile adiabatically, though their amplitude was slowly decaying because of some unavoidable damping. The Fourier analysis of the observed nonlinear wave field has clearly revealed the interplay between multiple solitons and dispersive radiation. Further analysis making use of the periodic scattering transform have been  implemented in ref. \cite{Redor:21} to discriminate linear wave motion states from integrable turbulence and soliton gas. Moreover the statistical properties of the soliton gas have been given in terms of probability density distribution, skewness, and kurtosis \cite{Redor:21}. 

The experiments reported in ref. \cite{Redor:19,Redor:20,Redor:21} have been made in the presence of an unavoidable slow damping but it has been shown that a stationary state typified by the interplay among random bidirectional solitons can be achieved because of the continuous energy input by the wavemaker. In these shallow water experiments, a route to integrable turbulence has been discovered through the disorganization of wave motion that is induced by the wave maker \cite{Redor:21}. This route has been shown to depend on the nonlinearity of the waves but also on the amplitude amplification and reduction due to the wavemaker feedback on the wave field \cite{Redor:21}.

\begin{figure*}[!t]\centering
	\includegraphics[width=16cm]{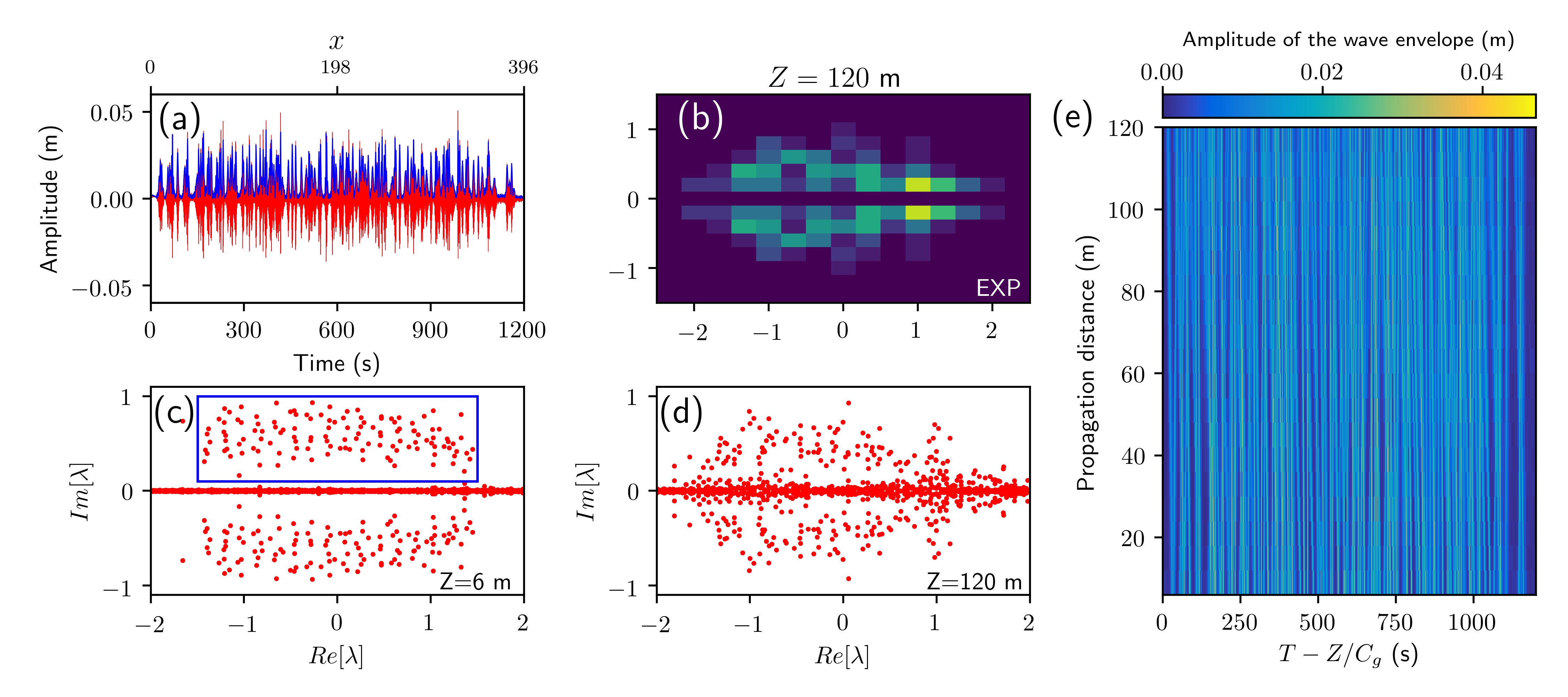}\\
	\caption{Gas of 128 solitons propagating in a 140-m long 1D water tank \cite{Suret:20}. (a) Water elevation (red line) and modulus of the wave envelope measured at Z=6 m, close to the wave maker. (b) DOS of the soliton gas measured at Z=120 m. (c) Discrete IST spectrum measured at Z = 6 m. (d) Discrete IST spectrum measured at Z=120 m. (e) Space-time evolution of modulus of the wave envelope recorded by the 20 gauges regularly spaced along the tank. Reproduced with permissions from~\cite{Suret:20}}
\label{fig:SG_exp}
\end{figure*}

Using an approach fully based on the IST method while also relying on the concept of DOS, a soliton gas has been generated in hydrodynamic experiments performed in the deep-water regime where wave propagation is described at leading order by the 1D fNLS equation \cite{Suret:20}. The experiment has been performed in a wave flume being 148 m long, 5 m wide, and 3 m deep. Unidirectional waves have been generated at one end of the tank with a computer assisted flap-type wave maker and the flume is equipped with an absorbing device strongly reducing wave reflection at the opposite end. In these experiments the space-time evolution of the generated wave packet is measured with 20 gauges uniformly distributed along the tank. 

The experiment reported in ref. \cite{Suret:20} starts from the numerical generation or synthesis of a soliton gas by using the methodology described in Sec. \ref{Sec:IST}. An ensemble of $128$ solitons having spectral parameters being distributed in a rectangular region of the spectral IST plane has been numerically generated. The solitons have the modulus of their norming constants being equal to unity while their phases are randomly distributed between $-\pi$ and $+\pi$. In the experiment, the generated soliton gas has the form of a random wave field spreading over $\sim 1200$ s, see Fig. \ref{fig:SG_exp}. It represents a {\it dense} soliton gas in where solitons are not isolated and not well separated like in a rarefied gas. 

In the experiments reported in ref. \cite{Suret:20}, a large number of discrete eigenvalues were distributed with some density within a limited region of the complex plane. This justifies the introduction of a statistical description of the spectral (IST) data. This represents a
key point for the analysis of the observed wave field in the framework of the SG theory. In ref. \cite{Suret:20}, the DOS of a homogeneous soliton gas has been measured for the first time in experiments, which provides an essential first step towards experimental verification of the
kinetic theory of nonequilibrium SGs. Nonlinear spectral analysis of the generated hydrodynamic soliton gas reveals that the density of states slowly changes under the influence of perturbative higher-order effects that break the integrability of the wave dynamics.

\section{Applications of soliton gases}
\label{sec:Applications}

Since the first paper of Zakharov~\cite{zakharov1971kinetic}, a peculiar interest has been ascribed to SG as a fundamental mathematical and physical concept. Importantly, it has been recently shown that SG theory can provide a powerful framework to describe theoretically the complex statistics underlying some well-known and fundamental nonlinear dispersive waves phenomena. It is indeed natural to expect that SG theory can be  used to describe some specific regimes of integrable systems (integrable turbulence, see Sec.~\ref{subsec:IntegrableTurbulence})~\cite{Agafontsev:15, Walczak2015Optical,Soto2016integrable,Akhmediev2016breather,Suret2016Single,Michel2020Emergence,Tikan2018Single}. In particular, by using numerical simulations, it has been shown in 2019 that the long-term statistical properties of the so-called {\it spontaneous modulation instability} coincides with those of a specifically-designed SG(see Sec.\ref{subsec:MI})~\cite{gelash2019bound}. Very recently, the general relationship between the DOS of a SG and the kurtosis (second-order moment) of the waves has been derived (see Sec.\ref{subsec:openStatistics}). This approach provides  the first theoretical description of the long-term evolution of the noise-induced modulation instability and paves the way of the description of integrable turbulence by using SG theory.

\subsection{Integrable turbulence}
\label{subsec:IntegrableTurbulence}

Wave turbulence can be generally defined as the ensemble of all the complex phenomena emerging in random nonlinear waves systems. The phrase {\it wave turbulence theory} is often used in a more restrictive sense and is then defined as the statistical mechanics of {\it weakly nonlinear} and dispersive waves~\cite{Nazarenko, Zakharov}. The standard wave turbulence theory applies to {\it non integrable} waves systems in which the Physics at long-time is dominated by resonant interactions~\cite{Picozzi:14}. A general feature of wave turbulence is the exchange of energy among all the scales induced by the nonlinearity. If there is a source of energy for some given spatial scales (for example the large scales) and if other scales (for example small scales) are damped by dissipation, wave turbulence theory predicts the possible existence of the so-called Kolmogorov-Zakharov cascade. This corresponds to an out-of-equilibirum phenomenon characterized by a constant flux of energy between large and small scales~\cite{Nazarenko, Zakharov, Picozzi:14}. On the other hand, in the absence of energy source and of losses, if the wave system is Hamiltonian, nonlinear random waves may reach a thermodynamical equilibrium state (characterized by the equipartition of energy and the Rayleigh-Jeans distribution)~\cite{Nazarenko, Zakharov, Picozzi:14,PourbeyramRayleighJeans2022,SuretRayleighJeans2022}. Note that, in common wave systems, both the Kolmogorov-Zakharov and the Rayleigh-Jeans are characterized by power-lawed spectra.

The Physics of {\it integrable} waves systems is of profoundly different nature because of the infinite number of constants of motion and of the absence of resonances. In particular, nonlinear random integrable waves cannot reach the thermodynamical Rayleigh-Jeans equilibrium~\cite{Picozzi:14,suret2011wave}. For this reason, Zakharov has introduced a new field of research, the {\it integrable turbulence} (IT), that is defined as the statistical description of integrable systems~\cite{Zakharov:09}. Since this seminal paper in 2009, integrable turbulence has received a growing interest both from the theoretical\cite{Zakharov:09, Zakharov:13,Pelinovsky:13, Agafontsev:15, Agafontsev:16, Randoux2016Nonlinear, zakharov2016non, dyachenko2016primitive, Akhmediev2016breather, Soto2016integrable, Suret:2017book, gelash2019bound, didenkulova2019numerical, xie2020integrable, agafontsev2020growing} and experimental~\cite{Walczak2015Optical, Suret2016Single, Koussaifi2017Spontaneous, Suret:2017book, Tikan2018Single, kraych2019statistical, Michel2020Emergence} points of view.\\

In practice, integrable turbulence corresponds to the propagation of random waves in  systems described by integrable equations such as the 1D NLS, the KdV or the Sine-Gordon equations. In this Sec.~\ref{sec:Applications}, we focus on recent results on the 1DNLS integrable turbulence. The one dimensional focusing NLS equation provides a bridge between nonlinear optics and hydrodynamics\cite{Chabchoub:15, Koussaifi2017Spontaneous}. The 1D focusing NLS equation describes  at leading order deep-water wave trains or optical fiber in anomalous dispersion regime and it plays a central role in the study of rogue waves \cite{Onorato:01, Onorato:13,  Akhmediev:13, Dudley:14,  dudley2019rogue}. The relevent of approach to study nonlinear random waves is a statistical description, including probability density functions (PDF) of wave amplitude $\psi$ or of intensity $|\psi|^2$ and moments such as the kurtosis $\kappa_4=\langle |\psi|^4\rangle/\langle |\psi|^2\rangle^2$. The last years, the statistical properties of integrable turbulence has been widely studied by using numerical simulation of the NLS equations. Preserving integrability in long-term simulations is a delicate and challenging task, but to the best of the knowledge, integrable turbulence is characterized by stationary statistical properties of the field for long time $t$.  This existence of a stationary statistical states in the long-time evolution of the waves system is the most fundamental known feature of integrable turbulence.\\

Integrable turbulence phenomena can be classified by considering the statistical properties of the initial conditions. Two classes of initial conditions have been extensively investigated : (i) the plane wave perturbed by a small random noise and, (ii) partially coherent waves. 

The homogeneous solution of the 1D focusing NLS equation (the plane wave or {\it condensate}) is unstable in the presence of long wave perturbation. When the perturbation is a random process, this dynamical mechanism, known as the ``noise-induced'' or {\it spontaneous modulation instability} (MI)~\cite{Agafontsev:15, Toenger:15,kraych2019statistical}, represents a prominent example of the integrable turbulence phenomena. Surprisingly, the long-term statistical state is characterized by a Gaussian local statistics of the field $\psi$ {\it i.e.} by a kurtosis $\kappa_4=2$~\cite{Agafontsev:15} while the other statistical properties such as the Fourier spectra or two-points correlations are not trivial~\cite{kraych2019statistical}. These statistical quantities have been quantitatively measured in experiments but up to very recent studies, no theoretical description was available. In the Sec.~\ref{subsec:MI}, we show that the soliton gas concept provides a powerful theoretical tool to predict quantitatively the statistical properties of the long term evolution of the spontanesous modulation instability.

Partially coherent waves are random waves characterized by a finite typical spatial scale (or identically a finite typical spectral width). Partially coherent waves made of numerous statistically independent modes represent the standard ansatz for initial conditions in the wave turbulence theory~\cite{Nazarenko, Picozzi:14} and exhibit Gaussian statistics. Such initial conditions have been extensively investigated in numerical simulations of defocusing and of focusing NLS equation and in experiments~\cite{Randoux:14, Walczak2015Optical,Suret2016Single,Randoux2016Nonlinear, Koussaifi2017Spontaneous, Michel2020Emergence, Tikan2018Single}.  Starting with partially coherent waves initial conditions, the statistics deviates from Gaussian statistics as the time evolves and eventually reaches a heavy or a low tailed PDF in the focusing and defocusing regime respectively~\cite{Randoux2016Nonlinear}. The deviation from Gaussianity in the stationary statistical state increases together with the strength of the nonlinearity and, importantly, in the focusing regime, the strongest deviation is characterized by a kurtosis $k_4=4$~\cite{agafontsev2021extreme}. Note that the higher is the kurtosis, the higher is the rate of emergence of extreme events (rogue waves). 

The evolution of partially coherent waves in (non integrable) weakly nonlinear dispersive systems corresponds to the fundamental question of wave thermalization that has been widely investigated in wave turbulence theory. The evolution of the statistical properties of partially coherent waves in the framework of 1D NLS integrable turbulence can be described by using  a non conventional wave turbulence theory approach~\cite{Janssen:03,suret2011wave}. This theoretical approach predicts the deviation from Gaussianity for a weak nonlinearity but is not valid in the high nonlinearity regime. In particular, up to now, the maximum value of the kurtosis $k_4=4$ was not understood. In Sec~\ref{subsec:openStatistics}, we summarize an extremely recent theoretical study which provides a demonstration of the maximum value of the kurtosis in the strongly nonlinear regime by using soliton gas theory.\\

It is important to note that {\it soliton gas is a peculiar case of integrable turbulence}. Indeed, in the framework of IST with zero boundary conditions, integrable turbulence can always be described by the combination of the discrete and of the continuous spectra. Our conjecture is that the high nonlinearity limit of integrable turbulence can always be described by purely solitonic solutions (soliton gas). We show in the Sec.~\ref{subsec:MI} and important example illustrating this conjecture.

\subsection{Spontaneous Modulation Instability}
\label{subsec:MI}

The MI appears in many physical systems, such as deep water waves~\cite{Osborne}, Bose-Einsteine condensates~\cite{Strecker:02} or nonlinear optical waves~\cite{Agrawal:2013}. If the plane wave is perturbed by  an initially small sinusoidal perturbation,  the nonlinear stage of MI is  described by homoclinic solutions of the 1D focusing NLS equation -- the Akhmediev Breathers~\cite{Akhmediev:85,akhmediev1986modulation,Akhmediev2009Waves, Grinevich:18}. As reminded above, in the case of random initial perturbation, single-point statistics evolves toward a stationary Gaussian distribution (and  $\kappa_4=\langle |\psi|^4\rangle/\langle |\psi|^2\rangle^2=2$.) despite the presence of  highly nonlinear breather-like structures ~\cite{Agafontsev:15,Soto2016integrable,Akhmediev2016breather}. The long-time (stationary) statistics is also typified by a quasi-periodic structure of the  autocorrelation function $g^{(2)}$ of the wave field intensity~\cite{kraych2019statistical}. \\

In this section, we review numerical simulations that proves that the nonlinear stage of the spontaneous MI in the focusing regime of the Eq.~\ref{eq:fNLS} ($\sigma=+1$) can be quantitatively described by a specifically-designed soliton gas~\cite{gelash2019bound}.\\

\begin{figure}[h!]
\centering
\includegraphics[width=3.4in]{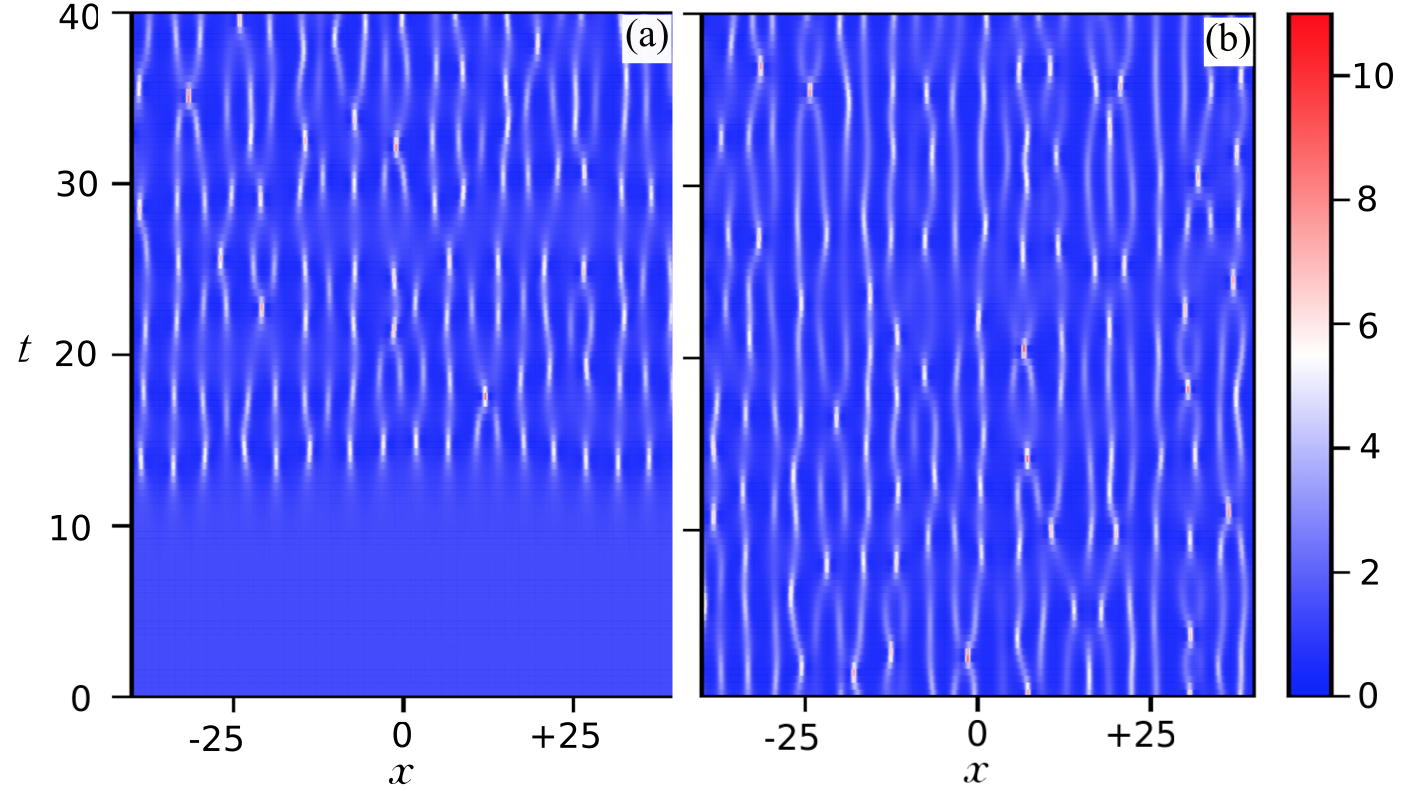}
\caption{\label{Fig:MIspatio}{\bf Numerical simulations of the one dimensional focusing NLS equation : Space-Time diagrams of $|\psi(x,t)|^2$} (a) Noise-induced Modulational Instability of a plane wave (periodic boundary conditions). (b) Dynamics of the random phase bound $N$-SS (only the central part of the $N$-SS having a total width $L_0\simeq 400$ is plotted).\\
Reproduced with permission from~\cite{gelash2019bound}}
\end{figure}

Without loss of generality, we consider the plane wave solution of Eq.\ref{eq:fNLS} -- the condensate --  of unit  amplitude $\psi_c(t,x) = \exp{i t}$. In the classical formulation of the spontaneous MI problem, the initial condition reads~\cite{Zakharov:MI:09,Agafontsev:15}:
\begin{equation}\label{eq:IC}
 \psi(t=0,x) = 1 + \eta(x)\,,
\end{equation}
where $\eta$ is a small noise, $\langle|\eta|^{2}\rangle\ll 1$, with zero average, $\langle\eta\rangle=0$. The destabilization of the condensate  with respect to long-wave perturbations was widely investigated, both numerically by using periodic boundary conditions in a box of large size ~\cite{Dudley:14, Toenger:15, Agafontsev:15,kraych2019statistical}, and experimentally~\cite{Narhi:16,kraych2019statistical, lebel2021single}. The typical spatio-temporal dynamics of the spontaneous MI can be seen in Fig.~\ref{Fig:MIspatio}.a. \\

In order to demonstrate that the statistical properties of the spontaneous MI coincide at long-time with the ones of SG, the first  step used in~\cite{gelash2019bound} is to modify the boundary conditions. The idea is that if one fixes the time at which  the nonlinear stage of MI is characterized (typically $t>30$ in Fig.~\ref{Fig:MIspatio}.a), the plane wave with periodic boundary condition can be replaced by a box with ZBC. The width of the box has to be sufficiently large to avoid any influence of the edges in the central part of the box at the considered time $t$. 

By using the same idea, one expects that any homogeneous SG can be locally modeled by a N-SS (with zero boundary conditions). Moreover, in order to model the long-time dynamics of a stochastic field, it is natural to assume random norming constants phases because  the phase rotations $-2i \lambda_n^2 t$ for large $t$  introduce an effective randomization.  Note that somehow, this random phases of the norming constant are similar to the so-called ``random phase approximation'' ({\it i.e.} random phases of the Fourier components) in wave turbulence theory~\cite{Zakharov, Nazarenko}.

Finally, the last step is to determine  the DOS. of the SG underlying the dynamics of the field. Here, the answer is rather simple because the discrete spectrum of a real-valued rectangular box of unit amplitude and width $L_{0}$ is known. In the limit $L_{0}\gg 1$,  the discrete spectrum of the semi-classical Zakharov-Shabat scattering problem is given by the Bohr-Sommerfeld quantization rule, see e.g.~\cite{Zakharov:1972Exact,novikov1984theory} and also~\cite{Lewis:85}:
\begin{equation}\label{eq:EIGfixed}
\lambda_n = i\, \beta_n = i\, \sqrt{1-\left[\frac{\pi(n-\frac{1}{2})}{ L_0} \right]^2}, \quad n=1, 2, \dots, N,
\end{equation}
where $N=\hbox{int}[L_0/\pi]$ (the density of the gas, {\it i.e.} the number of solitons per unit length is thus $1/\pi$). The continuus limit of Eq. \eqref{eq:EIGfixed} with $N \to \infty$, $\beta_n \to \beta$ gives the normalized distribution  $\varphi(\beta)$ of the IST eigenvalues:
\begin{equation}\label{eq:EIGdistr}
\varphi(\beta) = \frac{1}{N}\frac{dn}{d\beta}=\frac{\beta}{\sqrt{1-\beta^2}}
\end{equation}
which appears to be the so-called  Weyl distribution. Finally, here,  the DOS is simply:
\begin{equation}
f(\beta)=\frac{1}{L_0}\frac{dn}{d\beta}=\frac{1}{\pi} \varphi(\beta).
\end{equation}
This is nothing but the 1D focusing NLS bound state soliton gas DOS \eqref{cond_nls_dos} obtained in Section~\ref{sec_reductions} as the solution of the NDRs \eqref{dr_soliton_gas} in the limit $\sigma \to 0$ assuming the spectral support $\Gamma^+=[0, i]$.
In~\cite{gelash2019bound} a large number of realizations of N-SSs that fulfil the required  eigenvalues  distribution given by the Eq.~\ref{eq:EIGdistr} have been computed with random phase for the norming constants by using the procedure described in Sec.\ref{subsec:denseSGgeneration}. This realizations ensemble models a bound SG in the limit $N$ large ($N=128$ in~~\cite{gelash2019bound}). As the N-SS are bound states (Re $\lambda_n=0$), the expected dynamics of the wave field is identical to the dynamics observed in the nonlinear stage of the spontaneous MI. Indeed, the zero-velocity of the solitons prevent any dilution of the gas during the evolution.

The Fig.~\ref{Fig:MIspatio} displays the comparison between two NLS equation simulations made with different initial conditions: Fig.~\ref{Fig:MIspatio}.a corresponds to the dynamics of the plane wave (initially pertubed with noise) while Fig.~\ref{Fig:MIspatio}.b corresponds to the dynamics of one realization of the $N$-SS. The features characterizing the spatio-temporal dynamics of the long-time evolution of the plane wave (typically for time $t>20$) seams very similar to the one of the $N$-SS. As expected, the specifically-designed $N$-SS apparently is a very good model of the nonlinear stage of the spontaneous MI. \\

\begin{figure}
\centering
\includegraphics[width=0.95\linewidth]{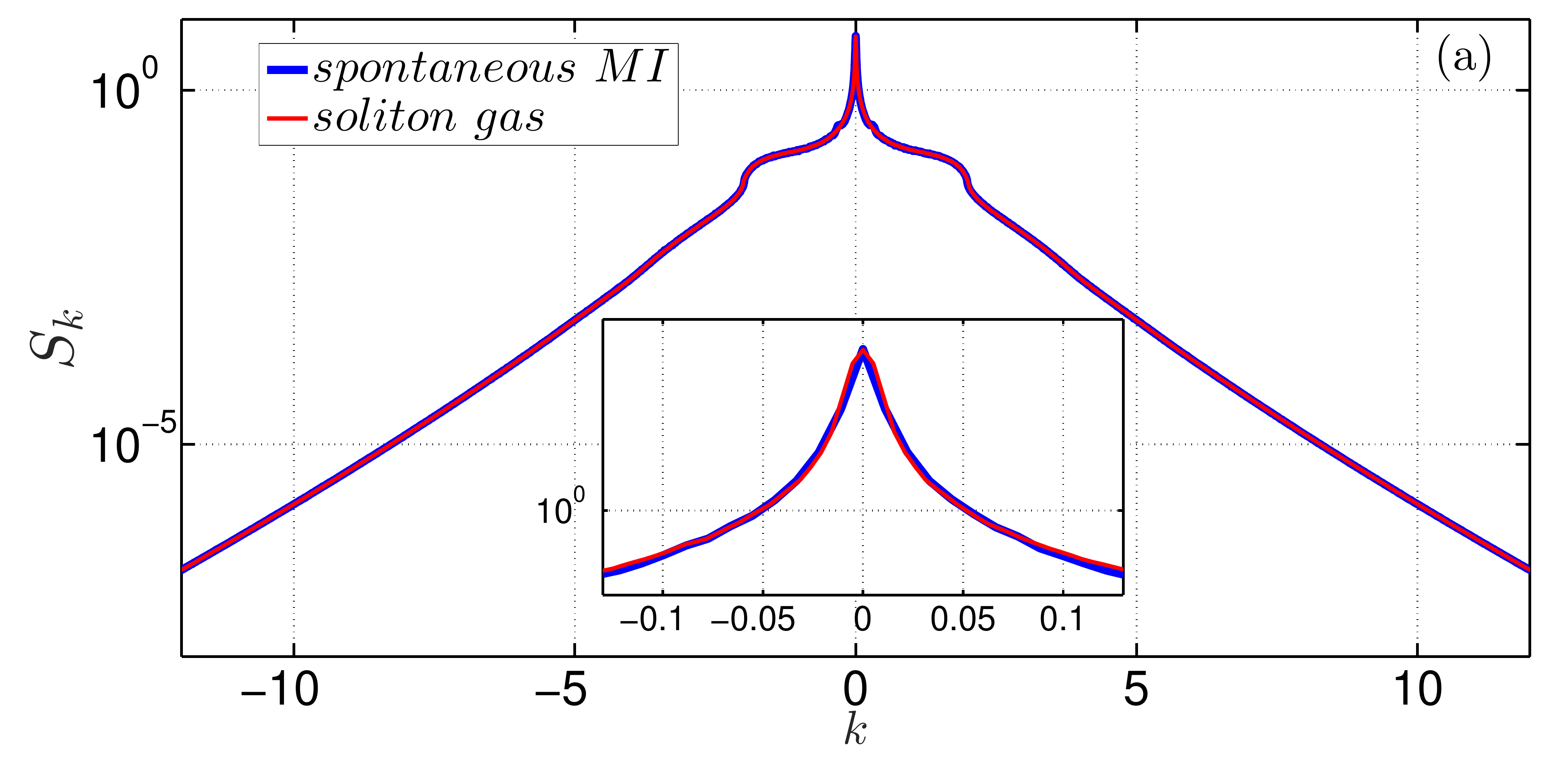}
\includegraphics[width=0.95\linewidth]{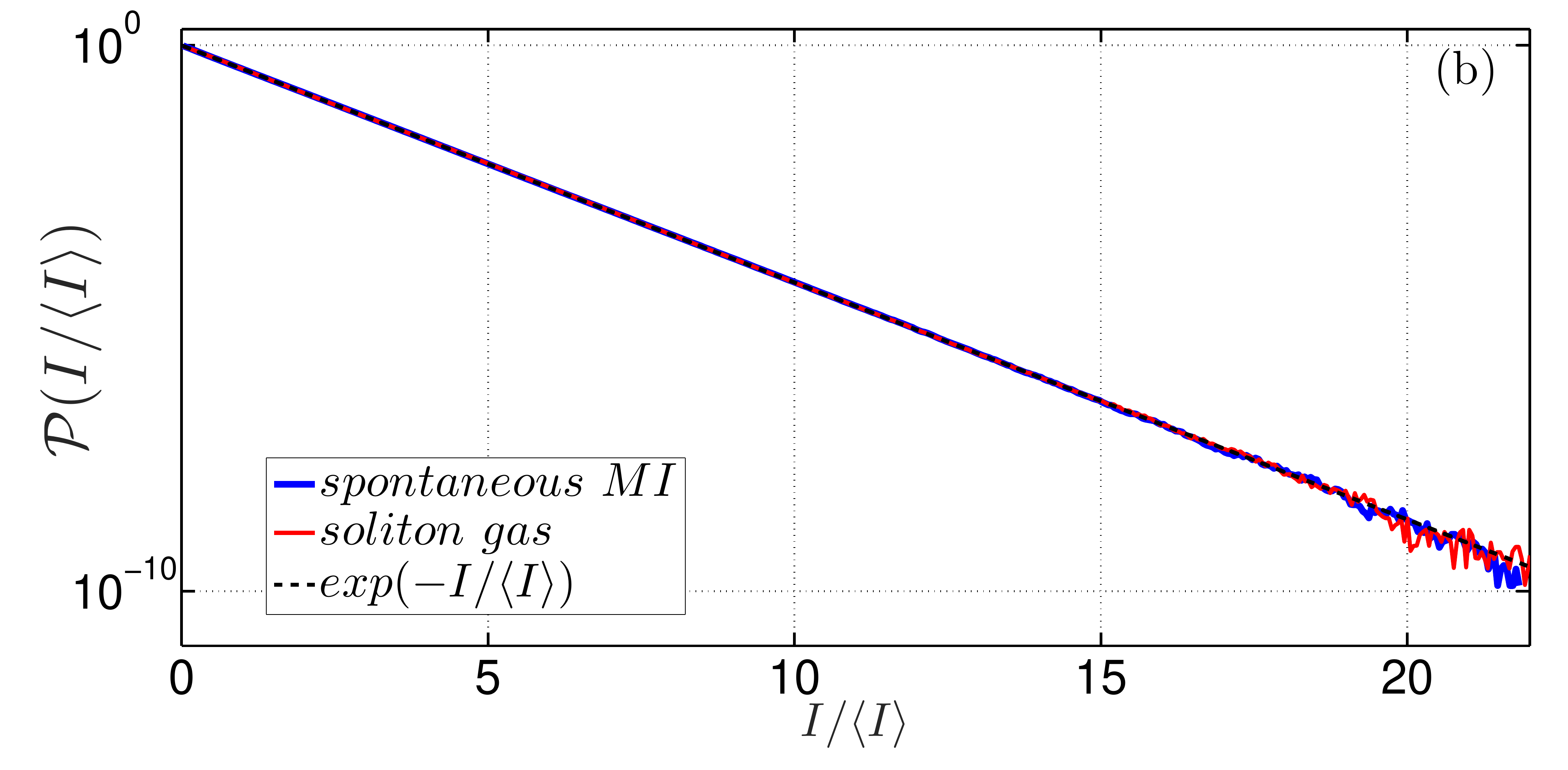}
\includegraphics[width=0.95\linewidth]{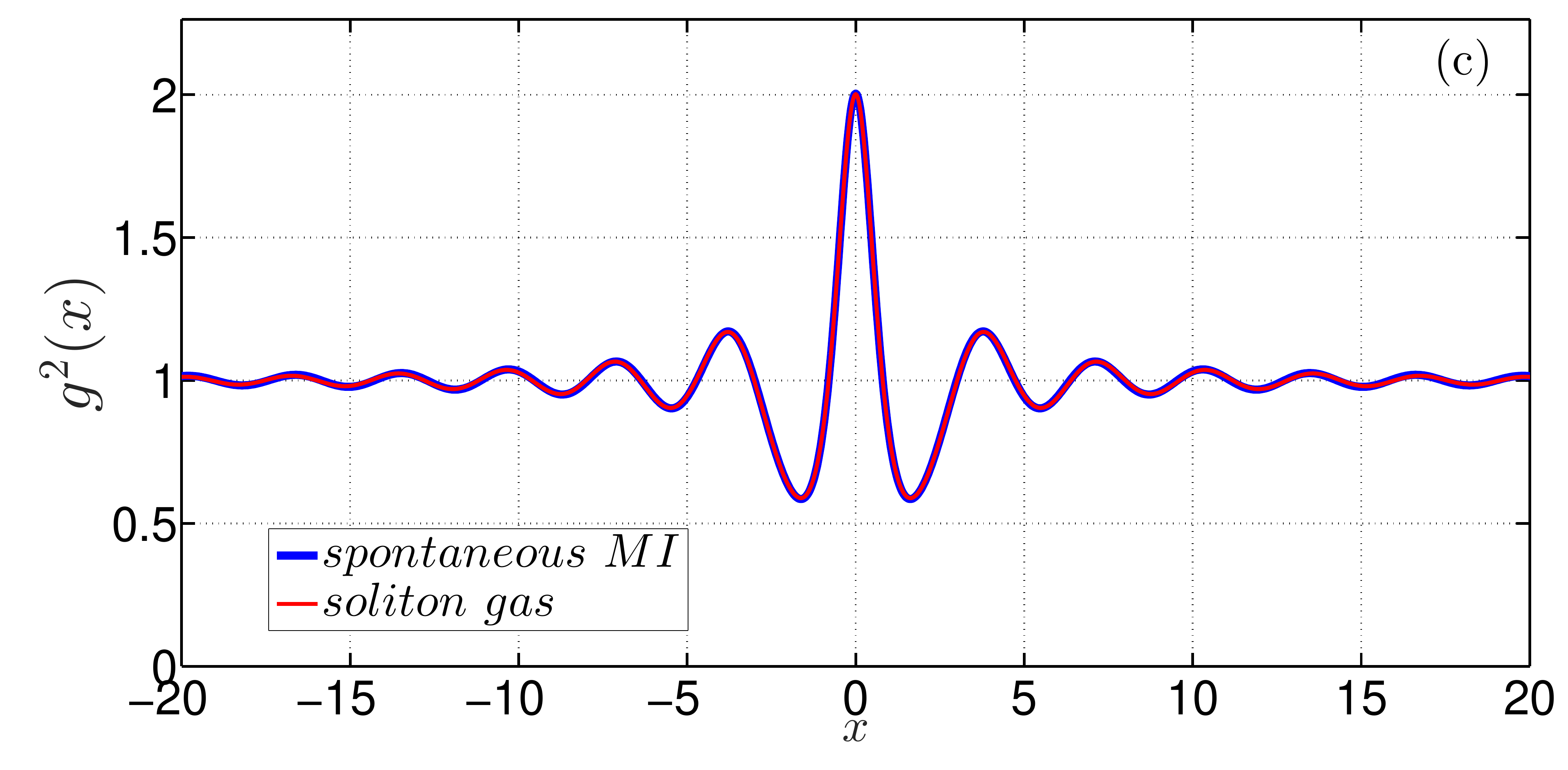}

\caption{\label{FigMIspectrum}
Comparison of ensemble averaged statistical characteristics of the asymptotic state of the MI development and of random phase $128$-SSs. 
(a) Wave action spectrum $S_k$. (b) The PDF $\mathcal{P}(I)$. (c) Autocorrelation function of intensity (second order degree of coherence) $g^{(2)}(x)$. Reproduced with permission from~\cite{gelash2019bound}.}
\end{figure}

More importantly, the statistical properties of SG coincide in a quantitative manner with those of the asymptotic stage of MI. For example, the long-term evolution of the noise-induced MI is characterized by stationary values the potential $H_{nl}$ and kinetic $H_{l}$ energy~\cite{Agafontsev:15},  $\langle H_{l}\rangle=0.5$ and $\langle H_{nl}\rangle=-1$ where the total  energy (Hamiltonian) $H$, which is one of the infinite constants of motion of the 1D-fNLS equation~\cite{novikov1984theory} reads
\begin{eqnarray}\label{eq:Energy}
&& H = H_{l} + H_{nl},\quad H_{l}=\frac{1}{2} \frac{1}{L}\int_{-L/2}^{L/2}|\psi_{x}|^{2}\,dx,\nonumber\\
&& H_{nl} = - \frac{1}{2}\frac{1}{L}\int_{-L/2}^{L/2}|\psi|^{4}\,dx.
\end{eqnarray}
The Fig.~\ref{FigMIspectrum} shows the comparison between three other statistical of both the long-time evolution of the spontaneous MI and of the ensemble of $N$-SS. The Fig.~\ref{FigMIspectrum}.a displays the wave-action spectrum, 
\begin{equation}\label{wave-action-spectrum}
S_{k} \propto \langle|\psi_{k}|^{2}\rangle,\quad \psi_{k} = \frac{1}{L}\int_{-L/2}^{L/2}\psi\, e^{-ikx}\,dx.
\end{equation}
The Fig.~\ref{FigMIspectrum}.b displays the probability density function (PDF) $\mathcal{P}(I)$ of the field intensity $I=|\psi|^{2}$  which is known to follow the exponential distribution in the asymptotic statistics of the unstable condensate~\cite{Agafontsev:15, kraych2019statistical}. Finally, the Fig.~\ref{FigMIspectrum}.c displays the  autocorrelation of the intensity $g^{(2)}(x)$ :
\begin{equation}
g^{(2)}(x)= \frac{\bigl\langle I(y,t) I(y-x,t) \bigr\rangle }{ \bigl\langle  I(y,t)  \bigr\rangle^2 }
\label{eq:g2}
\end{equation}
which represents the second-order degree of coherence.

Remarkably, all these statistical quantities computed in the asymptotic state of the MI  coincide with excellent accuracy with those of the considered SG. In addition, further studies revealed that extreme amplitude waves emerging in the asymptotic state of the MI and the soliton gas have identical dynamical and statistical characteristics \cite{agafontsev2021rogue}. Note that this agreement is weakly depend on the exact eigenvalues chosen for the $N$-SS because the key ingredient are the statistical distribution of the eigenvalues and the use of random phases for the norming constants (in other words, similar statistical results have been obtained in the case of soliton eigenvalues randomly distributed  according to the probability function~\eqref{eq:EIGdistr}~\cite{gelash2019bound}.\\

What are the main conclusions of this numerical study? {\it First, the asymptotic state of the spontaneous MI can be modeled as a specific SG -- the bound state soliton condensate}. This SG can be constructed with exact $N$-SSs of the one dimensional focusing NLS equation by using large values of $N$ and the Weyl's distribution of IST eigenvalues coinciding with the one predicted for the box potential in the semi-classical limit~\cite{Zakharov:1972Exact}. Moreover, the long-term statistical state of MI corresponds to a full stochastization of the phases of the norming constants {\it i.e.} the solitons' phases. Finally, note that for other distributions of eigenvalues tried by the authors, the statistical properties of the SG  do not coincide with the ones of the MI and/or are strongly non homogeneous in space~\cite{gelash2019bound}.

These results open a promising direction in the theory of integrable turbulence  by establishing a link between the MI and SGs dynamics. It is important to note that this quantitative link is possible in the case of MI because for a ``semi-classical'' box,  the contribution  of the ``non-soliton'' part of the  field, {\it i.e.} of the continuous IST spectrum,  decays exponentially with $L_0$ and so can be neglected~\cite{novikov1984theory}. As a consequence, this modeling of integrable turbulence by using SG can be {\it a priori} generalized to a broad class of IT problems  when the (random) wave field is strongly nonlinear, so that the impact of the non-solitonic content can be neglected in the asymptotic state ($t\to\infty$).  For such a case, the general strategy to study the asymptotic state should be to build $N$-soliton solutions with the distribution $\varphi(\lambda)$  of IST eigenvalues characterizing the field and {\it random phases} of the norming constants (see Sec.\ref{subsec:openStatistics}). We review in  Section~\ref{subsec:openStatistics} new results showing that this approach provides a a framework allowing to compute theoretically the observed value of $\kappa=2$ for the long term evolution in the spontaneous MI phenomenon~\cite{kraych2019statistical} and $\kappa=4$ for the semiclassical limit of partially coherent waves~\cite{agafontsev2021extreme}.

\section{Generalized hydrodynamics}
\label{sec:GHD}

\subsection{The perspective of emergent hydrodynamics}\label{ssect:prespectivehydro}

As mentioned, it is very natural to understand the theory of soliton gases within a kinetic perspective, as the fundamental objects -- the soliton DOS or phase-space density $f(\eta;x,t)$ introduced in Section \ref{sec:concept_sg}, its the kinetic equation, Section \ref{ssect:kinetic}, and the equation of state for the effective velocity $s(\eta;x,t)$, Eq.~\eqref{eq_state_gen} -- have a clear kinetic interpretation in terms of soliton propagation and scattering. This interpretation is mathematically accurate at low densities, but at high densities, although compelling, it remains nebulous. The kinetic viewpoint is in fact an a-posteriori interpretation: as reviewed in Section \ref{sec:spec_theory}, the soliton gas theory may be derived from an appropriate thermodynamic limit of finite-gap (quasi-periodic) solutions and its Whitham modulations.

Independently from the soliton gas theory, a framework for the emergent large-scale behaviours of quantum many-body integrable systems out of equilibrium was more recently developed, dubbed ``generalised hydrodynamics" (GHD) \cite{castro-alvaredo_emergent_2016,bertini_transport_2016}. In this context, the problem is to determine the dynamics of quantum systems, such as the Lieb-Liniger gas \cite{PhysRev.130.1605} and Heisenberg quantum spin chain \cite{hep-th/9605187}, out of equilibrium. One seeks, for instance, the full space-time profile of expectation values of local observables, from an initial state that present variations on large scales; or the full space-time profile of their correlation functions. As it was realised \cite{PhysRevLett.120.045301}, it turns out that the main objects of GHD -- the phase-space densities denoted in this context $\rho_{\rm p}(\eta;x,t)$, the kinetic equation reffered to as the ``GHD equation", and the equation of state for the effective velocity denoted $v^{\rm eff}(\eta;x,t)$ -- have exactly the same structure as in soliton gases. The spectral parameter $\eta$ is identified with the quasi momentum of the thermodynamic Bethe ansatz (TBA), and quite surprisingly the two-body scattering shift is simply identified with the semiclassical shift of quantum wavepackets, or the ``kernel" of the TBA equations. The TBA \cite{PhysRev.130.1605,Yang1969,ZAMOLODCHIKOV1990695,Takahashi1999,Mossel_2012} is a framework first developed at the beginning of the 1960's to construct the thermodynamics of Bethe-ansatz integrable systems.

However, by contrast to the theory of soliton gases, in GHD a different viewpoint is emphasised. Certainly, a kinetic perspective can be taken, as was done in one of the co-founding papers of GHD \cite{bertini_transport_2016}: ``Bethe quasiparticles" are the kinetic objects, and the effective velocity $v^{\rm eff}(\eta)$ had in fact been proposed earlier \cite{PhysRevLett.113.187203} as their emergent propagation velocity within finite-density states. However, in the quantum context it is more difficult to establish the validity of this perspective, even at low densities. Further, a quantum modulation theory has not yet been developed. Instead, the currently prevalent viewpoint, emphasised in the other co-found paper of GHD \cite{castro-alvaredo_emergent_2016}, is that of {\it the emergence of hydrodynamics at large spacetime scales}. This physical idea implies that the structure of GHD is in fact that of {\it Euler equations}, instead of a kinetic theory, only generalised to infinitely-many conservation laws.

Euler hydrodynamics is the idea that locally, within each ``mesoscopic" region of space and time (sometimes referred to as ``fluid cell"), the system's state looks as if it had relaxed. A mesoscopic spacetime region covers a length that is large as compared to the microscopic scales (the inter-particle scales and interaction distances) but small as compared to macroscopic scales (the length scales at which averages of local observables show variations); and a time that is likewise large compared to microscopic, and small compared to macroscopic, times. According to conventional physical wisdom, a state that has ``relaxed" is spacetime stationary and takes the Gibbs form. Thus, in Euler hydrodynamics, one assumes that at every point in space-time, the state looks like it is in Gibbs form. The Gibbs form arises from an entropy maximisation principle, so these are ``maximal entropy states", and we may therefore talk about ``local entropy maximisation". The local maximal entropy states depends on space-time, and upon imposing all the available local conservation laws, this gives the Euler equations for the system.

It has been worked out in the past 20 years (see the reviews \cite{1742-5468-2016-6-064002,1742-5468-2016-6-064007}) that the so-called Generalised Gibbs Ensembles (GGE), with density matrix
\begin{equation}
	\mbox{GGE:}\quad \rho\propto e^{-\sum_i\beta_iQ_i}
\end{equation}
where $Q_i= \int dx\,q_i(x,t)$, with $\frac{d Q_i}{dt} = 0$, are extensive conserved quantities, correctly describe relaxation in many-body integrable models. In the infinite-volume limit, infinitely many conserved quantities $Q_i$ must be considered (under some convergence condition). The emergent hydrodynamic perspective then simply states that the local MES are GGEs, so the local relaxation process is
\begin{equation}
	\langle o(x,t)\rangle_{\text{initial state}} \to
	\langle o\rangle_{{\rm GGE}(x,t)}
\end{equation}
for ``any" local observable $o(x,t)$. Here GGE$(x,t)$ is described by ``Lagrange parameters" $\beta_i(x,t)$ which depend on space-time. In the hydrodynamic approximation, the GGEs only depend slowly on space and time, and one then imposes the local conservation laws $\p_t q_i(x,t)+\p_x j_i(x,t)=0$ for GGE-average local densities $q_i(x,t)$ and their currents $j_i(x,t)$, in order to obtain the long-wavelength, slow dynamics,
\begin{equation}\label{qjGGE}
	\frac{\partial}{\partial t} \langle q_i\rangle_{{\rm GGE}(x,t)}
	+ \frac{\partial}{\partial x} \langle j_i\rangle_{{\rm GGE}(x,t)} =0.
\end{equation}
In principle, barring subtleties associated to hyperbolic systems of equations (see for instance \cite{Bressan2013}), these are enough equations to have a well-posed initial value problem. The crucial ingredient in \eqref{qjGGE} is the ``thermodynamic equations of state": the way the average currents $\langle j_i\rangle_{\rm GGE}$ are related to the average densities $\langle q_i\rangle_{\rm GGE}$. Once this is known explicitly, the hydrodynamic equation \eqref{qjGGE} is written explicitly.

Thus, the usual ideas of hydrodynamics are simply extended to the principle of ``generalised thermalisation" for many-body integrability; this is the hydrodynamic basis for GHD, justifying its name.

\subsection{The thermodynamic Bethe ansatz}\label{ssect:tba}

In the hydrodynamic perspective, no kinetic theory is invoked. This perspective emphasises not the kinetic interpretation of the equations, but rather their thermodynamic and hydrodynamic interpretations. But how does one deal with infinitely-many conservation laws, and a large space of maximal entropy states? And how does a formulation that looks like a kinetic theory emerge?

This is thanks to the structure of the TBA. In order to describe it, take the integrable model of Bose particles interacting with a delta-function potential, the repulsive Lieb-Liniger gas \cite{PhysRev.130.1605} (see the review \cite{Bouchoule_2022} where its GHD is explained),
\begin{equation}\label{HLL}
    H = -\sum_{n=1}^N \frc12 \frc{\partial^2}{\partial x_n^2} + \sum_{n<m =1}^N
    c \delta(x_n-x_m),\quad c>0.
\end{equation}
The fully symmetric $N$-particle Bethe ansatz eigenfunctions, parametrised by Bethe roots $\eta_n$'s, take the form
\begin{multline}\label{LLpsi}
	\Psi(\{x\}) \propto \sum_{\mathcal P:\text{permutations}}\ 
    \prod_{n<m} {\rm sgn}\,(x_{{\mathcal P}(n)}-x_{{\mathcal P}(m)})\,
	\exp \Big[\\{\rm i} \sum_n \eta_n x_{\mathcal P(n)}
	+ \frac{\rm i}4  \sum_{n\neq m} \phi(\eta_n-\eta_m)\,{\rm sgn}\,(x_{\mathcal P(n)}-x_{\mathcal P(m)})\Big].
\end{multline}
The quantity $\phi(\eta_n-\eta_m) = 2\,{\rm Arctan}\, \frac {\eta_n-\eta_m} c$ is the two-body quantum scattering phase shift occurring when a particle of Bethe root $\eta_n$ scatters with one of Bethe root $\eta_m$. Conserved quantities (including the Hamiltonian) take a simple form on these eigenfunctions:
\begin{equation}\label{LLeigenvalue}
	Q_i \Psi(\{x\}) = \sum_{n=1}^N h_i(\eta_n)\Psi(\{x\})
\end{equation}
where the functions $h_i(\eta)$ are the ``one-particle eigenvalue". These include the total number of particle $Q_0$ (with $q_0(x) = \sum_n \delta(x-x_n)$ and $h_0(\eta)=1$), the momentum $Q_1$ (with $q_1(x) = \frac12\sum_n -{\rm i} \{\p_{x_n},\delta(x-x_n)\}$ and $h_1(\eta)=\eta$), and the energy $Q_2 = H$ (with $h_2(\eta)=\eta^2/2$). In fact, local conserved charges -- those admitting a local density $q_i(x,t)$ -- have $h_i(\eta) \propto \eta^i$ for all $i\in\mathbb N$. In a system of finite length $L$, the values of $\eta_n$'s are quantised as is usual in quantum mechanics; however the quantisation condition is nontrivial: these are the Bethe ansatz equations, involving $\phi(\eta)$ (see for instance \cite{Bouchoule_2022}).

The TBA is based on the basic statistical mechanics principle of the equivalence of the microcanonical and macrocanonical ensembles, but generalised to all conserved charges, or equivalently all Bethe roots. Thus, the sum over eigenstates involved in a GGE concentrates on a {\it fixed distribution of Bethe roots} $\rho_{\rm p}(\eta)$, and one evaluates GGE averages of conserved densities by using this distribution, $\langle q_i\rangle_{{\rm GGE}} = \int d\eta\,\rho_{\rm p}(\eta) h_i(\eta)$, as follows from  \eqref{LLeigenvalue}.  The TBA gives an explicit map from $\beta_i$'s to $\rho_{\rm p}(\eta)$. This map is obtained by minimising a free energy functional that encodes the constraints on quasi-momenta arising from the Bethe ansatz equations. The result may be written in the suggestive form
\begin{multline}
	\varepsilon(\eta) = \sum_i \beta_i h_i(\eta) -  \int \frac{d \eta'}{2\pi}\,\varphi(\eta-\eta')\log(1+e^{-\varepsilon(\eta')}),\\\quad
	\rho_{\rm p}(\eta) = -2\pi \frac{\partial}{\partial \beta_0}
	\log(1+e^{-\varepsilon(\eta)})
\end{multline}
involving the {\it pseudoenergy} $\varepsilon(\eta)$, defined as the solution of the above non-linear integral equation, and the {\it differential scattering phase}, defined by
\begin{equation}\label{diffscat}
	\varphi(\eta) = \frac{d \phi(\eta)}{d\eta}
    = \frac{2c}{\eta^2+c^2}.
\end{equation}

In this sense, the phase-space density $\rho_{\rm p}(\eta;x,t)$ does not arise as a density for particle-like dynamical objects forming a gas, but rather as a way of characterising all averages of local conserved densities in the $x,t$-dependent GGE that arises from the Euler hydrodynamic principle,
\begin{equation}\label{qiGGE}
	\langle q_i\rangle_{{\rm GGE}(x,t)} = \int d\eta\,\rho_{\rm p}(\eta;x,t) h_i(\eta).
\end{equation}
Many-body integrable systems admit an infinite-dimensional space of conserved quantities $Q_i$, and the spectral parameter is just seen as a continuous parametrisation of this space (interpreted as a particular choice of a ``scattering basis", see e.g. the discussion in \cite{DeNardis_2022}).

As mentioned above, the crucial ingredient is the relation between GGE averages of currents and densities. Historically, this was in fact the main stumbling block in developing the hydrodynamics of integrable systems.

Average currents in GGE were first evaluated\cite{castro-alvaredo_emergent_2016} using the TBA and crossing symmetry of relativistic quantum field theory; they were later derived directly from the Bethe ansatz and other quantum integrability techniques \cite{10.21468/SciPostPhys.6.2.023,PhysRevLett.125.070602}, and then from ``self-conserved" currents \cite{PhysRevE.101.060103,10.21468/SciPostPhys.9.3.040}, using the symmetry of current-charge correlations \cite{castro-alvaredo_emergent_2016,10.21468/SciPostPhys.6.4.049,10.21468/SciPostPhys.6.6.068}; see the reviews \cite{Borsi_2021,Cubero_2021}. The result is striking: it takes the form
\begin{equation}
	\langle j_i\rangle_{{\rm GGE}(x,t)} = \int d\eta\,v^{\rm eff}(\eta;x,t)\rho_{\rm p}(\eta;x,t) h_i(\theta)
\end{equation}
where the effective velocity, here obtained, we recall, via Bethe ansatz calculations, satisfies the classical-looking collision rate ansatz \eqref{eq_state_gen}, with $G(\eta,\eta') = -\varphi(\eta-\eta')$ and $s_0(\eta) = \eta$
\begin{equation}\label{veffLL}
	v^{\rm eff}(\eta) = 
	\eta + \int d\mu\,\varphi(\eta-\mu)
	\rho_{\rm p}(\mu) (v^{\rm eff}(\mu)
	- v^{\rm eff}(\eta)).
\end{equation}
Its $x,t$ dependence $v^{\rm eff}(\eta)\to v^{\rm eff}(\eta;x,t)$ comes from the $(x,t)$-dependent GGE $\rho_{\rm p}(\eta)\to \rho_{\rm p}(\eta;x,t)$. Note how the differential scattering phase  $\varphi(\eta-\eta')$, Eq.~\eqref{diffscat}, arises: this is exactly the semiclassical scattering shift of Bethe ansatz wave packets. One then obtains, from \eqref{qjGGE} and assuming some completeness of the space of functions $h_i(\theta)$, the GHD equation
\begin{equation}\label{GHD}
	\frac{\partial}{\partial t} \rho_{\rm p}(\eta;x,t)
	+ \frac{\partial}{\partial x} \big(v^{\rm eff}(\eta;,x,t) \rho_{\rm p}(\eta;x,t)\big) = 0.
\end{equation}
This is, in this perspective, a Euler hydrodynamic equation, even though it looks like a kinetic equation.

We finally note that each value of the spectral parameter $\eta$ corresponds to a hydrodynamic normal mode -- a ``sound mode" or the like -- for the emergent Euler-scale equation, and $v^{\rm eff}(\eta;x,t)$ are the associated hydrodynamic velocities tangent to their characteristics. Riemann invariants can be explicitly constructed; indeed $\varepsilon(\eta;x,t)$, or any function of it, satisfies the diagonalised Euler-scale equation, $\p_t \varepsilon(\eta;x,t) + v^{\rm eff}(\eta;x,t)\p_x \varepsilon(\eta;x,t) = 0$, and so does the ``cumulative density" or height field $\int_{-\infty}^x dx'\,\rho_{\rm p}(\eta;x',t)$; likewise, for linear perturbations on top of a homogeneous stationary background, $\rho_{\rm p}(\eta) + \delta \rho_{\rm p}(\eta;x,t)$, we have $\p_t \delta \rho_{\rm p}(\eta;x,t) + v^{\rm eff}(\eta)\p_x\delta \rho_{\rm p}(\eta;x,t) = 0$.

\subsection{Universality of Euler hydrodynamics}

Note that, curiously, one obtains, using the above description, a re-interpretation of the Liouville equation of phase-space conservation in classical mechanics. Traditionally it is understood in kinetic theory as a ``collisionless" Boltzmann equation. Now take, for instance, the Tonks-Girardeau limit $c\to\infty$, where the Lieb-Liniger model becomes a model of non-integracting fermions, with $\varphi(\eta)=0$. The resulting GHD equation is the Liouville equation. But here, it is seen as a {\it hydrodynamic equation}, for a continuum of sound modes emerging at large scales in this system of non-interacting particles! The same hold for any system of non-interacting particles, quantum or classical.

This latter observation leads us to emphasise an important concept: {\it the hydrodynamic perspective has the advantage that it is indifferent to the precise nature of the underlying many-body system}. It has a large amount of universality.

This universality arises at two levels. First, the general structure of hydrodynamics at the Euler scale is always the same, no matter the underlying many-body system, under fairly general conditions (local interactions, and perhaps microscopic reversibility). The important point in establishing the Euler-scale hydrodynamic theory of a given many-body system is to characterise its full manifold of maximal entropy states. One expects that the ``extensive conserved quantities", widely studied in quantum many-body systems \cite{ilievski2016quasilocal}, span the tangent spaces to this manifold, and according to Euler hydrodynamics, their densities are the emergent dynamical degrees of freedom onto which the microscopic dynamics projects at large scales; at the linearised level, this phenomenon has been rigorously established in quantum spin chains \cite{doyon_projection} and lattices \cite{ampelogiannis_projection_2022}. Once the space of extensive conserved quantities is understood, the hydrodynamic principles -- local relaxation and the conservation laws -- are completely general, and do not require any strong dynamical assumptions such as chaos, or any particular structures for the underlying microscopic theory.

Second, within the family of many-body integrable models, the description of Section \ref{ssect:tba} is also completely universal. The microscopic system may be quantum or classical, composed of continuous fields, particles, solitons, spins, etc. -- the same structure emerges for its Euler-scale hydrodynamics. The model-dependent aspects are the phase space $\mathcal S$ of possible values of the spectral parameter $\eta$ (it is $\mathbb R_{>0}$ in the KdV soliton gas, $\mathbb R$ in the repulsive LL model, $\mathbb C$ in the soliton gas of focusing NLS, etc.), and the basic dynamical quantities, including the two-body shift $G(\eta,\mu)$ (it is $-\frac{2c}{(\eta-\mu)^2 + c^2}$ in the repulsive LL model, $\frac1\eta \log \big|\frac{\eta+\mu}{\eta-\mu}\big|$ in the KdV soliton gas, etc.), as well as a ``bare" velocity $s_0(\eta)$ entering as the source term in the equations of state, Eqs.~\eqref{eq_state_gen}, \eqref{eq_state_kdv}, \eqref{veffLL} ($\eta$ in the LL model, $4\eta^2$ in the KdV soliton gas, etc.).  Thus, in fact GHD is not only a theory for many-body quantum integrable systems, but also for classical systems, including soliton gases. The full equivalence between TBA quantities and those of soliton gases is given in \cite{bonnemain_2022}.

This universality of the hydrodynamic description of integrable models has its source in an important aspect of many-body integrability, that of {\it factorised, elastic scattering}. Factorised scattering for solitons was reviewed in Section \ref{sec:sol_int}, see Eq.~\eqref{delta_total}. It is also made apparent in the LL Bethe ansatz wave function \eqref{LLpsi}: the structure in the exponential implies that the phase of a full many-particle scattering is the sum of two-body scattering phases. If we put the LL model in a finite segment and let the particles expand in the vacuum, then $\eta_n$'s are the values of asymptotic momenta that will be seen at long times in this {\it time-of-flight ``gedenkenexperiment"}. In general, for both quantum and classical models, the spectral space is nothing else than the set of possible objects that emerge at long times (solitons, particles, bound states, waves, etc.), and a basic dynamical analysis will give the bare velocities and two-body scattering shifts for such objects. It turns out that TBA form of the thermodynamics then emerges quite generally solely from this scattering picture; in classical systems this was first observed \cite{doi:10.1063/1.5096892} in the Toda model  -- thus the TBA does not require the Bethe ansatz! In a fluid, a mesoscopic cell can be ``observed" by taking it out of the fluid and making a time-of-flight experiment on it, in order to determine the distribution of spectral parameters that characterise it. The manifold of GGEs is a manifold of distributions on the spectral space. In particular, the soliton gas is simply the case where we restrict the manifold of GGEs to be distributions of solitons only; and this restriction is stable under the Euler hydrodynamic evolution. This explains the general structure of GHD.

In fact, the scattering picture has far-reaching ramifications. One of them is the geometric viewpoint on GHD, whereby the GHD equations are seen as arising from a change of coordinates -- or a change of metric -- from the free-particle Liouville equations \cite{DOYON2018570,doyon_lecture_2020}. The change of metric is state-dependent (much like in Einstein's gravity!), and represents the map to the freely-propagating asymptotic coordinates. This leads to an integral-equation solution \cite{DOYON2018570}, a ``solution by characteristics" akin to the hodograph transform.

\section{Open questions}
\label{sec:open}

Over the last few years, various fundamental questions inspired by the exciting theoretical and experimental challenges have emerged in the growing fields of SGs and of GHD. We summarize here some of the most important of these open questions. 

\subsection{Spectral theory  and rigorous asymptotics}
\label{subsec:openSpectral}

The spectral theory of soliton gas outlined in Secton~\ref{sec:spec_theory} is based on the thermodynamic limit of  finite-gap potentials and their  Whitham modulation equations. At the core of this theory is the special  distribution (scaling) of finite-gap spectra ensuring appropriate balance of terms in the nonlinear dispersion relations.  Can this thermodynamic spectral scaling be obtained as a long-time asymptotics in some class of initial-value problems for integrable equations? One possible scenario to be explored was proposed in  \cite{El:16:dambreak}  where one considers a  chain of topological bifurcations of local invariant tori parametrized by slowly evolving finite-gap spectra  that emerge in the zero-dispersion (semi-classical) limit of the fNLS equation. This scenario resembles the classical Landau-Hopf transition to turbulence (see e.g. \cite{landau_fluid_1987})  realized in the framework of an integrable dispersive system. \\

A related major open question is a rigorous mathematical justification of the spectral kinetic theory. The  derivation of the kinetic equation via the thermodynamic limit of finite-gap modulation theory is formal in the sense that the question of the asymptotic validity of the kinetic equation in the framework of the original nonlinear dispersive PDE remains open. It would be highly desirable to have a rigorous asymptotic derivation of the kinetic equation for KdV, NLS and other integrable models.  An important step in this direction has been recently made in ref.~\cite{girotti_soliton_2022} where it was shown that kinetic equation for soliton gas describes the leading order asymptotic  behaviour of a special class of ``deterministic'' soliton gases for the modified KdV equation  constructed as an infinite-soliton limit of $N$-soliton solutions by invoking the theory of the so-called primitive potentials \cite{dyachenko2016primitive} (see also \cite{girotti_rigorous_2021}).  At the spectral level, the characterization of the gases studied in \cite{girotti_soliton_2022} coincides with that of soliton condensates \cite{congy_dispersive_2022} so the extension of the rigorous asymptotic theory to more general classes of inherently random soliton gases remains an outstanding problem. \\

Finally we mention that the spectral theory of soliton gas can be applied to any integrable dispersive PDE supporting finite-gap solutions associated with hyperelliptic Riemann surfaces. One can expect new interesting behaviours in integrable models qualitatively different from the already considered examples of the KdV and fNLS equations. These include the sine-Gordon equation (kink gas), the Camassa-Holm equation (peakon gas) and others. The theory of two-dimensional soliton gases (e.g. for the Kadomtsev-Petviashvili or Davey-Stewartson equations) is another completely uncharted territory yet to be explored.

\subsection{Thermodynamics and Statistics}
\label{subsec:openStatistics}

The statistical description of random waves in integrable systems represents a fundamental application of the SG theory. We have reviewed several important recent steps achieved in this challenging direction of research. Generalized Hydrodynamics provides a framework to establish a thermodynamic description of soliton gases. However, up to now, there is no existing comparison between SGs experiments and GHD theoretical results. On the other hand, the possible correspondence between SGs and natural phenomena stimulates the study of statistical properties of SGs (for example numerical simulations show that the so-called spontaneous modulation instability is with high accuracy by a specifically-designed SG, see Sec.~\ref{subsec:MI}).

Very recently, some of the authors of this paper and their collaborators have derived a general formula for the kurtosis for an homogeneous SG. Derived in the framework of SG theory, the kurtosis is then expressed as a function of the DOS:
\begin{equation}
  \label{eq:kurtosis}
  \kappa = \frac{\aver{|\psi|^4}}{\aver{|\psi|^2}^2} =  \frac{{\rm Im}(\overline{\lambda^2s(\lambda)}- \frac{4}{3} \overline{\lambda^3})}{2 \, {\rm Im}(\overline{\lambda})^2},
\end{equation}
  where the averaging procedure is  $\overline{h(\lambda)} = \int h(\lambda) f(\lambda) \rmd \lambda$ and $f(\lambda)$ is the DOS of the homogeneous SG ($f$ dos not depends on $x$ and $t$).
Applying the Eq.~\ref{eq:kurtosis} to the Weyl's bound state SG corresponding to the long-term evolution of the spontaneous MI (see Sec.\ref{subsec:MI}), it is easy to show that $k_4=2$. This  corresponds to the value of $k_4$ for the exponential distribution of $|\psi|^2$ empirically found in numerical simulations and experiments devoted to the spontaneous MI~\cite{Agafontsev:15,Walczak2015Optical, kraych2019statistical}. Note that this result is consistent with the virial theorem ($H_{nl}=2H_l$) known in the context of zero boundary conditions in NLS~\cite{zakharov1986virial}. Beyond the MI problem, it is possible to compute the DOS of any soliton gas generated by the propagation of a semiclassical field (if $H_{nl}\gg H_l$ initially). Using this approach, one can also show that the corresponding value of the kurtosis is $k_4=4$ in the case of partially coherent waves. Remarkably, this corresponds to the largest value found recently in numerical simulation in the case of the strongly nonlinear regime of partially coherent waves~\cite{agafontsev2021extreme}.

These recent results pave the way to a general statistical description of nonlinear random waves naturally found in various physical systems. However, it is important to note that the evaluation of the kurtosis is only the first step toward a general statistical theory. Among the various open questions, one finds the formal evaluation of the probability density functions (of the field or its amplitude for example) and of correlation functions (such as the $g^{(2)}$, see Eq.~\ref{eq:g2}).

The spectral power density (Fourier spectrum) $\langle|\widetilde{\psi}(k,t)\rangle|^2$ is a key measurable variable of turbulence, allowing for example to characterize Kolmogorov cascade. Moreover the spectrum can be easily and directly measured in optical experiments devoted to the observation of SG. The analysis and the understanding of Fourier spectra of SGs thus represents an important direction of research. The natural framework of the the soliton gas theory is the IST and the relationship between the IST spectrum and the Fourier spectrum is highly nontrivial from the mathematical point of view.\\

It is important to emphasize again that, the SG theory provides a promising framework to describe and understand the statistics of wave systems close to integrability. The spontaneous modulation instability in the focusing regime of the one dimensional focusing NLS equation is the first example of  physical phenomenon quantitatively described by a SG (see Sec.\ref{subsec:MI}) and ~\cite{gelash2019bound}). One natural question is the possible link between natural phenomena and breather gases. In particular, as Akhmediev breather is the exact solution associated with the sinusoidal perturbation of a plane wave, one might expect that the spontaneous modulation instability can also be described by a breather gas. \\

The general description of integrable turbulence (random waves in integrable systems) is still an open question. Any random waves in integrable systems can be decomposed into radiative waves and solitons, the former being associated to the continuous spectrum and the latter being associated to the discrete spectrum in the framework of IST (see Sec.~\ref{sec:sol_int}). SGs thus correspond to the peculiar case of integrable turbulence having no continuous spectrum. The study of nonlinear random waves phenomena by using a SG description is based on the conjecture that continuous spectrum can be neglected in the strongly nonlinear regime. One can naturally ask : what happens for example with partially coherent wave for weaker nonlinearity ?

The general description of integrable turbulence thus requires the development of a statistical theory involving both discrete and continuous spectrum. In principle, the general case can be described in the framework of the finite gap theory (see Sec.~\ref{sec:spec_theory}).  On the other hand, by taking into account the non resonant interactions, a non standard wave kinetic theory (developed in the basis of Fourier components) describes the statistical behavior and the Fourier spectrum of integrable turbulence~~\cite{Janssen:03, suret2011wave, Picozzi:14}. One of the fundamental and interesting open question is the IST formulation in the weakly nonlinear regime when the wave system is dominated by radiation components (continuous spectrum). Investigations of this question may build a bridge between finite gap theory and wave turbulence theory.

\subsection{Experimental challenges}

Experiments devoted to the study of SGs can be classified by waves generation techniques and by data analysis types. While solitons can simply be generated one by one in diluted SGs, the experimental realization of dense SG is highly non trivial. One possible approach is the use of dynamical phenomenon such as the soliton fission~\cite{Redor:19} in which the DOS is not controlled. Another strategy has been recently demonstrated in order to achieve a controlled generation of dense SG~\cite{Suret:20}: by using the numerical procedure described in~\ref{subsec:denseSGgeneration}, N-solitons solution with random parameters are computed and then used to build the experimental SG.\\

It is important to note that, up to now, the procedure based on the N-solitons allows the generation of {\it homogeneous} dense SG having an arbitrary DOS $f(\lambda)$. The generation of a {\it non-homogeneous} dense SG with a space-dependant DOS $f(\lambda,x)$ is an open problem. In the context of the focusing NLSE, this extremely challenging task will require a deep theoretical understanding of the link between the positions of the solitons and the amplitude of the norming constants in the N-soliton solution with $N\gg 1$. Solving this problem would be a fundamental milestone in the study of SG. Indeed, the most intriguing and complex phenomena are expected to emerge in the context of non-homogeneous SG whose non-equilibrium, macroscopic dynamics are described by the non trivial continuity equation (\ref{FNLS_kin}). The experimental test of this continuity equation requires first the generation of non-trivial DOS $f(x,\lambda)$.\\

On the other hand, the study of non-homogeneous SG will also require the development of new tools for the data analysis. The measurement of a space-dependant DOS is not trivial; one will have first to define the local DOS of a measurable field. One of the difficulties is the scale separation: in the theory, the DOS evolves spatially very slowly and the number of solitons in one fluid cell $dx$ tends to infinity. In experiments, the number of solitons is limited and thus, the measurement of the local DOS is a complex and challenging task.\\

The experimental test of the GHD is another open exciting challenge. This includes for example the measurement of space-time correlation in soliton gases, the measurement of GGEs, \dots

\subsection{Breakdown of integrability}

In the ``real-world'' experiments, integrable equations such as the 1DNLS or KdV, only describe the systems at leading order; This means that in any experiments, at long time (or long propagation distance), high order effects break integrability and play a role in the dynamics and in the statistics of the wave field. Integrability can be broken by linear effects (losses or high order dispersion for example) or non linear (stimulated Raman scattering in optical fiber for example). These effects induce non elastic collisions of solitons (for example, two interacting solitons do not recover their initial amplitudes and velocities over large time). The study of the influence of high order effects on SG is of fundamental and practical importance. This includes also the influence of external forces on solitons (induced for example by some potential).

In various systems, the high order effects can be considered as small perturbations of the integrable system. As a consequence, IST spectra can be seen as slow varying quantities that evolve adiabatically. The IST perturbation theory of nearly integrable systems is well elaborated for simple wave field patterns, such as single and two-soliton pulses \cite{Kivshar1989Dynamics}; meanwhile, the collective multi-soliton dynamics under the influence of weak external forces now is treated only with numerical simulations \cite{slunyaev2015wave,agafontsev2023bound}. Building a theory of SG including perturbative effects is an open and fundamental problem. GHD is a promising framework to investigate perturbative effects (see Sec.~\ref{subSec:breakingGHD}).

\subsection{Lessons from GHD: correlations, external forces, diffusion, integrability breaking}
\label{subSec:breakingGHD}

Communities working on soliton gases, and on quantum and classical many-body systems and statistical mechanics, have been mostly disconnected until recently. Certainly, making a better connection between the ideas that have arisen in both communities would be fruitful.

For instance, the metric transform from the Liouville equation to the GHD equation \cite{DOYON2018570} is nothing else but a generalisation of the transformation from free particles to hard rods, used extensively in addressing the hard rod gas \cite{boldrighini_hr,Doyon_2017}. In this transformation, each quasiparticle is given a precise location, and occupies a certain momentum-dependent space that, if taken away, reduces the quasiparticles' dynamics to that of free particles. A similar transformation exists in the box-ball system (a certain cellular automaton) \cite{croydon_GHDboxball}, where it allows one to identify the precise position of each soliton within a dense soliton gas. Can something like this be achieved in KdV or NLS soliton gases?

Further, the hydrodynamic viewpoint on GHD has been extremely powerful. It has allowed for the extension of known structures of hydrodynamics to the realm of integrability. Taking and developing the full hydrodynamic perspective in soliton gases should lead to interesting new result, and this is still at its infancy. Here we briefly mention four directions: correlation functions, the inclusion of external forcing, the diffusive and higher-order corrections, and the inclusion of small integrability-breaking effects via Boltzmann-like equations.

Correlation functions in space-time are natural objects to be studied by hydrodynamics. The basic idea is that the propagation of hydrodynamic modes gives the leading large-scale correlations between local observables. Technically, one studies the linearised Euler equation for small variations $\delta \langle q_i\rangle$ on top of a homogeneous, stationary state. This gives the following form for the Fourier transform of connected correlation functions $S_{ij}(k,t) = \int dx\,e^{{\rm i}kx}\langle q_i(x,t)q_j(0,0)\rangle^{\rm c}$ in that state:
\begin{multline}
	S_{ij}(k,t) \sim \Big(\exp\Big[{\rm i}kt \mathsf A\Big]\mathsf C\Big)_{ij},\quad \mathsf A_{ij} = \frac{\partial \langle j_i\rangle}{\partial\langle q_j\rangle},\\ \quad \mathsf C_{ij} = -\frac{\partial \langle q_i\rangle}{\partial\beta_j}\quad (k\to0,\,t\to\infty,\,kt\ \mbox{fixed}).
\end{multline}
The flux Jacobian $\mathsf A$ and static covariance $\mathsf  C$ can be written in terms of TBA quantities, giving rather explicitly \cite{10.21468/SciPostPhys.3.6.039,10.21468/SciPostPhys.5.5.054}
\begin{equation}
	S_{ij}(k,t) \sim 
	\int d\eta\,\rho_{\rm p}(\eta) f_{\rm stat}(\varepsilon(\eta))\, \frac{\p \varepsilon(\eta)}{\p\beta_i} \frac{\p \varepsilon(\eta)}{\p\beta_j} e^{{\rm i} k t v^{\rm eff}(\eta)}
\end{equation}
where $f_{\rm stat}(\varepsilon)$ encodes the {\it statistics} of the fundamental particles (the asymptotic objects), e.g.~$f_{\rm stat}(\varepsilon) = \frac1{1+e^{-\varepsilon}}$ in the LL model, and $f_{\rm stat}(\varepsilon) = 1$ in the KdV soliton gas \cite{bonnemain_2022}. This formula was verified numerically in various integrable models; see the review \cite{DeNardis_2022} and more recent results in the KdV soliton gas \cite{bonnemain_2022} and the Toda model \cite{Mazzuca_correlations_2022}. In soliton gases, the formula would still need to be understood from the IST perspective.

One can go much further and obtain two-point correlation functions not just of conserved densities, but also of currents, and in fact of arbitrary observables, by the use of hydrodynamic projections \cite{10.21468/SciPostPhys.5.5.054,doyon_projection}; as well as, quite surprisingly, two-point correlation functions in non-stationary backgrounds \cite{10.21468/SciPostPhys.5.5.054,10.21468/SciPostPhysCore.3.2.016}. Going beyond, based on similar ideas, the Euler-scale large-deviation theory of integrated currents and other extensive quantities \cite{10.21468/SciPostPhys.8.1.007,Doyon_largedeviation}, and non-linear response functions \cite{doi:10.1073/pnas.2106945118}, have been obtained. More generally, the ballistic macroscopic fluctuation theory \cite{doyon_bmft_2022}, which has in particular been applied to GHD, gives a complete framework where many-point correlation functions and Euler-scale large-deviation theory can be evaluated, predicting novel long-range spatial correlations in moving fluids \cite{doyon_longrange_2022}. All these results apply, in principle, to soliton gases as well -- but, in this context, numerical verifications and a full theoretical underpinning are still lacking.

Generalised external forces may be written as external fields coupled to conserved densities. These change the Hamiltonian to $H + V$ where $V = \sum_i \int dx\,V_i(x)q_i(x)$. Although generically $V$ breaks the integrability of $H$, with $V_i(x)$ slowly varying in space, Euler hydrodynamic equations with generalised force terms remain valid for all original conservation laws -- indeed, for conventional gases, Euler equations can be written within external force fields, even when such fields break momentum conservation. Within GHD, the corresponding force terms have been obtained \cite{SciPostPhys.2.2.014}, with \eqref{GHD} modified to
\begin{multline}
	\frac{\partial}{\partial t} \rho_{\rm p}(\eta;x,t)
	+ \frac{\partial}{\partial x} \big(v^{\rm eff}(\eta;,x,t) \rho_{\rm p}(\eta;x,t)\big)\\ + \frac{\partial}{\partial \eta} \big(a^{\rm eff}(\eta;,x,t) \rho_{\rm p}(\eta;x,t)\big) = 0.
\end{multline}
Quite surprisingly, the effective acceleration $a^{\rm eff}(\eta;,x,t)$ satisfies a ``collision rate ansatz" as \eqref{veffLL} but with the bare velocity $\eta$ replaced by the bare acceleration $a(\eta;x) = -\sum_i V_i'(x) h_i(\eta)$. It is this GHD equation, for the LL model and with a simple external force field, was verified experimentally in cold atomic gases restrained to one dimension of space \cite{PhysRevLett.122.090601,doi:10.1126/science.abf0147,PhysRevLett.126.090602}, see the review \cite{Bouchoule_2022}. This is the simplest situation of externally changing parameters: the more general situation was worked out \cite{PhysRevLett.123.130602}, including time dependence, and varying the coupling strength $c\to c(x,t)$ in \eqref{HLL}, something which is crucial for comparison with some experiments. External force fields and slowly-varying couplings also naturally occur in many situations where soliton gases emerge. The theory from GHD is in principle fully applicable to soliton gases; however, again up to now, the application to soliton gases and the IST perspective on such GHD results are still completely missing.

Hydrodynamics is a derivative expansion, and as such, one may wonder about the higher-derivative corrections. At second derivative, this is the {\em diffusive correction}, such as the viscosity term in Navier-Stokes equations. Again, an exact expression of the diffusive matrix -- or diffusive operator on spectral space -- has been evaluated in GHD \cite{PhysRevLett.121.160603,Gopalakrishnan2018,10.21468/SciPostPhys.6.4.049}, with convincing comparisons against numerical results, see the review \cite{DeNardis_2022}. The form obtained is
\begin{multline}
	\frac{\partial}{\partial t} \rho_{\rm p}(\eta;x,t)
	+ \frac{\partial}{\partial x} \big(v^{\rm eff}(\eta;,x,t) \rho_{\rm p}(\eta;x,t)\big) \\= \frac12\frc{\partial}{\partial x}\Big(\int d\eta'\,\mathcal D_{\eta,\eta'}[\rho_{\rm p}(\cdot;x,t)]\frac{\partial}{\partial x} \rho_{\rm p}(\eta';x,t)\Big).
\end{multline}
The diffusion kernel $\mathcal D_{\eta,\eta'}[\rho_{\rm p}]$ is evaluated from the Kubo formula involving space-time integrated current two-point functions, using form factor methods of quantum integrability \cite{10.21468/SciPostPhys.6.4.049,Cubero_2021}. The general formula, applicable to quantum and classical models alike, is conjectured by comparison with the diffusion kernel obtained in the 1980's for the classical hard rod gas \cite{Boldrighini_hr_diffusive}. Again, the general formula involves the statistical factor $f(\varepsilon)$. The combination of diffusion with external forces has also been evaluated \cite{Durnin_2021}. The third-order, {\em dispersive} correction was proposed recently \cite{denardis_higher_2022}, although much work is still needed to fully establish it.

Is there diffusion in soliton gases? If so, is it correctly described by the GHD formula? Further, can we evaluate the exact 3rd-order, dispersion term? A natural conjecture concerns the condensate limit; in the GHD of quantum integrable models, the condensate limit had been studied earlier, and is known as zero-entropy GHD \cite{PhysRevLett.119.195301}. The connection between soliton-gas condensate limit and zero-entropy GHD was partially made in \cite{congy_dispersive_2022}. Do dispersive terms of GHD / soliton gases reproduce, in the zero-entropy / condensate limit, dispersive terms of the fundamental dynamical equations (e.g. the KdV equation)?

Finally, the effects of small perturbations that break integrability has been studied. The development is still in its infancy, with various approaches and different physical situations proposed, see the review \cite{Bastianello_2021}. The perspective taken in GHD is different from that taken in soliton gases, and it would be fruitful to make a better connection. One important point that has been emphasised\cite{PhysRevLett.127.130601} generalises the viewpoint discussed above, whereby the Liouville equation -- the kinetic equation for free particles -- is seen as a Euler-scale hydrodynamic equation. It is possible to modify the Euler-scale hydrodynamic equation to account for terms that break the conservation laws on which it is based. There are general Kubo-like formulas this modification, and when applied to GHD, these give terms that can be written, at least in quantum models, in a form-factor expansion. Specialised to the GHD of free particles, these terms are nothing else but Botlzmann collision terms from the Boltzmann equation; form factors of interacting integrable models generalise Boltzmann collision terms. Is there a parallel notion of form factors that can be used to evaluate Boltzmann collision terms in soliton gases? Thus, again, we obtain a different viewpoint: the Boltzmann equation, a kinetic equation, is re-interpreted as a hydrodynamic equation, with terms that break the infinitely-many conservation laws admitted by free particles. This re-interpretation has, potentially, far-reaching consequences, which still need to be addressed.


\section*{Acknowledgments}
This work has been partially supported by the Agence Nationale de la Recherche through the LABEX CEMPI project (ANR-11-LABX-0007), the SOGOOD project (SOGOOD ANR-21-CE30-0061), the Ministry of Higher Education and Research, Hauts de France council and European Regional Development Fund (ERDF) through the Nord-Pas de Calais Regional Research Council and the European Regional Development Fund (ERDF) through the Contrat de Projets Etat-R\'egion (CPER Wavetech). 
The authors would like to thank the Isaac Newton Institute for Mathematical Sciences for support and hospitality during the programme ``Dispersive hydrodynamics: mathematics, simulation and experiments, with applications in nonlinear waves''  when part of the work on this paper was undertaken. 
This work was  supported by EPSRC  Grant Number EP/R014604/1.
GE's  work was also supported by EPSRC  Grant Number EP/W032759/1, and BD's work was supported by EPSRC Grant Number EP/W010194/1. 
This work also has received funding from the European Union's Horizon 2020 research and innovation programme under the Marie Skłodowska-Curie grant agreement No. 101033047 (to AG). 
The work of DA was supported by the state assignment of IO RAS, Grant FMWE-2021-0003. 
The authors thank Giacomo Roberti, Thibault Bonnemain, Thibault Congy, Alex Tovbis and Francois Copie for fruitful discussions. The authors also thank Francois Copie for the Fig. 1.


%

\end{document}